\documentclass[twocolumn,times,numberedappendix,appendixfloats]{aastex61}

\usepackage{times,graphicx}
\usepackage{color}
\usepackage{tabularx}

\DeclareGraphicsExtensions{.pdf,.png,.jpg}

\newcommand{\besancon}{Besan{\c c}on }
\newcommand{\euclid}{\emph{Euclid}}
\newcommand{\wfirst}{\emph{WFIRST}}
\newcommand{\afta}{\emph{AFTA}}
\newcommand{\kepler}{\emph{Kepler}}
\newcommand{\hst}{\emph{HST}}

\newcommand{\mabuls}{{\sc{gulls}}}

\newcommand{\thetae}{\theta_{\mathrm{E}}}
\newcommand{\re}{r_{\mathrm{E}}}

\newcommand{\dl}{D_{\mathrm{l}}}
\newcommand{\ds}{D_{\mathrm{s}}}

\newcommand{\tein}{t_{\mathrm{E}}}
\newcommand{\murel}{\mu_{\mathrm{rel}}}
\newcommand{\tzero}{t_{\mathrm{0}}}

\newcommand{\uzero}{u_{\mathrm{0}}}

\newcommand{\zone}{z_{\mathrm{1}}}

\newcommand{\blendfs}{f_{\mathrm{s}}}

\newcommand{\texp}{t_{\mathrm{exp}}}
\newcommand{\tohead}{t_{\mathrm{ohead}}}

\newcommand{\msun}{M_{\odot}}
\newcommand{\mearth}{M_{\earth}}
\newcommand{\rearth}{R_{\earth}}

\newcommand{\dd}{\mathrm{d}}

\newcommand{\kms}{km~s$^{-1}$}

\newcommand{\mpl}{M_{\mathrm{p}}}

\newcommand{\percent}{\%}

\defcitealias{Penny2013}{P13}
\defcitealias{Penny2014b}{P14b}
\defcitealias{Green2012}{DRM report}
\defcitealias{Spergel2013}{AFTA report}

\begin{document}

\title{Predictions of the {\it WFIRST} Microlensing Survey I: Bound Planet Detection Rates}

\author[0000-0001-7506-5640]{Matthew T. Penny}
\affiliation{Department of Astronomy, The Ohio State University, 140 W. 18th Avenue, Columbus, OH 43210, USA}

\author[0000-0003-0395-9869]{B. Scott Gaudi}
\affiliation{Department of Astronomy, The Ohio State University, 140 W. 18th Avenue, Columbus, OH 43210, USA}

\author{Eamonn Kerins}
\affiliation{Jodrell Bank Centre for Astrophysics, Alan Turing Building, University of Manchester, Manchester M13 9PL, UK}

\author[0000-0001-5069-319X]{Nicholas J. Rattenbury}
\affiliation{Department of Physics, University of Auckland, Private Bag 92019, Auckland, New Zealand}

\author[0000-0001-8317-2788]{Shude Mao}
\affiliation{Physics Department and Tsinghua Centre for Astrophysics, Tsinghua University, Beijing 100084, China}
\affiliation{National Astronomical Observatories, Chinese Academy of Sciences, A20 Datun Rd., Chaoyang District, Beijing 100012, China}
\affiliation{Jodrell Bank Centre for Astrophysics, Alan Turing Building, University of Manchester, Manchester M13 9PL, UK}

\author[0000-0001-8654-9499]{Annie C. Robin}
\affiliation{Institut Utinam, CNRS UMR 6213, OSU THETA, Universit\'e Bourgogne-Franche-Comt\'e, 41bis avenue de l'Observatoire, F-25000 Besan{\c c}on, France}

\author[0000-0002-7669-1069]{Sebastiano Calchi Novati}
\affiliation{IPAC, Mail Code 100-22, Caltech, 1200 East California Boulevard, Pasadena, CA 91125, USA}

\date{\today}

\begin{abstract}
The {\it Wide Field InfraRed Survey Telescope} (\wfirst) is the next NASA astrophysics flagship mission, to follow the {\it James Webb Space Telescope} ({\it JWST}). 
The \wfirst\ mission was chosen as the top-priority large space mission of the 2010 astronomy and astrophysics decadal survey in order to achieve three primary goals: to study dark energy via a wide-field imaging survey, to study exoplanets via a microlensing survey, and to enable a guest observer program. 
Here we assess the ability of the several \wfirst\ designs to achieve the goal of the microlensing survey to discover a large sample of cold, low-mass exoplanets with semimajor axes beyond roughly one AU, which are largely impossible to detect with any other technique.
We present the results of a suite of simulations that span the full range of the proposed \wfirst\ architectures, from the original design envisioned by the decadal survey, to the current design, which utilizes a 2.4-m telescope donated to NASA.
By studying such a broad range of architectures, we are able to determine the impact of design trades on the expected yields of detected exoplanets.
In estimating the yields we take particular care to ensure that our assumed Galactic model predicts microlensing event rates that match observations, consider the impact that inaccuracies in the Galactic model might have on the yields, and ensure that numerical errors in lightcurve computations do not bias the yields for the smallest mass exoplanets.
For the nominal baseline \wfirst\ design and a fiducial planet mass function, we predict that a total of ${\sim}1400$ bound exoplanets with mass greater than ${\sim}0.1~M_{\oplus}$ should be detected, including ${\sim}200$ with mass ${\lesssim}3~M_{\oplus}$. 
\wfirst\ should have sensitivity to planets with mass down to ${\sim}0.02~M_{\oplus}$, or roughly the mass of Ganymede.
\end{abstract}

\section{Introduction}\label{intro}
The study of the demographics of exoplanets, the end result of the planet formation process, has entered a statistical age. Large samples of transiting planets from \kepler~\citep[e.g.,][]{Thompson2018}, massive planets at small to moderate separations from ground-based radial velocity surveys of planetary systems in the solar neighborhood~\citep[e.g.,][]{Udry2007,Winn2015}, and direct imaging studies of young planets at large separations~\citep[e.g.,][]{Bowler2016}, are beginning to reveal the complex distribution of exoplanets as a function of mass and separation from their host stars, and the properties of the host stars themselves. 

Data from the \kepler\ mission has revealed a sharp rise in the occurrence rate of hot and warm planets as radius decreases down to about $2.8\rearth$, before leveling off~\citep[e.g.,][]{Howard2010,Fressin2013}. Precise spectroscopic measurements of \kepler's super Earth hosts has revealed a radius dichotomy between large and small super Earths~\citep{Fulton2017} that is likely due to atmospheric stripping~\citep{Owen2017}.
At large orbital distances ${\gtrsim}10$~AU, direct imaging searches have found young, massive planets to be present, but rare~\citep[e.g.,][]{Nielsen2010,Chauvin2015,Bowler2015}.
However, there remains a large area of the exoplanet parameter space -- orbits beyond ${\sim}1$~AU and masses less than that of Jupiter -- that remains relatively unexplored by transit, radial velocity, and direct imaging techniques.

Indeed, if every planetary system resembled our own, only a handful of planets would have been discovered by the radial velocity, transit, or direct imaging techniques to date. This fact begs the question: Is our solar system architecture rare?  If so, why?

To obtain a large sample of exoplanets beyond 1~AU and across a large range of masses requires a different technique. Gravitational microlensing enables a statistical survey of exoplanet populations beyond $1$~AU, because its sensitivity peaks at the Einstein radius of its host stars~\citep{Mao1991,Gould1992,Bennett1996}. For stars along the line of sight to the Galactic bulge (where the microlensing event rate is highest) the physical Einstein radius is typically $2$--$3$~AU~\citep[see, e.g., the review of][]{Gaudi2012}. Thanks to the fact that microlensing is sensitive directly to a planet's mass and not its light or effect on a luminous body, the techniques sensitivity extends out to all orbital radii beyond ${\sim}1$~AU~\citep{Bennett2002}.

Perhaps the most important reason to perform a large exoplanetary microlensing survey is that it opens up a large new region of parameter space. 
The history of exoplanet searches has been one of unexpected discoveries. At every turn, when a new area of parameter space has been explored, previously unexpected planetary systems have been found. This process began with the pulsar planets~\citep{Wolszczan1992} and relatively short-period giant planets discovered by the first precision radial velocity searches sensitive to planets~\citep{Mayor1995, Campbell1988, Latham1989}. As radial velocity surveys' sensitivities and durations grew, highly eccentric massive planets and low mass Neptunes and ``Super Earth'' planets were discovered~\citep[e.g.,][]{Naef2001, Butler2004, Rivera2005}.
When originally conceived, and with the Solar System as a guide, the \kepler\ mission aimed to detect potentially habitable Earth-sized planets in ${\sim}1$~AU orbits around Solar-like stars~\citep{Borucki2003}. 
Were all exoplanet systems like our own, \kepler\ would have found few or no planets~\citep{Burke2015}, and those that it did would have been at the limit of its signal-to-noise ratio. This result was obviously preempted by the discovery of hot Jupiters, which demonstrated conclusively that {\it not} all planetary systems have architectures like our own. 
\kepler\ itself has gone on to discover thousands of planetary systems very unlike ours, including tightly-packed multiplanet systems \citep[e.g.,][]{Lissauer2011} and circumbinary planets \citep[e.g.,][]{Doyle2011}, to name but a few examples. Even moving into unprobed areas of the host mass parameter space has revealed unexpected systems such as TRAPPIST-1~\citep{Gillon2016} and KELT-9~\citep{Gaudi2017}.
Direct imaging searches have revealed young, very massive planets that orbit far from their hosts~\citep[e.g.,][]{Chauvin2004}, the most unusual (from our solar-system-centric viewpoint) being the four-planet system around HR8799~\citep{Marois2008,Marois2010}.

Despite finding the planetary system that is arguably most similar to our own \citep[the OGLE-2006-BLG-109 Jupiter-Saturn analogs][]{Gaudi2008, Bennett2010}, ground-based microlensing surveys too have discovered unexpected systems. 
For example, microlensing searches have found several massive planets around M-dwarf stars~\citep[e.g.][]{Dong2009a} that appear to at least qualitatively contradict the prediction of the core accretion theory that giant planets should be rare around low-mass stars~\citep{Laughlin2004}. Measurements of planet occurrence rates from microlensing also superficially appear to contradict previous radial velocity results, although a more careful analysis indicates that the microlensing and radial velocity results are consistent~\citep{Montet2014, Clanton2014, Clanton2014a}. Other notable microlensing discoveries include circumbinary planets~\citep{Bennett2016}, planets on orbits of ${\sim}1$--$10$~AU around components of moderately wide binary stars~\citep[e.g.,][]{Gould2014,Poleski2014}, planets on wide orbits comparable to Uranus and Neptune~\citep[e.g.,][]{Poleski2017}, and planets orbiting ultracool dwarfs~\citep[e.g.][]{Shvartzvald2017}.

Having spent over a decade conducting two-stage survey-plus-follow-up planet searches~\citep[see, e.g.,][for a review]{Gould2010}, microlensing surveys have entered a second generation mode, that relies only on survey observations. The OGLE~\citep{Udalski2015-ogleiv}, MOA~\citep{Sako2007} and three KMTNet telescopes~\citep{Kim2016} span the southern hemisphere and provide continuous high-cadence microlensing observations over tens of square degrees every night that weather allows. Such global, second generation, pure survey-mode microlensing surveys will enable the initial promise of microlensing to provide the large statistical samples of exoplanets necessary to study demographics~\citep{Henderson2014-kmt}, and have begun to deliver~\citep{Shvartzvald2016,Suzuki2016}. It has long been recognized~\citep[e.g.,][]{Bennett2002,Beaulieu2008,Gould2009ds} that exoplanet microlensing surveys are best conducted from space, thanks to the greater ability to resolve stars in crowded fields and to continuously monitor fields without interruptions from weather or the day-night cycle.

\wfirst~\citep{Spergel2015} is a mission conceived of by the 2010 decadal survey panel~\citep{nwnh} as its top priority large astrophysics mission. It combines mission proposals to study dark energy with weak lensing, baryon acoustic oscillations, and supernovae~\citep[Joint Dark Enegry Mission-Omega, JDEM-Omega, ][]{Gehrels2010}, with a gravitational microlensing survey~\citep[Microlensing Planet Finder,][]{Bennett2010mpf}, a near infrared sky survey~\citep[Near Infrared Sky Surveyor,][]{Stern2010} and a significant guest observer component~\citep{nwnh}. The later addition of a high-contrast coronagraphic imaging and spectroscopic technology demonstration instrument~\citep{Spergel2015} addresses a top medium scale 2010 decadal survey priority as well. In this paper we examine only the microlensing survey component of the mission.

The structure of the paper is as follows. In \autoref{wfirstintro} we describe the \wfirst\ mission, and each of its design stages. In \autoref{sims} we describe the simulations we have performed. \autoref{baselineyield} presents the yields of the baseline simulations, while \autoref{tradeoffs} considers the effects of various possible changes to the mission design. \autoref{discuss} discusses the uncertainties that affect our results and how they might be mitigated by future observations, modeling, and simulations. \autoref{conclusions} gives our conclusions.

\section{{\it WFIRST}}\label{wfirstintro}

\subsection{Goals of the \wfirst\ Microlensing Survey}\label{goals} 

A primary science objective of the \wfirst\ mission is to conduct a statistical census of exoplanetary systems, from 1~AU out to free-floating planets, including analogs to all of the Solar System planets with masses greater than Mars, via a microlensing survey. It is in the region of ${\sim}1$--$10$~AU that the microlensing technique is most sensitive to planets over a wide range of masses~\citep{Mao1991,Gould1992}, and where other planet detection techniques lack the sensitivity to detect low-mass planets within reasonable survey durations or present-day technological limits. However, the $1$--$10$~AU region is perhaps the most important region of protoplanetary disks and planetary systems for determining their formation and subsequent evolution, and can have important effects on the habitability of planets. 

The enhancement in surface density of solids at the water ice line, ${\sim}1.5$--$4$~AU from the star, is thought to be critical for the formation of giant planets~\citep{Hayashi1981,Ida2004,Kennedy2006}. Nevertheless, all stars do not produce giant planets that survive~\citep[e.g.,][]{Winn2015}. It remains to be seen whether this is due to inefficient production of giant planets, or a formation process that is ${\sim}100$\% efficient followed by an effective destruction mechanism, such as efficient disk migration~\citep[e.g.,][]{Goldreich1980}, or ejection or host star collisions caused by dynamics~\citep[e.g.][]{Rasio1996}. In the core accretion scenario~\citep[e.g.,][]{Goldreich1973,Pollack1996}, runaway gas accretion onto protoplanet cores to produce giant planets is an inevitability, {\it if} the cores grow large enough before the gas dissipates from the protoplanetary disk~\citep{Mizuno1980}. If core growth rate is the rate limiting step in the production of giant planets, and the process is indeed inefficient, then we can expect a population of ``failed cores'' of various masses with a distribution that peaks near the location of the ice line. Conversely, if giant planet formation and subsequent destruction is efficient, we can expect the formed giant planets to clear their orbits of other bodies, and thus would expect to see a deficit of low mass planets in \wfirst's region of sensitivity.

It is clear that planetary systems are not static, and the orbits of planets can evolve during and after the planet formation process, first via drag forces while the protoplanetary disk is in place~\citep{Lin1996}, and subsequently by $N$-body dynamical processes once the damping effect of the disk is removed~\citep{Rasio1996}. In addition to rearranging the orbits of planets that remain bound, the chaotic dynamics of multiplanet systems can result in planets being ejected~\citep[e.g.,][]{Safronov1972}. The masses and number of ejected planets from the system will be determined by the number of planets in their original systems, their masses and orbital distribution~\citep[e.g,][]{Papaloizou2001,Juric2008,Chatterjee2008,Barclay2017}. So, the mass function of ejected, or free-floating, planets can be an important constraint on the statistics of planetary systems as a whole. Because they emit very little light, only microlensing observations can be used to detect rocky free-floating planets. For masses significantly below Earth's, only space-based observations can provide the necessary combination of photometric precision, cadence, and total number of sources monitored in order to collect a significant sample of events.

For both bound and free-floating objects it is valuable to extend the mass sensitivity of an exoplanet survey down past the characteristic mass scales of planet formation theory, where the growth behavior of forming planets changes. This is particularly the case for boundaries in mass between low and high growth rates, as these should be the locations of either pile-ups or deficits in the mass function, depending on the sense of the transition. In the core accretion scenario of giant planet formation, moving from high to low masses, characteristic mass scales include the critical core mass for runaway gas accretion at ${\sim}10\mearth$~\citep{Mizuno1980}, the isolation mass of planetary embryos at ${\sim}0.1\mearth$~\citep{Kokubo2002}, and the transition core mass for pebble accretion at ${\sim}0.01\mearth$~\citep{Lambrechts2012}. The detection of features due to these characteristic mass scales would be strong evidence in support of current planet formation theory. Additionally, a statistical accounting of planets on wider orbits more generally will be a valuable test of models of planet formation developed to explain the large occurrence rate of super-Earths closer than 1~AU.

An estimate of the occurrence rate of rocky planets in the habitable zones~\citep[e.g.,][]{Kasting1993,Kopparapu2013} of solar-like stars, $\eta_{\oplus}$, is an important ingredient for understanding the origins and evolution of life, and the uniqueness of its development on Earth. However, it is precisely this location that it is both hardest to detect ${\sim}1\mearth$ planets around ${\sim}1\msun$ stars, while also remaining tantalizingly achievable. Tiny signals recurring on approximately year timescales mean that transit, radial velocity, and astrometric searches must run for multiple years to make robust detections. Only the transit technique has demonstrated the necessary precision to date, and even so, \kepler\ fell just short of the mission duration necessary to robustly measure $\eta_{\oplus}$~\citep{Burke2015}. Direct imaging of habitable exoplanets will require significant technological advances~\citep[e.g.,][]{Mennesson2016,Bolcar2017,Wang2018}, several of which \wfirst's coronagraphic instrument will demonstrate, in order to reach the contrast and inner working angles required, and the observing time required to perform a blind statistical survey to measure $\eta_{\oplus}$ may be prohibitively expensive. The typical inner sensitivity limit for a space-based microlensing survey, which is proportional to the host mass $M^{1/2}$, crosses the habitable zone, which scales as ${\sim}M^{3.5}$, at ${\sim}1\msun$. Nevertheless the detection efficiency for low-mass-ratio planets inside the Einstein ring ($R_{\rm E}\sim 3$~AU) is very small, and solar-mass stars make up only a small fraction of the lens population, so large, long-duration microlensing surveys from space are required to robustly measure $\eta_{\oplus}$ with microlensing. In all likelihood, no one technique will prove sufficient, and it will be necessary to combine measurements from multiple techniques to be confident in the accuracy of $\eta_{\oplus}$ determinations. If the habitable zone is extended outward~\citep[e.g.,][]{Seager2013}, by volcanic outgassing of of H$_2$~\citep{Ramirez2017} or some other process, the number of habitable planets space-based microlensing searches are sensitive to increases significantly.

Each of the goals described above can be addressed in whole or in part by studying the statistics of a large sample of planets with orbits in the range of $1$--$10$ AU, and a similar sample of unbound planets. Such a sample can only be delivered by a space-based microlensing survey. Astrometry from {\it Gaia} can be used to discover a large sample of giant planets in similar orbits, but it will not have the sensitivity to probe below ${\sim}30\mearth$~\citep{Perryman2014}. Space-based transit surveys have sensitivity to very small, and low-mass planets, but would be required to observe for decades to cover the same range of orbital separations as does microlensing. Ground-based microlensing searches have sensitivity to low-mass exoplanets down to ${\sim}1\mearth$, as recently demonstrated by \citet{Bond2017} and \citet{Shvartzvald2017}, but are limited from gathering large samples of such planets or extending their sensitivity to masses significantly smaller than this by a combination of the more limited photometric precision possible from the ground, the larger angular diameter of source stars for which high precision is possible, and the lower density of such sources on the sky.

A critical element in measuring the mass function of planets from microlensing events is actually measuring planets masses. The lightcurve of a binary microlensing event alone only reveals information about the mass {\it ratio} $q$, unless the event is long enough to measure the effect of annual microlensing parallax on the lightcurve, or the event is observed from two widely separated observers, e.g., a spacecraft such as {\it Spitzer}~\citep[e.g.,][]{Udalski2015-spitzer} or {\it Kepler}~\citep[see][for a review]{Henderson2016-k2c9}. It is also necessary to measure finite source effects in the lightcurve, but this is routinely achieved for almost all planetary microlensing events observed to date, and will likely be possible for the majority of \wfirst's planetary events~\citep[e.g.,][]{Zhu2014}. High resolution imaging enables an alternative method to measure the host and planet mass. Over time, the source and lens star involved in a microlensing event will separate, and the lens, if bright enough will be detectable, either as an elongation in the combined source-lens image, as a shift in the centroid of the pair as a function of color, or if moving fast enough, the pair will become resolvable~\citep[e.g.,][]{Bennett2007}. The measured separation between the stars, and the color and magnitude of the lens star can be combined with the measurement of the event timescale to uniquely determine the mass of the lens. A principle requirement of the \wfirst\ mission is the ability to make these measurements routinely for most events. This is made possible by the resolution achievable from space, and is also greatly aided by the fainter source stars that space-based observations enable. For \wfirst\ to make these measurements it is necessary that it observe the microlensing fields over a time baseline of $4$ or more years.

In this paper we will only address \wfirst's ability to measure the mass function of bound planets. The challenges and opportunities to detect free-floating planets, and planets in the habitable zone, differ somewhat from those for the general bound planet population. We have therefore elected to give them the full attention that they deserve in subsequent papers, rather than provide only the limited picture that would be possible in this paper.

\subsection{Evolution of the {\it WFIRST} Mission: Design Reference Missions, AFTA, and Cycle 7}

The \wfirst\ mission is in the process of ongoing design refinement, and has gone through four major phases so far. This paper presents analysis of each of these missions, even though some of these designs are no longer under active development. This is important for two reasons. First, we are documenting the quantitative simulations that have informed the \wfirst\ microlensing survey design process from the first science definition team. Second, each design represents a internally self consistent set of mission design parameters, that when evolved to a new mission design necessarily captures the majority of covariance between all of the possible design choices. These covariances are difficult to account for in simulations that might aim to investigate variations in individual parameters that by isolating their effects.

The first \wfirst\ design, the Interim Design Reference Mission (IDRM) was based directly on the \wfirst\ mission proposed by the decadal survey and described in the first \wfirst\ Science Definition Team's (SDT) interim report~\citep{Green2011}. This in turn was based on the design for the JDEM-Omega mission~\citep{Gehrels2010}. The IDRM consisted of an unobstructed 1.3-m telescope with a $0.294$~deg$^2$ near-infrared imaging channel with broadband filters spanning ${\sim}0.76-2.0$~$\mu$m, including a wide $1-2$~$\mu$m filter for the microlensing survey, and two slitless spectroscopic channels. 

The final report of the first \wfirst\ SDT~\citep{Green2012} presented two Design Reference Missions (DRM1 and DRM2). DRM1 was an evolution of the IDRM, adhering to the recommendation of the decadal survey to only use fully developed technologies. It improved on the IDRM by increasing the upper wavelength cutoff of the detectors to $2.4$~$\mu$m, and removing the two spectroscopic channels. The detectors and prism elements were added to the imaging channel to increase its field of view to $0.377$~deg$^2$.

DRM2 was a design intended to reduce the cost of the mission. This was done by reducing the size of the primary mirror to 1.1-m (in order to fit onto a less costly launch vehicle). It also switched to a larger format 4k$\times$4k detector to reduce the number of detectors while increasing the field of view to $0.585$~deg$^2$, at the cost of additional detector development. The larger field of view also allowed the mission duration to be shortened to 3 years instead of 5.

The \wfirst\ design process was disrupted in 2012 when NASA was gifted two $2.4$-m telescope mirrors and optical tube assemblies by another government agency. 
The value of these telescopes to the \wfirst\ mission was initially assessed in a report by \citet{Dressler2012}, and the mission designed around one of the $2.4$-m telescopes was dubbed \wfirst-AFTA (AFTA standing for Astrophysically Focussed Telescope Assets). A new SDT was assembled to produce an \afta\ DRM, which added a coronagraphic instrument channel to the mission~\citep{Spergel2013}. The design of the wide field instrument (WFI) also changed, requiring a finer pixel scale to sample the smaller point spread function (PSF) of the $2.4$-m telescope, and hence a smaller field of view of $0.282$~deg$^2$. Unlike the previous designs, the telescope has an obstructed pupil, so the PSF has significant diffraction spikes that the previous versions did not. The design also required a shorter wavelength cutoff of $2.0$~$\mu$m due to concerns about the ability to operate the telescope at low temperature required for $2.4 \mu$m observations. The final results of this design process were presented by \citet{Spergel2015}.

\wfirst\ entered the formulation phase (phase A) in early 2016. The AFTA design was adopted, and the mission reverted to its simpler naming of \wfirst. Extensive design and testing work has been conducted since formulation began. This includes a large amount of detector development, validation of the ability of the telescope to operate cooled, redesign of the wide field instrument, and consideration of various mission descopes. Very recently, WFIRST ended into the second phase of mission development (phase B).

The most significant update to the design affecting the microlensing survey is a more pessimistic accounting of the observatory's slew time performance compared to the AFTA design. Another important change was a rotation of the elongated detector layout by 90$\degr$. While the mission design continues to evolve, we present simulations here that most closely match the Cycle 7 design, and so throughout the paper we will refer to the design as \wfirst\ Cycle 7.

\newcommand{\scol}[1]{\multicolumn{2}{c}{#1}}

\begin{table*}
\caption{Adopted parameters of each mission design}
\begin{tabular}{lcccccccccc}
\hline
& \scol{IDRM} & \scol{DRM1} & \scol{DRM2} & \scol{AFTA} & \scol{\bf \wfirst\ Cycle 7}\\ 
\hline
Reference & \scol{\citet{Green2011}} & \scol{\citet{Green2012}} & \scol{\citet{Green2012}} & \scol{\citet{Spergel2015}} & \scol{---$^{1,2}$}\\
Mirror diameter (m) & \scol{1.3} & \scol{1.3} & \scol{1.1} & \scol{2.36} & \scol{\bf 2.36} \\
Obscured fraction (area, \%) & \scol{0} & \scol{0} & \scol{0} & \scol{13.9} & \scol{\bf 13.9} \\
Detectors & \scol{$7{\times}4$ H2RG-10} & \scol{$9{\times}4$ H2RG-10} & \scol{$7{\times}2$ H4RG-10} & \scol{$6{\times}3$ H4RG-10}  & \scol{$\mathbf{6{\times}3}$ {\bf H4RG-10}} \\
Plate scale (``/pix) & \scol{0.18} & \scol{0.18} & \scol{0.18} & \scol{0.11} & \scol{\bf 0.11} \\
Field of view (deg$^2$) & \scol{0.294} & \scol{0.377} & \scol{0.587} & \scol{0.282} & \scol{\bf 0.282} \\
Fields & \scol{7} & \scol{7} & \scol{6} & \scol{10} & \scol{\bf 7} \\
Survey area (deg$^s$) & \scol{2.06} & \scol{2.64} & \scol{3.52} & \scol{2.82} & \scol{\bf 1.97} \\
Avg. slew and settle Time (s) & \scol{38} & \scol{38} & \scol{38} & \scol{38} & \scol{\bf 83.1} \\
\hline
Orbit & \scol{L2} & \scol{L2} & \scol{L2} & \scol{Geosynchronous} & \scol{L2} \\
Total Survey length (d) & \scol{432} & \scol{432} & \scol{266} & \scol{411$^{\ast\ast}$} & \scol{\bf 432}\\
Season length (d) & \scol{72} & \scol{72} & \scol{72} & \scol{72} & \scol{\bf 72} \\
Seasons & \scol{6} & \scol{6} & \scol{3.7} & \scol{6} & \scol{\bf 6}\\
Baseline mission duration (yr) & \scol{5} & \scol{5} & \scol{3} & \scol{6} & \scol{\bf 5} \\
Primary bandpass ($\mu$m) & \scol{$1.0$--$2.0$ (W149)} & \scol{$1.0$--$2.4$ (W169)} & \scol{$1.0$--$2.4$ (W169)} & \scol{$0.93$--$2.00$ (W149)} & \scol{\bf 0.93--2.00 (W149)} \\
Secondary bandpass ($\mu$m)  & \scol{$0.74$--$1.0$ (Z087)} & \scol{$0.74$--$1.0$ (Z087)} & \scol{$0.74$--$1.0$ (Z087)} & \scol{$0.76$--$0.98$ (Z087)} & \scol{\bf 0.76--0.98 (Z087)} \vspace{10pt}\\
 & W149 & Z087 & W169 & Z087 & W169 & Z087 & W149 & Z087 & {\bf W149} & {\bf Z087} \\
\hline
Zeropoint$^{\ast}$ (mag) & 26.315 & 25.001 & 26.636 & 24.922 & 25.990 & 24.367 & 27.554 & 26.163 & {\bf 27.615} & {\bf 26.387} \\
Exposure time (s) & 88 & 116 & 85 & 290 & 112 & 412 & 52 & 290 & {\bf 46.8} & {\bf 286} \\
Cadence & 14.98 min & 11.89 hr & 14.35 min & 12.0 hr & 15.0 min & 12.0 hr & 15.0 min & 12.0 hr & {\bf 15.16 min} & {\bf 12.0 hr} \\
Bias (counts/pix) & 380 & 380 & 1000 & 1000 & 1000 & 1000 & 1000 & 1000 & {\bf 1000} & {\bf 1000} \\
Readout noise$^{\star}$ (counts/pix) & 9.1 & 9.1 & 7.6 & 4.2 & 9.1 & 9.1 & 8.0 & 8.0 & {\bf 12.12} & {\bf 12.12} \\
Thermal + dark$^{\dagger}$ (counts/pix/s) & 0.36 & 0.36 & 0.76 & 0.76 & 0.76 & 0.76 & 1.30 & 0.05 & {\bf 1.072} & {\bf 0.130} \\
Sky background$^{\ddagger}$ (mag/arcsec$^2$) & 21.48 & 21.54 & 21.53 & 21.48 & 21.52 & 21.50 & 21.47 & 21.50 & {\bf 21.48} & {\bf 21.55} \\
Sky background (counts/pix/s) & 2.78 & 0.79 & 3.57 & 0.77 & 1.99 & 0.45 & 3.28 & 0.89 & {\bf 3.43} & {\bf 1.04} \\
Error floor (mmag) & 1.0 & 1.0 & 1.0 & 1.0 & 1.0 & 1.0 & 1.0 & 1.0 & {\bf 1.0} & {\bf 1.0} \\
Saturation$^{\mathsection}$ ($10^3$ counts/pix) & 65.5 & 65.5 & 80 & 80 & 80 & 80 & 679 & 2037 & {\bf 679} & {\bf 679} \\
\hline
\end{tabular}
\\{\bf Notes:} Parameters listed in the table are those used in the main simulations whose results are described in Sections~\ref{baselineyield} and \ref{tradeoffs} and are not necessarily the same as described in the relevant \wfirst\ reports. Where parameters are incorrect, the impact they would have was judged to be too insignificant to justify a repeated run of the simulations with the correct parameters (see text for further justifications). For correct parameter values the reader should refer to the appropriate Science Definition Team (SDT) report, or the reference information currently listed on the \wfirst\ websites below.\\
$^1$\url{https://wfirst.gsfc.nasa.gov/science/WFIRST_Reference_Information.html}\\
$^2$\url{https://wfirst.ipac.caltech.edu/sims/Param_db.html}\\
$^{\ast}$Magnitude that produces $1$ count per second in the detector.\\
$^{\star}$Effective readout noise after multiple non-destructive reads. All values are inaccurate, as they depend on the chosen readout scheme. However, the readout noise will not be larger than the correlated double sampling readout noise of ${\sim}20$~e$^{-}$, which is still sub-dominant relative to the combination of zodiacal light and blended stars.\\
$^{\dagger}$Sum of dark current and thermal backgrounds (caused by infrared emission of the telescope and its support structures etc.).\\
$^{\ddagger}$Evaluated using zodiacal light model at a season midpoint; in our simulations we use a time dependent model of the Zodiacal background (see \autoref{mabulsimprovements} for details).\\
$^{\mathsection}$Effective saturation level after full exposure time. For the designs preceding AFTA, we assumed saturation would occur when the pixel's charge reached the full-well depth. For AFTA we assume that, thanks to multiple reads, useful data can be measured from pixels that saturate after two reads, so for a constant full well depth, the saturation level increases with exposure time.\\
$^{\ast\ast}$Accounts for time lost to moon angle constraints.
\label{designparams}
\end{table*}

\begin{figure}
\includegraphics[width=\columnwidth]{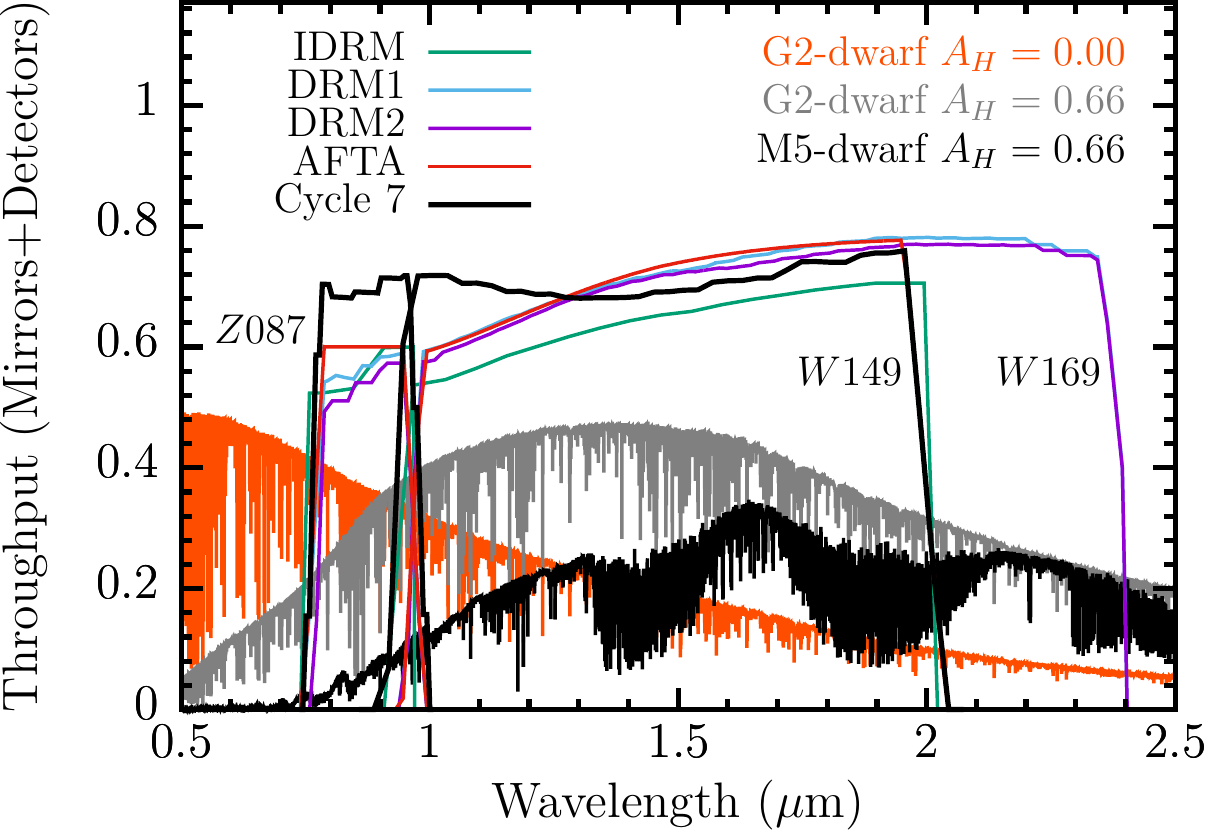}
\caption{Total throughput curves for each of the mission designs, compared to spectra of stars of different spectral types, suffering differing amounts of extinction. The spectrum of a $T_{\mathrm{eff}}=5800$~K, $\log g=4.5$ G-dwarf taken from the {\sc NEXTGEN} grid \citep{Hauschildt1999} is plotted with no extinction (orange line) and with $A_H=0.66$ (gray line), which is typical for the expected \wfirst\ fields shown in \autoref{fields}. G-dwarfs will be the bluest stars that will act as source stars in significant numbers, because more massive stars have evolved off the main sequence in the old bulge population. The $y$-axis units of the spectra are proportional to the photons per unit wavelength ($\dd N/\dd\lambda$), but each is arbitrarily normalized. The throughput curves show the total system throughput including detector quantum efficiency, and are only shown for the $Z087$ and $W149$ filters. The \wfirst\ microlensing survey will likely use a wider selection of filters than this.}
\label{bandpasses}
\end{figure}

\autoref{designparams} summarizes the parameters of each mission design that we study. We use these parameters for the results presented in \autoref{baselineyield} and \autoref{lowmass}. In \autoref{tradeoffs} we present the results of ``trade-off'' simulations that were conducted during the design process, when many of the mission parameters changed regularly. Between any given set of trade-off simulations, the exact values of many of the simulation parameters changed by small amounts. Rather than tediously detail each of these parameter changes that have little effect on the absolute yields, we will only indicate changes to parameters where they are important to each study. Invariably, these will be the independent variables of each study, or parameters closely related to these. As the unimportant parameters do not change internal to each trade-study, we compute differential yield measurements, i.e., yields relative to a fiducial design.

\subsection{The {\it WFIRST} Microlensing Survey}

The full operations concept for the \wfirst\ mission must ensure that the spacecraft can conduct all the observations necessary to meet its primary mission requirements, while maintaining sufficient flexibility to conduct a significant fraction of potential general observer observations. These considerations must feed into the spacecraft hardware requirements while simultaneously being constrained by practical design considerations in an iterative process. An example of an observing time line that results from this process in given in the \citet{Spergel2015} report.

While constructing a sample observing schedule for \wfirst\ is a complicated optimization task, requirements set by the nature of microlensing events significantly simplify the process for the microlensing survey. First, the microlensing event rate is highly concentrated towards the Galactic bulge, close to the ecliptic. This means that a spacecraft with a single solar panel structure parallel to its telescope's optical axis can only perform microlensing observations twice per year when the Galactic bulge lies perpendicular to the Sun-spacecraft axis. Second, microlensing events last roughly twice the microlensing timescale $\tein$, with $2\tein{\sim}60$~d, and planetary deviations last between an hour and a day~\citep[see, e.g.,][]{Gaudi2012}. This means that a microlensing survey must observe for at least $60$~days in order to characterize the whole microlensing event, while also observing at high cadence continually for periods ${\gtrsim}1$~day in order to catch and characterize microlensing events. To operate at maximum efficiency it should continuously observe for the entire duration of its survey windows. Finally, the survey requirement to detect ${\sim} 100$ Earth-mass planets combined with a detection efficiency of ${\sim}0.01$ per event and a microlensing event rate of a ${\sim}$few$\times10^{-5}$~yr$^{-1}$~star$^{-1}$ imposes a requirement of monitoring a ${\sim}$few$\times10^8$ star years over the duration of the survey.

The duration of planetary deviations places a requirement on the cadence of the microlensing observations. The timescale of planetary deviations is comparable to the Einstein crossing timescale of an isolated lens of the same mass, $\tein\approx 2$~hr~$\sqrt{M/\mearth}$, and the deviation must be sampled by several data points in order to robustly extract parameters. Furthermore, the duration of finite source effects in the lightcurve, which carry information about the angular Einstein radius $\thetae$, and also need to be resolved by several data points are ${\sim}1$~hr. Combined, these require an observing cadence of ${\sim}15$~min. If stars are well resolved, accurate photometry is possible for much of the bulge main sequence in exposures ${\sim}1$~minute on ${\sim}1$~m-class telescopes, and so it should be possible to observe between 5 and 10 fields within the cadence requirements if the observatory can slew fast enough.

The microlensing event rate is highest within a few degrees of the Galactic center, but these regions are also affected by a large amount of extinction. Observations in the near infrared drastically reduce the effect of the extinction. The \wfirst\ microlensing survey maximizes its photometric precision by using a wide (1--2~$\mu$m) bandpass for most of its observations, shown in \autoref{bandpasses}. Combining the wide filter with its wide ${0.28}$~deg$^{2}$ field of view, the current design of \wfirst\ can monitor a sufficient number of microlensing events with sufficient precision with ${\sim}400$~days of microlensing observations. Less frequent observations will be taken in more typical broadband filters, here we assume $Z087$ in order to measure the colors of microlensing source stars and to measure color-dependent centroid shifts for luminous lenses when the source and lens separate; the range of intrinsic $Z087-W149$ colors of stars is shown in \autoref{cmd} in \autoref{customfilters}. Note that \wfirst\ magnitudes are on the AB system~\citep{Oke1983}, and all magnitudes in this paper will be expressed in this system, unless denoted by a subscript Vega.

\begin{table}
  \caption{The \wfirst\ Microlensing Survey at a Glance}
  \begin{tabular}{ll}
    \hline
    Area & 1.96~deg$^2$\\
    Baseline & 4.5 years\\
    Seasons & $6\times 72$~days\\
    $W149$ Exposures & ${\sim}41,000$ per field\\
    $W149$ Cadence & $15$~minutes\\
    $W149$ Saturation & ${\sim}14.8$\\
    Phot. Precision & 0.01 mag @ $W149{\sim}21.15$\\
    $Z087$ Exposures & ${\sim}860$ per field\\
    $Z087$ Saturation & ${\sim}13.9$\\
    $Z087$ Cadence & $\lesssim12$~hours\\
    Stars ($W149<15$) & ${\sim}0.3\times10^{6}$\\
    Stars ($W149<17$) & ${\sim}1.4\times10^{6}$\\
    Stars ($W149<19$) & ${\sim}5.8\times10^{6}$\\
    Stars ($W149<21$) & ${\sim}38\times10^{6}$\\
    Stars ($W149<23$) & ${\sim}110\times10^{6}$\\
    Stars ($W149<25$) & ${\sim}240\times10^{6}$\\
    Microlensing events $|\uzero|<1$ & ${\sim}27,000$\\
    Microlensing events $|\uzero|<3$ & ${\sim}54,000$\\
    Planet detections ($0.1$--$10^{4}\mearth$) & ${\sim}1400$\\
    Planet detections ($<3\mearth$) & ${\sim}200$\\
    \hline
  \end{tabular}
      {\bf Notes}: Assumes the Cycle 7 design. Saturation estimates assumes the brightest pixel accumulates $10^5$ electrons before the first read. Star counts have been corrected for the \besancon\ model's under-prediction (see \autoref{starcounts}). The exposure time and cadence of observations in the $Z087$ and other filters has not been set; we have assumed a $12$ hour cadence here, but observations in the other filters are likely to be more frequent.
      \label{wfirstsummary}
\end{table}

\wfirst's microlensing survey therefore looks similar across all designs of the spacecraft, with $72$~continuous days of observations occurring around vernal and autumnal equinoxes. Six of these seasons are required, with three occurring at the start of the mission and three at the end in order to maximize the baseline over which relative source-lens proper motion can be measured~\citep[see, e.g.,][]{Bennett2007}. The 2.4~m telescope designs of \wfirst\ have a smaller field of view than the ${\sim}1$~m class designs, but can monitor significantly fainter stars at a given photometric precision due to their smaller PSF and larger collecting area, resulting in a similar number of fields being required to reach the same number of stars, despite the difference in field of view. After these two effects cancel, the designs with larger diameter mirrors come out as significantly more capable scientifically due to their improvement in ability to measure relative lens-source proper motions. \autoref{wfirstsummary} summarizes the parameters of the latest iteration of the \wfirst\ microlensing survey design and the survey yields that we will describe in later sections.

\section{Simulating the {\it WFIRST} microlensing survey}\label{sims}
We performed our simulations using the \mabuls\ code, of which we only give a brief overview here, and refer the reader to \citet[][hereafter \citetalias{Penny2013}]{Penny2013} for full details.\footnote{Note that in \citetalias{Penny2013} the software was called {\sc MaB$\mu$LS}, but was renamed to disambiguate it from the MaB$\mu$LS online tool~\citep{Awiphan2016}.} In order to fully simulate \wfirst\ we have made a number of upgrades to \mabuls, which are described in the Appendix.

\begin{figure*}
\includegraphics[width=\textwidth]{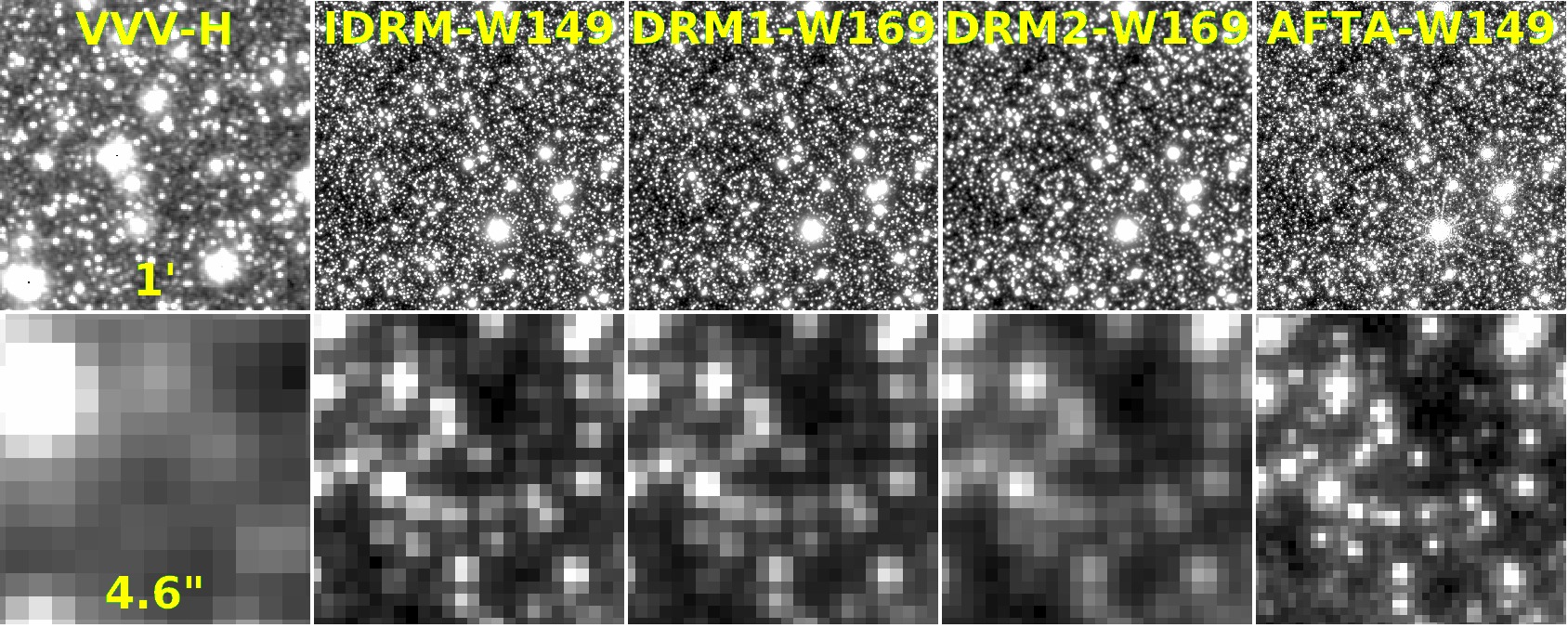}
\caption{\emph{Left column:} Section of a VVV $H$ band image \citep{Saito2012} from near $(\ell,b)=(1\fdg1,-1\fdg2)$, which lies close to the center of the expected \wfirst\ fields. \emph{Right four columns:} Simulated images in the primary wide band of the IDRM, DRM1, DRM2, and AFTA \wfirst\ designs of the same mock star field drawn from the \besancon\ model sight line at $(\ell,b)=(1\fdg1,-1\fdg2)$. The top panels show a $1{\times}1$~arcmin$^2$ region and the bottom panels show a $4.6{\times}4.6$ arcsec$^2$ ($\approx 13\times$) zoom-in. The pixel sizes are $0\farcs339$, $0\farcs18$, $0\farcs18$, $0\farcs18$ and $0\farcs11$ from left to right respectively. Note that the apparent dark, tenuous, serpentine feature on the left side of the simulated images is a result of random fluctuations in the stellar density, and is not due to spatially varying extinction (e.g., a dust lane). The VVV image based on data products from observations made with ESO Telescopes at the La Silla or Paranal Observatories under ESO programme ID 179.B-2002.}
\label{images}
\end{figure*}

\begin{figure}
\includegraphics[width=\columnwidth]{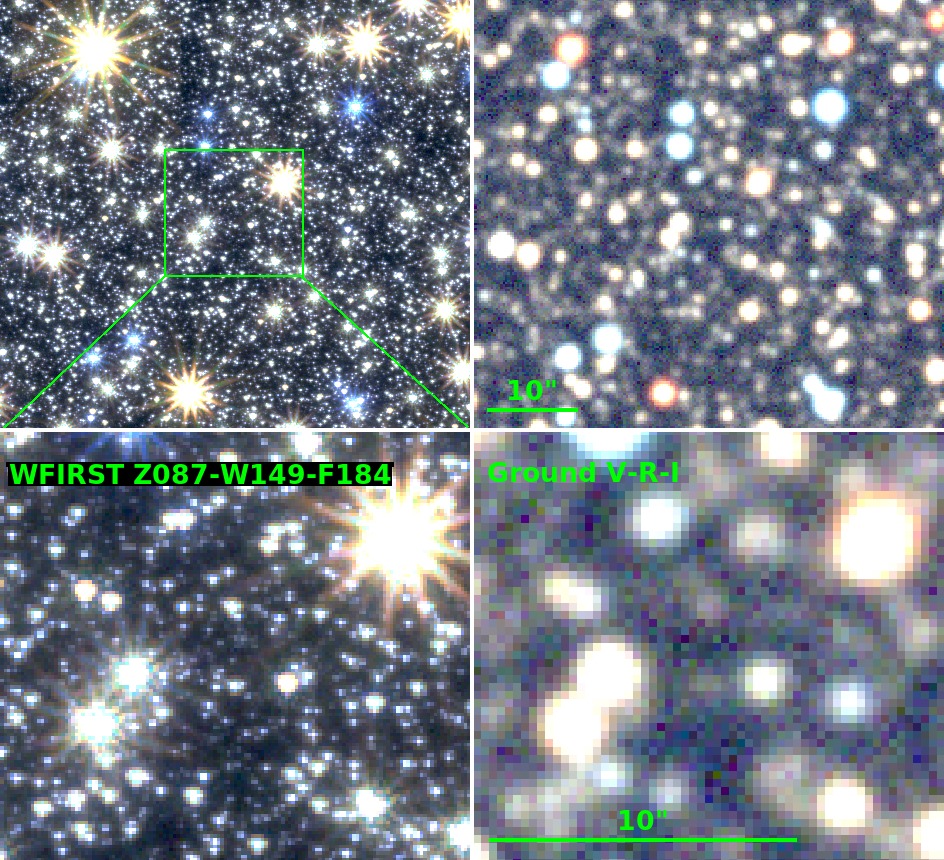}
\caption{Simulated color images of an example bulge field at $(\ell,b)=(0\fdg0,-1\fdg5)$ imaged using \wfirst's Cycle 7 detector, compared with a ground-based observatory based on OGLE's 1.3-m telescope \citep[e.g.,][]{Udalski2015-ogleiv} in optical filters. The \wfirst\ image is built from a single simulated exposure of $290$, $52$, and $145$~s in $Z087$, $W149$, and $F184$ filters, respectively; the OGLE image was built from single simulated exposures of $150$, $125$, and $100$~s in $V$, $R$, and $I$ filters, respectively, i.e., typical of the standard OGLE survey exposures. Note the different sizes of the images compared to the previous figure, and that at least some of the \wfirst\ fields will be amenable to observations with ground-based optical telescopes.}
\label{cycle7image}
\end{figure}

\mabuls\ simulates large numbers of individual microlensing events involving source and lens stars that are drawn from star catalogs produced by a population synthesis Galactic model. Source stars are drawn from a catalog with a faint magnitude limit (here $H_{\rm Vega}=25$), and lens stars from a catalog with no magnitude limit; source lens pairs where the distance of the source is less than the distance of the lens are rejected. Each catalog is drawn from a small solid angle $\delta\Omega$, but represents a larger $0\fdg25\times0\fdg25$ sight line at its specified Galactic coordinates $(\ell,b)$. The impact parameter $\uzero$ and time of the event $\tzero$ are drawn from uniform distributions with limits $[-u_{0,{\rm max}},+u_{0,{\rm max}}]$ and $[0,T_{\rm sim}]$, respectively, where $u_{0,{\rm max}}=3$ is the maximum impact parameter and $T_{\rm sim}$ is the simulation duration.

Each simulated event $i$ is assigned a normalized weight $w_i$ proportional to its contribution to the total event rate in the sight line
\begin{equation}
w_i = 0.25^2\text{deg}^{2} f_{1106,WFIRST} \Gamma_{{\rm deg}^2} T_{\rm sim} u_{0,{\rm max}} \frac{2 \mu_{{\rm rel},i} \theta_{{\rm E},i}}{W},
\label{rateweight}
\end{equation}
where $\Gamma_{{\rm deg}^2}$ is the event rate per square degree computed via Monte Carlo integration of the event rate using the source and lens catalogs~(see \citetalias{Penny2013} and \citealt{Awiphan2016} for details), $f_{1106,WFIRST}$ is the event rate scaling factor that we use to scale the event rate computed from the Galactic model to match measured event rates (see \autoref{ratecorr} for details),  $\mu_{{\rm rel},i}$ is the relative lens-source proper motion of simulated event $i$, $\theta_{{\rm E},i}$ is the angular Einstein radius of event $i$, and 
\begin{equation}
W = \sum_i 2 \mu_{{\rm rel},i} \theta_{{\rm E},i}
\end{equation}
is the sum of un-normalized ``event rate weights'' for all simulated events in a given sight line.
As such, the sum of $w_i$ for all events is simply the number of microlensing events we expect to occur in the sight line during the simulation duration with source stars matching the source catalog's selection criteria. Similarly, the prediction for the number of events matching a given criteria (e.g., a $\Delta\chi^2>160$ detection threshold due to a planetary deviation) is simply the sum of normalized weights of events that pass the cut, e.g.,
\begin{equation}
N(\Delta\chi^2>160) = \sum_i w_i H(\Delta\chi_i^2-160),
\end{equation}
where $H(x)$ is the Heaviside step function.

Binary (planetary) microlensing lightcurves are computed using a combination of the hexadecapole approximation~\citep{Pejcha2009,Gould2008}, contour integration~\citep{Gould1997, Dominik1998} and rayshooting~\citep[when errors are detected in the contour integration routines,][]{Kayser1986}. Realistic photometry of each event is simulated by constructing images of star fields (drawn from the same population synthesis Galactic model) for each observatory and filter considered, such that the same stars populate images with different pixel scales and filters. The PSF of the baseline source and lens stars are added at the same position on the image. As the event evolves, the source star brightness is updated and photometry is performed on the image for each data point. For some of the simulations we have implemented a faster photometry scheme that bypasses the need to create a realization of an image for each data point, and which is described in the Appendix. \autoref{images} shows examples of simulated images for IDRM, DRM1, DRM2, and AFTA designs compared to a ground-based IR image, and \autoref{cycle7image} shows an example simulated color image comparing \wfirst's performance to a simulated ground-based optical telescope in a field typical for the \wfirst\ microlensing survey.

Each star is added using realistic numerical PSFs that are integrated over the detector pixels for a range of sub-pixel offsets and stored in a lookup table for rapid access. For IDRM, DRM1 and DRM2, which all have unobstructed apertures, we used an Airy function averaged over the bandpass of the filter. For AFTA we used numerical PSFs produced using the {\sc ZEMAX} software package (D. Content priv. comm.). For the Z087 filter we used the monochromatic PSF computed at $1$~$\mu$m and for the wide filter we averaged the PSFs computed at $1.0$, $1.5$ and $2.0$~$\mu$m with equal weights. This crude integration procedure insufficiently samples the changing size of the Airy rings as a function of wavelength, so the resulting PSFs have much more prominent higher-spatial frequency rings than the actual PSF. The real PSF will be much smoother in the wings \citep[see, e.g.,][]{Gould2014}. The spacing of the unrealistic rings is smaller than the photometric aperture we use, so the inaccurate PSF will have little effect on our results because maxima and minima will average out over the aperture). For the Cycle 7 design we used well sampled numerical PSFs generated using the {\sc WebbPSF} tool~\citep{WebbPSF} with parameters from Cycle 5; while the diffraction spikes of these PSFs are rotated 90 degrees relative to the Cycle 7 design, this has no practical effect on our simulated results.

\begin{figure}
\includegraphics[width=\columnwidth]{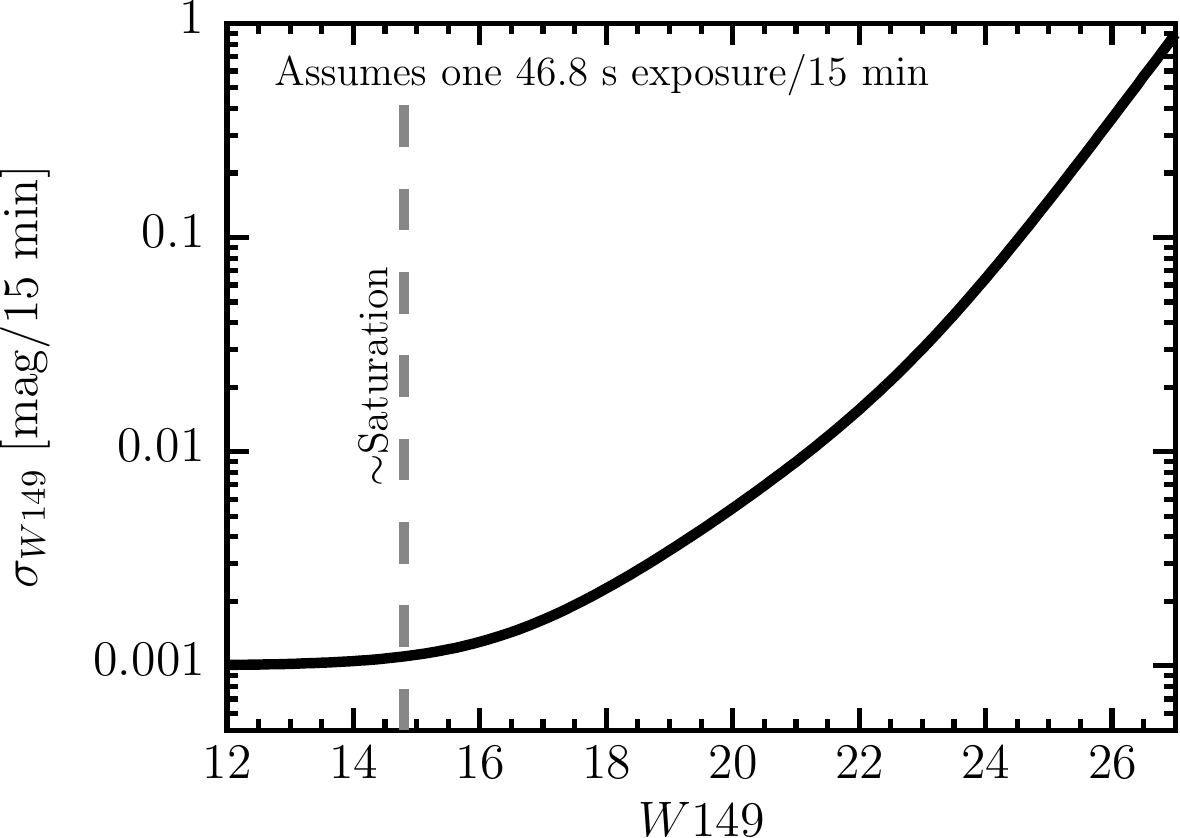}
\caption{Single epoch photometric precision for isolated point sources as a function of magnitude for the Cycle 7 design's assumed exposure time ($46.8$~s) assuming no blending. The vertical dashed line indicates the approximate point of saturation in a single read.}
\label{snrplot}
\end{figure}

The capabilities of crowded field photometric techniques are approximated by performing aperture photometry on the image with fixed pointing. We found that a $3{\times}3$~pixel square aperture produced the best results in the crowded \wfirst\ fields.
This simple photometry scheme enables us to accurately simulate all the sources of photon and detector noise that arise in the conversion of photons to data units on an image of minimal size. Aperture photometry is sub-optimal in crowded fields, but we can use this fact to compensate for the effect of any un-modeled causes of additional photometric noise resulting from imperfect data analysis (the precision of any method can only asymptotically approach the theoretically possible photon noise, and sometimes may be far from it), systematic errors (e.g., we do not simulate pointing shifts or variations in pixel response), or data loss. The resultant photometric precision as a function of magnitude is shown in \autoref{snrplot}, assuming no blending. Properly simulating all sources of systematic or red noise would require a more detailed simulation of the photometry pipeline (for example, by performing difference imaging on images that suffer pointing shifts) and would be significantly more computationally expensive as a significantly larger image would need to be recomputed for each data point. Rather than do this, we simply add in quadrature a Gaussian systematic error floor to the photometry we measure.

We note that we have not correctly simulated the read out schemes employed by the HAWAII HgCdTe detectors \wfirst\ will use. Our simulations simulate the CCD readout process, i.e., an image is exposed for a time $t_{\rm exp}$ before being read out pixel by pixel by a one or a small number of amplifiers in a destructive process. Individual pixels in an infrared HgCdTe array have their own amplifier and can be read out non destructively multiple times per exposure at a chosen rate. HgCdTe amplifiers typically have higher read noise than CCD amplifiers, and so multiple non-destructive reads are employed to reduce the effective read noise in the image. Multiple image ``frames'' can potentially be stored and downlinked, or processed on-board the spacecraft, enabling retrieval of useful data from pixels that would saturate in the full exposure time or that get hit by cosmic rays.

Our photometry simulations assume a single readout at the end of an exposure with a gain of $1$~e$^{-}$/ADU. The actual gain value will be different, but any digitization uncertainty will be small compared to the readout noise. We approximate the effective readout noise in \wfirst\ images using the erratum correction of the formula given by \citet{Rauscher2007}, based on the correlated double sampling readout noise requirements of each design, an assumed readout rate, and the exposure time. The full well depth parameter in our simulations is applied to the full exposure time, so we increase the detector full well depth requirement by a factor of $t_{\rm exp}/t_{\rm read}$, where $t_{\rm read}$ is the time interval between reads of a given pixel, to simulate the ability to extract a measurement from a pixel that does not saturate before the first read. This workaround results in an underestimate of the Poisson noise component of photometry, but the addition of a 0.001~mag systematic uncertainty in quadrature to the final photometric measurement involving 9 pixels prevents a severe underestimate of the uncertainty in such situations. We note that it should be possible to extract accurate photometry from any pixels that do not saturate before the first read~\citep[see][]{Gould2015}.

To assess whether a simulated event contains a detectable planet we use a simple $\Delta\chi^2$ selection criteria
\begin{equation}
\Delta\chi^2 \equiv \chi_{\rm FSPL}^2 - \chi_{\rm true}^2 > 160,
\end{equation}
where $\chi_{\rm FSPL}$ is the $\chi^2$ of the simulated data lightcurve relative to the best fitting finite source single point (FSPL) lens model lightcurve~\citep{Witt1994}, and $\chi_{\rm true}^2$ is the $\chi^2$ of the simulated data relative to the true simulated lightcurve. In practice we only fit a FSPL model if a point source point lens model fit produces a $\Delta\chi^2$ above the detection threshold. We do not consider whether the lightcurve can be distinguished from potentially ambiguous binary lens models or binary source models.

Our choice of $\Delta\chi^2$ threshold is the de facto standard among microlensing simulations~\citep[e.g.,][]{Bennett2002,Bennett2003,Penny2013,Henderson2014-kmt}. \citet{Yee2012} and \citet{Yee2013} discussed the issue of the detection threshold in survey data for high magnification events, and concluded that for one particular event a clear planetary anomaly in the full data set might be marginally undetectable in a truncated survey data set at $\Delta\chi^2\approx 170$. For a uniform survey data set, and a search that included low-magnification events, \citet{Suzuki2016} used a $\Delta\chi^2$ threshold of $100$. We expect systematic errors for a space based survey to be lower than for a ground-based survey, therefore our choice of $\Delta\chi^2=160$ should be reasonably conservative. Additionally, except near the edges of its survey sensitivity, the number of planet detections \wfirst\ can detect is only weakly dependent of $\Delta\chi^2$ as discussed in \autoref{tradeoffs}. This means that our yields will be relatively insensitive to any innaccuracies in our simulations or models that affect $\Delta\chi^2$. Additionally, because we have chosen relatively conservative assumptions for the systematic noise floor and the $\Delta\chi^2$ threshold, it is possible that the yield of the hardest-to-detect planets could be significantly larger than we predict.

For the smallest mass planets we do not expect binary lens ambiguity be an issue for many events. Most low-mass planet detections will come from planetary anomalies in the wings of low-magnification events. In such events the caustic location is well constrained and hence also the projected lens source separation $s$. Once $s$ is constrained, the caustic size and anomaly duration scales only with the mass ratio $q$ of the lens as $q^{1/2}$~\citep[e.g.,][]{Han2006}. Binary source stars with extreme flux ratios can potentially produce false positives for low-mass planetary microlensing~\citep{Gaudi1998}. As \wfirst\ will observe source stars much closer to the bottom of the luminosity function, and the near infrared luminosity function is shallower than the optical luminosity function, we can expect a smaller fraction of \wfirst's binary source stars to have the properties required to mimic a planetary microlensing event. We leave a detailed reassessment of the importance of binary source star false positives for \wfirst's microlensing survey to future work.

\subsection{Galactic Model}

In this work we use version 1106 of the \besancon\ Galactic model, hereafter BGM1106. This version of the model is described in full detail by \citetalias{Penny2013} and references therein. It is intermediate to the original, publicly available version~\citep{Robin2003}, and a more recent version~\citep{Robin2012}. It also differs from the model versions used by \citet{Kerins2009} and \citet{Awiphan2016} to compute maps of microlensing observables.

As the model has been detailed in other papers we only give an overview of the most important features here. The BGM1106 bulge is a boxy triaxial structure following the \citet{Dwek1995} $G_2$ model with scale lengths of $(1.63,0.51,0.39)$~kpc and orientated with the long axis $12\fdg5$ from the Sun-Galactic center line. The thin disk uses the \citet{Einasto1979} density law with a scale length of $2.36$~kpc for all but the youngest stars, which have a scale length of $5$~kpc. The disk has a central hole with a scale length of 1.31~kpc, except for the youngest stars where the hole scale length is $3$~kpc. The disk scale height is set by self consistency requirements between kinematics and Galactic potential~\citep{Bienayme1987}. The model also has thick disk and halo components, but they do not provide a significant fraction of sources or lenses. The full form of the density laws are given in \citet{Robin2003} Table 3. 

Stellar magnitudes are computed from stellar evolution models and model atmospheres based on stellar ages determined by separate star formation histories and metallicity distributions of the different components. Stellar masses are drawn from an initial mass function (IMF) that differs between the disk and bulge. Each is a broken power law, $\dd N/\dd M\propto M^{\alpha}$, where $M$ is the stellar mass, with $\alpha=-1.6$ for $0.079<M<1\msun$ and $\alpha=-3$ for $M>1\msun$ in the disk, and $\alpha=-1$ for $0.15<M<0.7\msun$ and $\alpha=-2.35$ for $M>0.7\msun$ in the bulge. Extinction is determined by the 3-d extinction map of \citep{Marshall2006}, expressed as measurements of $E(J-K)$ reddening at various distances and with a resolution of $0\fdg25\times 0\fdg25$ on the sky. Reddening is converted to extinction in other bands using the \citet{Cardelli1989} extinction law with a value of total to selective extinction $R_V=3.1$.

\subsection{Normalizing the event rate}\label{ratecorr}

\mabuls\ computes microlensing event rates by performing Monte Carlo integration of star catalogs produced by the population synthesis Galactic model. In \citetalias{Penny2013} we found that BGM1106 under-predicted the microlensing optical depth by a factor of $f_{\rm od,P13}=1.8$ and star counts in Baade's Window by a factor of $f_{\rm sc,P13}=1.3$. To account for this we applied a correction factor, 
\begin{equation}
f_{\rm 1106,P13}=f_{\rm od,P13}f_{\rm sc,P13}=1.8\times1.3=2.33
\end{equation}
to the BGM1106 event rates. Here we will update this event rate correction factor by making comparisons of the Galactic model predictions to new star count and microlensing event rate measurements.

\subsubsection{Comparison to Star Counts}\label{starcounts}

\citetalias{Penny2013} used a comparison between the BGM1106 and \hst\ star counts in Baade's window at $(\ell,b)=(1\fdg00,-3\fdg90)$ as measured by \citet{Holtzman1998} to derive a partial correction to the event rate of $f_{\rm sc,P13}=1.30$. This field lies more than 2 degrees further away from the Galactic plane than the center of the likely \wfirst\ fields at $b\approx-1\fdg7$. The Sagittarius Window Eclipsing Extrasolar Planet Search \citep[SWEEPS][]{Sahu2006} field, originally studied by \citet{Kuijken2002}, lies at $(\ell,b)=(1\fdg25,-2\fdg65)$ and is significantly closer to the \wfirst\ fields, but still slightly outside the nominal survey area. The field has been observed by \hst\ multiple times over a long time baseline, enabling extremely deep proper motion measurements~\citep{Clarkson2008,Calamida2015} and now star counts~\citep{Calamida2015}. By comparing star counts closer to the \wfirst\ fields we can hope to reduce the impact of any extrapolation errors when estimating an event rate correction.

\begin{figure}
\includegraphics[width=\columnwidth]{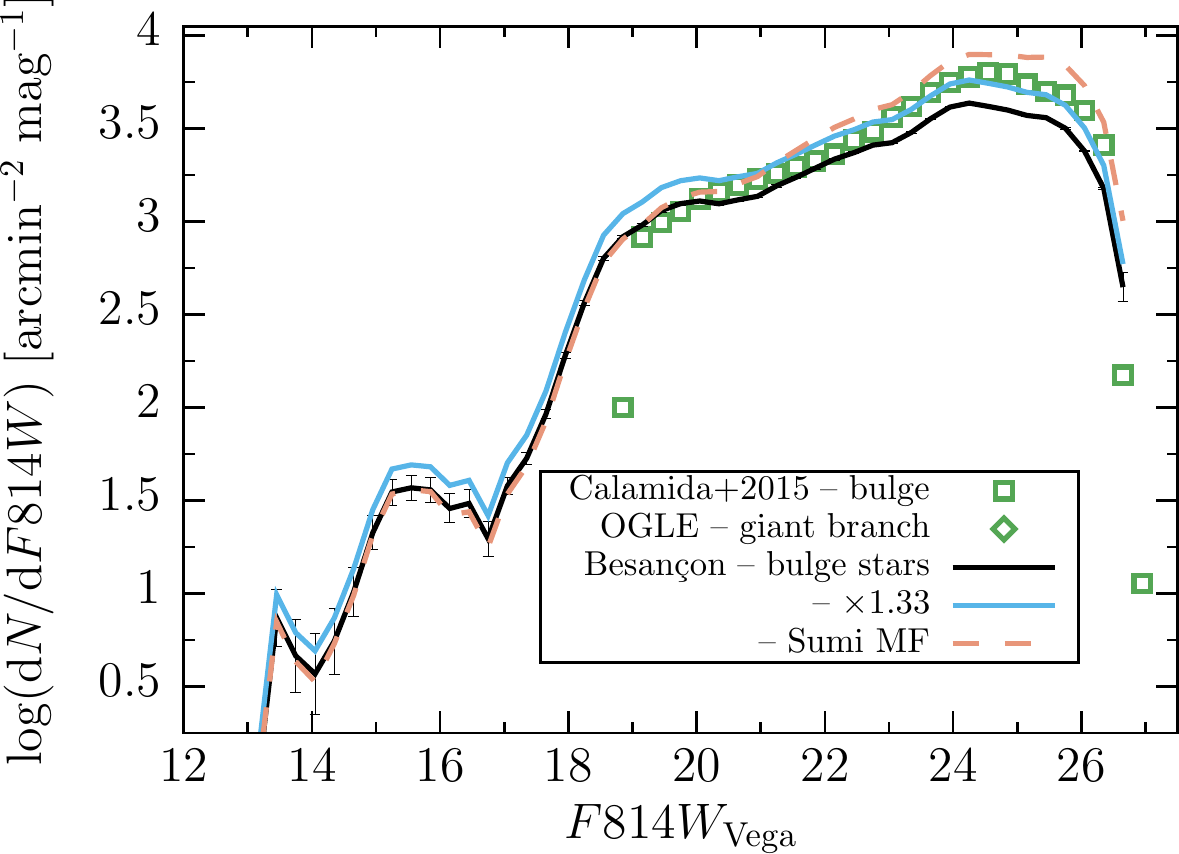}
\caption{Comparison of \besancon\ model star counts for the bulge population as a function of magnitude $F814W_{\rm Vega}$ at $(\ell,b)=(1\fdg35,-2\fdg70)$ to those measured in the {\it HST} SWEEPS field at $(\ell,b)=(1\fdg25,-2\fdg65)$, which lies close to the expected \wfirst\ fields. Green squares show bulge-only star counts from {\it HST}~\citet{Calamida2015} and diamonds show counts of red giant branch stars in the same area from OGLE-III~\citep{Szymanski2011}. The {\it HST} stars were selected to be bulge stars by proper motion cuts, and have been corrected for the approximate efficiency of this cut. The solid black line shows the BGM1106 prediction, with error bars denoting the Poisson uncertainty of the catalogs. While there are differences in the detailed shape of the star count distribution, integrated over the range $F814W=19$--$26.5$, BGM1106 under-predicts the total number of stars by $33$\%; the blue line shows the BGM1106 scaled up by this factor. The dashed black line shows the BGM1106 model star counts if the mass function is changed in the bulge to match mass function 1 of \citet{Sumi2011}, namely a broken power law with slopes of -1.3 and -2.0 ($\dd N/\dd M$) each side of a break at $0.7\msun$.}
\label{calamida}
\end{figure}

\citet{Calamida2015} measured the magnitude distribution of bulge stars by selecting stars with a proper motion cut designed to exclude disk stars. \citet{Calamida2015} correct for completeness using artificial star tests, but we add additional corrections for the efficiency of the proper motion cut ($34$~percent, A. Calamida, priv. comm.) and the field area ($3.3\arcmin\times3.3\arcmin$), in order to plot the absolute stellar density as a function of magnitude in \autoref{calamida}. We do not consider the bins at the extremes of the magnitude distribution which are likely affected by saturation or large incompleteness. To compare to the observed distribution, we computed the magnitude distribution of bulge stars in the BGM1106 sight line at $(\ell,b)=(1\fdg35,-2\fdg70)$, which is the closest to the SWEEPS field. 

BGM1106 matches the measured magnitude distribution reasonably well between $F814W_{\rm Vega}=19.5$ and $23$, though with minor differences in shape. BGM1106 starts to significantly underpredict the number of stars fainter than $F814W_{\rm Vega}=23$. Brighter than $F814W_{\rm Vega}=22.9$, \citet{Calamida2015} find 10~percent more stars than the BGM1106 predicts. Integrated over the magnitude range $F814W_{\rm Vega}=19$--$26.5$ BGM1106 under-predicts star counts by $33$~percent. The magnitude of the discrepancy is very similar to that we found between the BGM1106 and the \citet{Holtzman1998} luminosity function, giving us some confidence that there is no significant gradient in the BGM1106's star count discrepancy. We adopt the star count scaling factor of $f_{\rm sc}=1.33$.

The cause of the discrepancy between model and data can be partially explained by the BGM1106's choice of initial mass function (IMF) in the bulge, $\dd N/\dd M \propto M^{-1.0}$. Adopting a more reasonable mass function \citep[e.g., mass function number 1 from ][$\dd N/\dd M \propto M^{-2.0}$ for $M>0.7\msun$ and $\dd N/\dd M \propto M^{-1.3}$ for $0.08<M<0.7\msun$]{Sumi2011}, and assuming that the BGM1106 star counts were normalized using turn-off stars of $1.0\msun$, produces the luminosity function prediction shown by the dashed line in the plot. This mass function over-predicts the star counts fainter than $F814W_{\rm Vega}\approx 20$, but better matches the shape of the entire observed luminosity function between $I_{\rm Vega}\approx 19$--$26$. Both the original and modified BGM1106 mass functions slightly under-predict the number of giant branch star counts from the same sight line detected by OGLE~\citep{Szymanski2011}, which have not been corrected for incompleteness. We note here that we do not adopt an alternative mass function \citep[e.g, mass function 1 from][]{Sumi2011}, but discuss the impact of the mass function on our results in section~\ref{discuss}.

\subsubsection{Comparison to Microlensing Event Rates}\label{eventrates}

Since writing \citetalias{Penny2013}, \citet{Sumi2013} published measurements of the microlensing event rate towards the bulge, in addition to optical depth measurements. Measurements of the event rate per source star allow a more direct route to estimating any corrections to the model's predicted event rates, so here we only perform a comparison to the event rates and not the optical depths. For the comparison we use the event rates from \citet{Sumi2016}, which corrected the \citet{Sumi2013} event rates and optical depths for a systematic error in estimates of the number of source stars monitored.

\citet{Sumi2013} present event rates for two samples of events, the ``extended red clump'' (ERC) sample composed of events with source stars brighter than $I=17.5$ and colors selected to only include the bulge giant branch, and the ``all stars'' (AS) sample composed of all events with $I<20$ and no color cut. We selected star catalogs from the BGM1106 to match these samples, and computed event rates per source $\Gamma$ by Monte Carlo integration over this source catalog and a lens catalog with no magnitude or color cuts~\citep[see][for a detailed description of such calculations]{Awiphan2016}. The small angle of the Galactic bar to the line of sight in the BGM1106 (${\sim}12\degr$) results in the bulk of our AS sample source stars lying in front of most of the bulge stars. This leads to significantly smaller event rates per source than would be expected for a more reasonable bar angle of ${\sim}30\degr$~\citep[e.g.][]{Wegg2013,Cao2013}, which could lead us to over-correcting the event rates. We therefore only compare to the ERC sample event rates.

\begin{figure}
\includegraphics[width=\columnwidth]{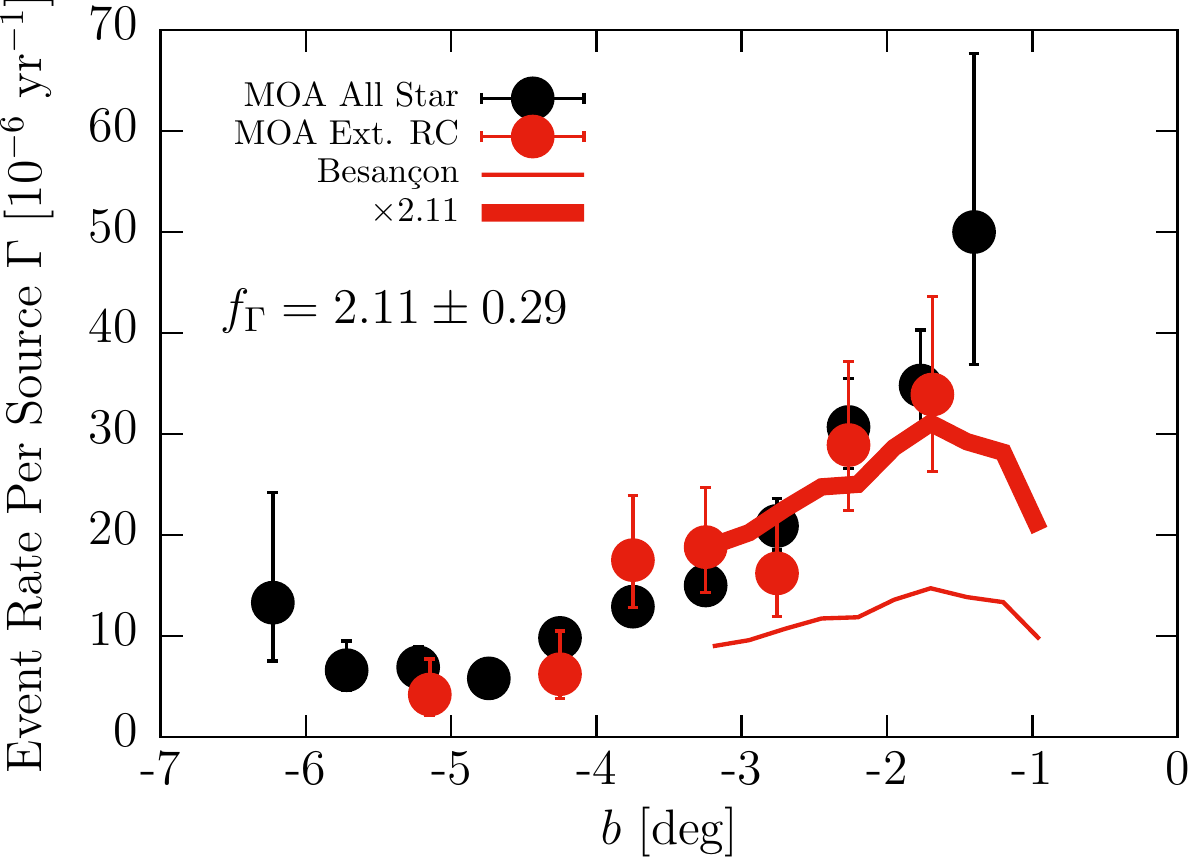}
\caption{Comparison of the microlensing event rate per source predicted by the Besan{\c c}on model to the \citep{Sumi2016} revision of measurements by \citep{Sumi2013}. Black data points show measurements for all source stars, while red data points show measurements for the extended red clump source stars (see text for details). The thin line shows the BGM1106's prediction of extended red clump event rates, and the thick red line shows this prediction after multiplication by the best fit scaling parameter $f_{\Gamma}=2.11\pm0.29$.}
\label{gammacomp}
\end{figure}

\autoref{gammacomp} shows the predicted model event rates, averaged over the range $-0.53<\ell\le +2.73$, and the data from \citet{Sumi2016}, which was averaged over the range $|\ell|<5$. BGM1106 predicts a lower event rate than is measured. At latitudes $|b|<1.5$ the predicted ERC event rate turns over likely because extinction begins to limit the range of distances over which significant numbers of bulge giants pass the ERC color and magnitude cuts; at more negative latitudes, where the observations we compare to were made, the extinction likely has a smaller impact. We find that multiplying the BGM1106 event rates by a constant scaling factor $f_{\Gamma} = 2.11 \pm 0.29$ yields a good match to the observed ERC rates, with $\chi^2=1.58$ for $3$ degrees of freedom. Although the model predictions cover a smaller range of $\ell$ than the measurements, \citet{Sumi2016} results binned by $\ell$ indicate only a relatively weak dependence of $\Gamma$ on $|\ell|$.

\subsubsection{Adopted Event Rate Scaling}\label{scaling}

In \citetalias{Penny2013} we scaled the microlensing event rates computed using the BGM1106 by making the assumption that all of the relevant distributions (e.g., kinematics, mass, and density) were reasonable, but that there could be errors in the normalization of the numbers of source and lens stars. To make a correction for the number of source stars we directly compared the BGM1106 predictions to deep star counts measured by HST. To estimate the correction for the number of lens stars, we compared model predictions to measurements of the microlensing optical depth. This has the advantage that, should the density distribution and mass function of stars in the model be reasonable, the necessary correction to the event rate due to lenses should scale with the optical depth discrepancy between model and data. However, as we have described in the preceding subsections and will expand on in \autoref{discuss}, the density distribution (specifically the bar angle of the bulge), the bulge mass function, and the kinematics of bulge stars in the BGM1106 are inconsistent with current measurements. These will affect the event rate and optical depth in different ways that are not trivial to calculate. This makes any simple scaling of the event rate based on optical depth comparisons suspect. In contrast, a scaling based on measured event rates is far more direct with fewer assumptions. Therefore, to correct the event rates predicted by the Galactic model, we adopt the event rate scaling factor
\begin{equation}
f_{1106,WFIRST} = f_{\rm sc} f_{\Gamma},
\end{equation}
where
\begin{equation}
f_{\rm sc} = 1.33, \text{and} f_{\Gamma}=2.11,
\end{equation}
for a total event rate correction of
\begin{equation}
f_{1106,WFIRST} = 2.81,
\end{equation}
which is about $20$\% larger than the scaling adopted in \citetalias{Penny2013}. We will discuss the impact of uncertainties and innaccuracies in the Galactic model beyond the event rate scaling in \autoref{discuss}.

We apply the $f_{1106,WFIRST}$ scaling throughout the paper as our fiducial event rate normalization. However, we will also present our main results with the scalings used for each of the \wfirst\ reports in order to aid comparison to these earlier works; these results using obsolete scalings are presented in \autoref{cassan_table_report} in \autoref{baselineyield}. We note, however, that when applying the obsolete scaling, we did not include a factor of $1.475$ in the scaling that was used in the \citet{Green2012}. This factor was used to account for a factor of 2.2 discrepancy in the microlensing detection efficiency of our \mabuls\ simulations and simulations performed by D. Bennett, based on \citet{Bennett2002} and updated for simulations of \wfirst~\citep{Green2011}; 1.475 was the geometric mean of the relative detection efficiencies for planets of $1\mearth$ with a period of $2$~years. The cause of the difference in detection efficiencies was not conclusively tracked down. However, at fixed period, the projected separation $s\propto M^{-5/6}$, and the detection efficiency is a strong function of $s$, so a difference between the host mass function of the simulations (see \autoref{starcounts} and \autoref{bulgeimf}) is likely to cause a significant difference in the detection efficiency at fixed period. Averaged over a range of semimajor axis, as we have done in the simulations presented in \autoref{baselineyield}, we can expect any difference in the detection efficiency at fixed planet mass to be significantly smaller.

\subsection{The {\it WFIRST} fields}\label{fields}

\begin{figure}
\includegraphics[width=\columnwidth]{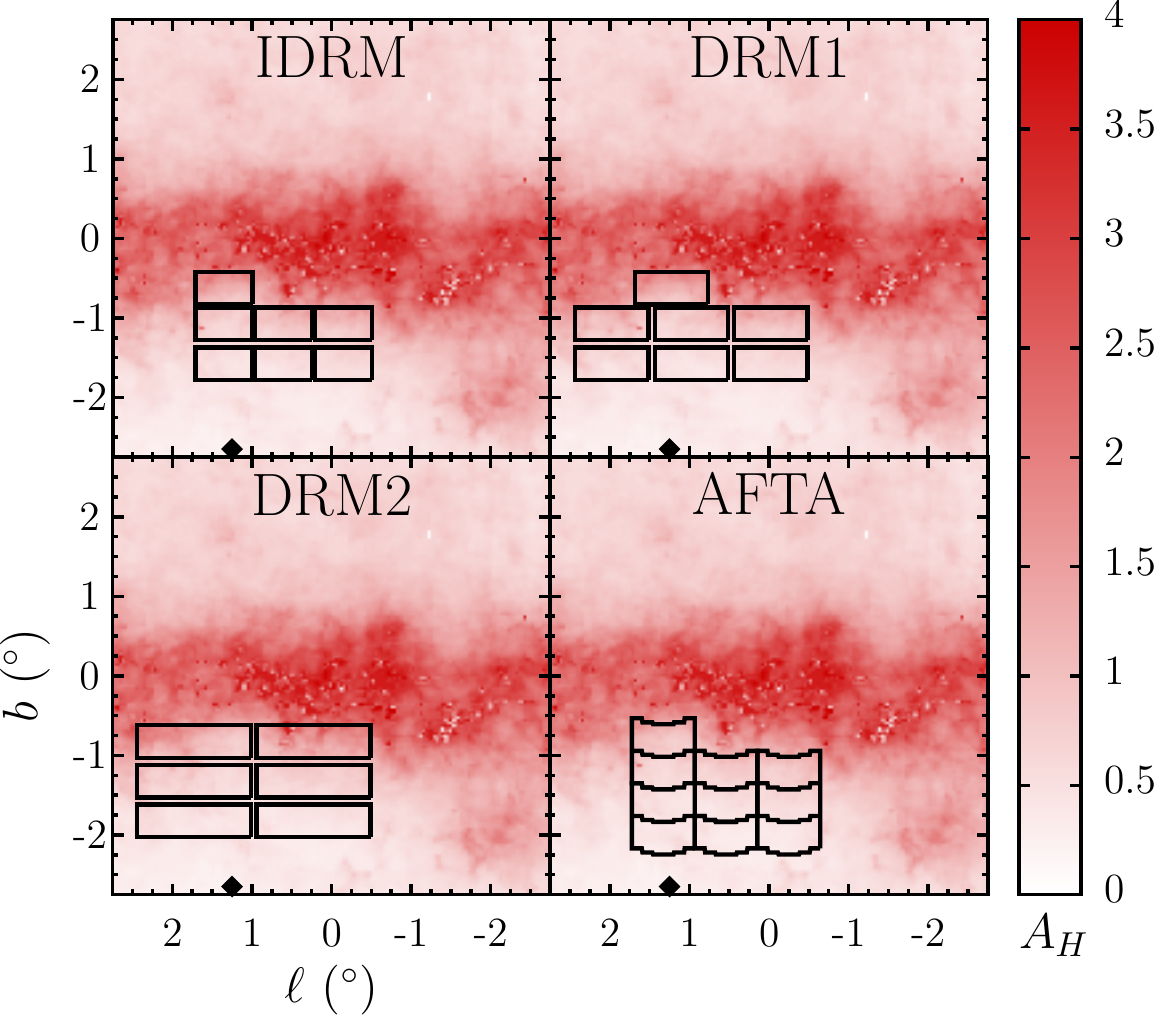}\\
\includegraphics[width=\columnwidth]{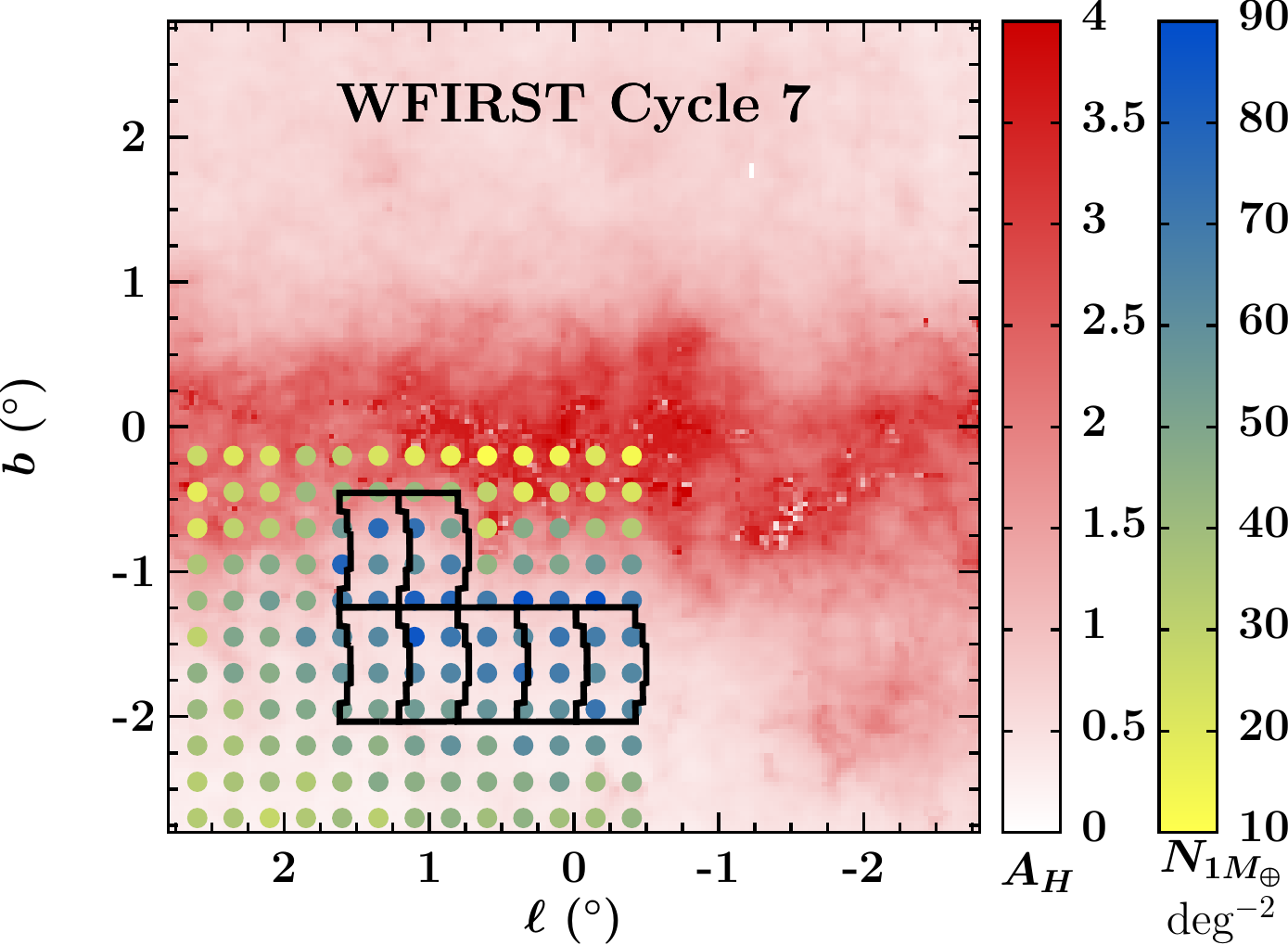}
\caption{Assumed field placement for each \wfirst\ design, plotted over a map of $H$-band extinction~\citep{Gonzalez2012}. The gaps between IDRM, DRM1 and DRM2 fields were included to mimic the effects of gaps between detectors. For the AFTA and Cycle 7 simulations we accounted for the individual detector placement within each field more carefully, so fields are close-butted (note the curved focal plane). Note also that in reality the 1-m-class designs would also likely have curved detector layouts and that, unlike the AFTA and Cycle 7 designs, the fields would probably not be orientated with their principle axes aligned with Galactic coordinates. The black diamond in the top panels shows the location of the \hst\ SWEEPS field. In the Cycle 7 panel, colored dots show the detection rate of $1\mearth$ planets per square degree as a function of position for the Cycle 7 design. A version of the Cycle 7 plot is available at \url{https://github.com/mtpenny/wfirst-ml-figures}.}
\label{fieldplacement}
\end{figure}

For the IDRM, DRM1, DRM2, and AFTA simulations, the field placement was not rigorously optimized. We show the fields we adopted for each design in \autoref{fieldplacement}. For IDRM, DRM1 and DRM2, the field placement is significantly different than what it would be in reality if each design were flown. This is due to uncertainties in the orientation of the detectors in the instrument bay. We therefore chose the simplest field orientation we could, aligning the principle axes of the fields with Galactic latitude and longitude. Note however, that this is an optimistic assumption, as the extinction and event rate at zeroth order depend strongly on $b$ but weakly on $\ell$, so detector orientations that align the long axis with $\ell$ are likely to be close to optimal. For the IDRM, DRM1 and DRM2 field layouts we accounted for gaps between detectors in the focal plan by placing the twice the sum of all chip gaps between each of the fields.

For AFTA we considered the field layout more carefully. The telescope instrument bay already exists, setting the orientation of the field and constraining the layout of detectors within it. Coincidently the orientation of the WFI focal plane for AFTA is within a couple of degrees of aligned with Galactic coordinates. From spring to fall seasons the orientation of the detector will be rotated by $180\degr$, which means that with the curved geometry of the active focal plane, the fields observed will not be exactly the same. For the layout shown this results in ${\sim}90$\% of stars that fall on a chip in spring seasons also falling on a chip in the fall seasons. Occasional gap filling dithers could be used to ensure some observations in both spring and fall for all events.

Between the AFTA design and Cycle 7, the \wfirst\ wide field instrument was redesigned and consequently the field orientation was rotated by $90\degr$. Additionally, the spacecraft's slew and settle time estimates were updated, and were more than twice that we had assumed for the AFTA design. These two changes led us to conduct an optimization of the field layouts. This optimization is described in \autoref{fieldopt}. With this field layout we can expect to detect ${\sim}27,000$ microlensing events with $|\uzero|<1$ and roughly twice this with $|\uzero|<3$ during the course of the mission. While there are three times as many events with $|\uzero|<3$ compared to $|\uzero|<1$, the maximum magnification of a \citet{Paczynski1986} single lens lightcurve at $|\uzero|=3$ is only $1.017$, compared to $1.34$ at $|\uzero|=1$, so only on brighter stars will it be possible for \wfirst\ to detect these low-magnification events.

\section{Baseline {\it WFIRST} planet yields}\label{baselineyield}
To assess the performance of each \wfirst\ design we ran a series of simulations to investigate the number of planets that would be detected during the \wfirst\ microlensing survey. To assess the performance of the survey over a broad range of masses and orbits we simulated single planet events with fixed masses in the range $0.1\le M\le 10^4$~$\mearth$ with semimajor axes distributed logarithmically in the range $0.3\le a< 30$~AU -- roughly a factor of $10$ either side of the typical Einstein radius ($2$--$3$~AU).

\newcolumntype{R}{>{\raggedleft\arraybackslash}X}
\newcolumntype{L}{>{\raggedright\arraybackslash}X}
\begin{table*}
\caption{Raw Simulation Planet Yields--$\log$ mass function}
\begin{tabularx}{\textwidth}{lR@{}c@{}LR@{}c@{}LR@{}c@{}LR@{}c@{}LR@{}c@{}L}
\hline \\[-5.0pt]
{\bf Mission} & \multicolumn{3}{c}{IDRM} & \multicolumn{3}{c}{DRM1} & \multicolumn{3}{c}{DRM2} & \multicolumn{3}{c}{AFTA} & \multicolumn{3}{c}{{\bf WFIRST Cycle 7}}\\
{\bf Duration} & \multicolumn{3}{c}{432 d} & \multicolumn{3}{c}{432 d} & \multicolumn{3}{c}{266 d} & \multicolumn{3}{c}{357 d}  & \multicolumn{3}{c}{{\bf 432 d}}\\
{\bf Area} & \multicolumn{3}{c}{2.06 deg$^{2}$} & \multicolumn{3}{c}{2.64 deg$^{2}$} & \multicolumn{3}{c}{3.52 deg$^{2}$} & \multicolumn{3}{c}{2.82 deg$^{2}$} & \multicolumn{3}{c}{{\bf 1.97 deg}$^{\mathbf{2}}$}\\
{\bf Rate Norm.} & \multicolumn{3}{c}{2.81} & \multicolumn{3}{c}{2.81} & \multicolumn{3}{c}{2.81} & \multicolumn{3}{c}{2.81} & \multicolumn{3}{c}{{\bf 2.81}}\\
{\bf Mass} ($\mathbf{M_{\oplus}}$) &&& &&& &&& &&& &&& \\
\\[-5.0pt] \hline \\[-5.0pt]
$\mathbf{0.1}$ & $   8.1$ &  \phantom{0}$\pm$\phantom{0} & $0.6$ & $  11.1$ &  \phantom{0}$\pm$\phantom{0} & $0.4$ & $   7.0$ &  \phantom{0}$\pm$\phantom{0} & $0.2$ & $  18.0$ &  \phantom{0}$\pm$\phantom{0} & $0.3$ & $\mathbf{   9.9}$ & \phantom{0}$\mathbf{\pm}$\phantom{0} & $\mathbf{0.4}$ \\
$\mathbf{1}$ & $  79.8$ &  \phantom{0}$\pm$\phantom{0} & $3.6$ & $  87.5$ &  \phantom{0}$\pm$\phantom{0} & $2.1$ & $  66.1$ &  \phantom{0}$\pm$\phantom{0} & $1.5$ & $   138$ &  \phantom{0}$\pm$\phantom{0} & $  2$ & $\mathbf{  87.5}$ & \phantom{0}$\mathbf{\pm}$\phantom{0} & $\mathbf{2.6}$ \\
$\mathbf{10}$ & $   366$ &  \phantom{0}$\pm$\phantom{0} & $ 15$ & $   496$ &  \phantom{0}$\pm$\phantom{0} & $  9$ & $   350$ &  \phantom{0}$\pm$\phantom{0} & $  6$ & $   643$ &  \phantom{0}$\pm$\phantom{0} & $  6$ & $\mathbf{   439}$ & \phantom{0}$\mathbf{\pm}$\phantom{0} & $\mathbf{  8}$ \\
$\mathbf{100}$ & $  1610$ &  \phantom{0}$\pm$\phantom{0} & $ 47$ & $  2110$ &  \phantom{0}$\pm$\phantom{0} & $ 39$ & $  1500$ &  \phantom{0}$\pm$\phantom{0} & $ 26$ & $  2440$ &  \phantom{0}$\pm$\phantom{0} & $ 51$ & $\mathbf{  1780}$ & \phantom{0}$\mathbf{\pm}$\phantom{0} & $\mathbf{ 84}$ \\
$\mathbf{1000}$ & $  5480$ &  \phantom{0}$\pm$\phantom{0} & $150$ & $  6610$ &  \phantom{0}$\pm$\phantom{0} & $110$ & $  4790$ &  \phantom{0}$\pm$\phantom{0} & $ 80$ & $  7670$ &  \phantom{0}$\pm$\phantom{0} & $130$ & $\mathbf{  5210}$ & \phantom{0}$\mathbf{\pm}$\phantom{0} & $\mathbf{ 86}$ \\
$\mathbf{10^4}$ & $ 12700$ &  \phantom{0}$\pm$\phantom{0} & $230$ & $ 15400$ &  \phantom{0}$\pm$\phantom{0} & $190$ & $ 11400$ &  \phantom{0}$\pm$\phantom{0} & $130$ & $ 17500$ &  \phantom{0}$\pm$\phantom{0} & $200$ & $\mathbf{ 11300}$ & \phantom{0}$\mathbf{\pm}$\phantom{0} & $\mathbf{150}$ \\
\\[-5.0pt] \hline
\end{tabularx}

\\{\bf Notes:} The table presents planet yields for each simulated survey, with the uncertainties due to Poisson shot noise in the drawn event parameters; this uncertainty does not include any systematic component, due, e.g., to the normalization of event rates; see \autoref{discuss} for a discussion of the magnitude of these errors. The survey duration, total field area and event rate normalization are shown in the header lines of the table. The yields assume a planet of fixed mass and an occurrence rate of one planet per decade of semimajor axis in the range $0.3\le a<30$~AU per star.
\label{raw_table}
\end{table*}

\autoref{raw_table} shows the raw results of these simulations for each \wfirst\ design, and the same numbers are plotted in \autoref{yieldplot} below. The table is not particularly useful for assessing the number of planets that \wfirst\ will detect, because the mass function that it implies (one planet per decade of mass and semimajor axis, we will call this the log mass function) is a significant overestimate at larger masses. It is however the most convenient form from which other mass functions can be applied. The results for IDRM1, DRM1 and DRM2 were first presented in \citetalias{Green2012}, though using a different set of event rate corrections (see \autoref{ratecorr} above for a description of the event rate corrections) and included a small number of additional detections of planets with semimajor axis between $0.03\le a<0.3$~AU. The results for DRM2 differ further from the report due to a correction factor used to the correction of a mistake in the number of fields and the exposure times that was made in the simulations presented in the \citet{Green2012} report. The AFTA simulations were rerun completely to incorporate the changes introduced at Cycle 4 of the \wfirst\ design. In hindsight these yields are overly optimistic due to an unrealistic assumption of the spacecraft's slew performance. The yields for Cycle 7 stand as the most realistic and most up-to-date estimates, and so we have made these bold in the table.

\begin{table*}
\caption{Best-estimate planet yields--fiducial mass function}
\begin{tabularx}{\textwidth}{lR@{}c@{}LR@{}c@{}LR@{}c@{}LR@{}c@{}LR@{}c@{}L}
\hline \\[-5.0pt]
{\bf Mission} & \multicolumn{3}{c}{IDRM} & \multicolumn{3}{c}{DRM1} & \multicolumn{3}{c}{DRM2} & \multicolumn{3}{c}{AFTA} & \multicolumn{3}{c}{{\bf WFIRST Cycle 7}}\\
{\bf Duration} & \multicolumn{3}{c}{432 d} & \multicolumn{3}{c}{432 d} & \multicolumn{3}{c}{266 d} & \multicolumn{3}{c}{357 d}  & \multicolumn{3}{c}{{\bf 432 d}}\\
{\bf Area} & \multicolumn{3}{c}{2.06 deg$^{2}$} & \multicolumn{3}{c}{2.64 deg$^{2}$} & \multicolumn{3}{c}{3.52 deg$^{2}$} & \multicolumn{3}{c}{2.82 deg$^{2}$} & \multicolumn{3}{c}{{\bf 1.97 deg}$^{\mathbf{2}}$}\\
{\bf Rate Norm.} & \multicolumn{3}{c}{2.81} & \multicolumn{3}{c}{2.81} & \multicolumn{3}{c}{2.81} & \multicolumn{3}{c}{2.81} & \multicolumn{3}{c}{{\bf 2.81}}\\
{\bf Mass} ($\mathbf{M_{\oplus}}$) &&& &&& &&& &&& &&& \\
\\[-5.0pt] \hline \\[-5.0pt]
$\mathbf{0.1}$ & $  16.7$ &  \phantom{0}$\pm$\phantom{0} & $1.3$ & $  22.8$ &  \phantom{0}$\pm$\phantom{0} & $0.8$ & $  14.4$ &  \phantom{0}$\pm$\phantom{0} & $0.5$ & $  37.1$ &  \phantom{0}$\pm$\phantom{0} & $0.6$ & $\mathbf{  20.5}$ & \phantom{0}$\mathbf{\pm}$\phantom{0} & $\mathbf{0.8}$ \\
$\mathbf{1}$ & $   164$ &  \phantom{0}$\pm$\phantom{0} & $  8$ & $   180$ &  \phantom{0}$\pm$\phantom{0} & $  4$ & $   136$ &  \phantom{0}$\pm$\phantom{0} & $  3$ & $   284$ &  \phantom{0}$\pm$\phantom{0} & $  3$ & $\mathbf{   180}$ & \phantom{0}$\mathbf{\pm}$\phantom{0} & $\mathbf{  5}$ \\
$\mathbf{10}$ & $   455$ &  \phantom{0}$\pm$\phantom{0} & $ 19$ & $   615$ &  \phantom{0}$\pm$\phantom{0} & $ 11$ & $   434$ &  \phantom{0}$\pm$\phantom{0} & $  7$ & $   799$ &  \phantom{0}$\pm$\phantom{0} & $  8$ & $\mathbf{   545}$ & \phantom{0}$\mathbf{\pm}$\phantom{0} & $\mathbf{  9}$ \\
$\mathbf{100}$ & $   371$ &  \phantom{0}$\pm$\phantom{0} & $ 11$ & $   488$ &  \phantom{0}$\pm$\phantom{0} & $  9$ & $   346$ &  \phantom{0}$\pm$\phantom{0} & $  6$ & $   563$ &  \phantom{0}$\pm$\phantom{0} & $ 12$ & $\mathbf{   412}$ & \phantom{0}$\mathbf{\pm}$\phantom{0} & $\mathbf{ 19}$ \\
$\mathbf{1000}$ & $   236$ &  \phantom{0}$\pm$\phantom{0} & $  6$ & $   284$ &  \phantom{0}$\pm$\phantom{0} & $  5$ & $   206$ &  \phantom{0}$\pm$\phantom{0} & $  4$ & $   330$ &  \phantom{0}$\pm$\phantom{0} & $  6$ & $\mathbf{   224}$ & \phantom{0}$\mathbf{\pm}$\phantom{0} & $\mathbf{  4}$ \\
$\mathbf{10^4}$ & $   101$ &  \phantom{0}$\pm$\phantom{0} & $  2$ & $   124$ &  \phantom{0}$\pm$\phantom{0} & $  2$ & $  91.3$ &  \phantom{0}$\pm$\phantom{0} & $1.1$ & $   141$ &  \phantom{0}$\pm$\phantom{0} & $  2$ & $\mathbf{  90.7}$ & \phantom{0}$\mathbf{\pm}$\phantom{0} & $\mathbf{1.2}$ \\
\\[-5.0pt] \hline
{\bf Total (0.1--}$\mathbf{10^4 M_{\oplus})}$
& \multicolumn{3}{c}{  1294}& \multicolumn{3}{c}{  1653}& \multicolumn{3}{c}{  1183}& \multicolumn{3}{c}{  2084}& \multicolumn{3}{c}{{\bf   1428}}\\
\hline
\end{tabularx}

\\{\bf Notes:} In this table the yields presented in \autoref{raw_table} have been multiplied by our fiducial mass function (\autoref{fiducialmf}). The total survey yield is found by integration of the tabulated values with the extended trapezoidal rule.
\label{cassan_table}
\end{table*}

\begin{table*}
\caption{{\it Obsolete} planet yields from \wfirst-AFTA final report--fiducial mass function}
\begin{tabularx}{\textwidth}{lR@{}c@{}LR@{}c@{}LR@{}c@{}LR@{}c@{}LR@{}c@{}L}
\hline \\[-5.0pt]
{\bf Mission} & \multicolumn{3}{c}{IDRM} & \multicolumn{3}{c}{DRM1} & \multicolumn{3}{c}{DRM2} & \multicolumn{3}{c}{AFTA} & \multicolumn{3}{c}{{\bf WFIRST Cycle 7}}\\
{\bf Duration} & \multicolumn{3}{c}{432 d} & \multicolumn{3}{c}{432 d} & \multicolumn{3}{c}{266 d} & \multicolumn{3}{c}{357 d}  & \multicolumn{3}{c}{{\bf 432 d}}\\
{\bf Area} & \multicolumn{3}{c}{2.06 deg$^{2}$} & \multicolumn{3}{c}{2.64 deg$^{2}$} & \multicolumn{3}{c}{3.52 deg$^{2}$} & \multicolumn{3}{c}{2.82 deg$^{2}$} & \multicolumn{3}{c}{{\bf 1.97 deg}$^{\mathbf{2}}$}\\
{\bf Rate Norm.} & \multicolumn{3}{c}{$2.41\times1.475$} & \multicolumn{3}{c}{$2.46\times1.475$} & \multicolumn{3}{c}{$2.42\times1.475$} & \multicolumn{3}{c}{$2.46\times1.475$} & \multicolumn{3}{c}{{\bf $\mathbf{2.46\times1.475}$}}\\
{\bf Mass} ($\mathbf{M_{\oplus}}$) &&& &&& &&& &&& &&& \\
\\[-5.0pt] \hline \\[-5.0pt]
$\mathbf{0.1}$ & $  21.1$ &  \phantom{0}$\pm$\phantom{0} & $1.6$ & $  29.4$ &  \phantom{0}$\pm$\phantom{0} & $1.0$ & $  18.3$ &  \phantom{0}$\pm$\phantom{0} & $0.6$ & $  47.8$ &  \phantom{0}$\pm$\phantom{0} & $0.8$ & $\mathbf{  25.9}$ & \phantom{0}$\mathbf{\pm}$\phantom{0} & $\mathbf{1.0}$ \\
$\mathbf{1}$ & $   208$ &  \phantom{0}$\pm$\phantom{0} & $ 10$ & $   232$ &  \phantom{0}$\pm$\phantom{0} & $  6$ & $   173$ &  \phantom{0}$\pm$\phantom{0} & $  4$ & $   367$ &  \phantom{0}$\pm$\phantom{0} & $  4$ & $\mathbf{   228}$ & \phantom{0}$\mathbf{\pm}$\phantom{0} & $\mathbf{  7}$ \\
$\mathbf{10}$ & $   575$ &  \phantom{0}$\pm$\phantom{0} & $ 24$ & $   793$ &  \phantom{0}$\pm$\phantom{0} & $ 14$ & $   551$ &  \phantom{0}$\pm$\phantom{0} & $  9$ & $  1030$ &  \phantom{0}$\pm$\phantom{0} & $ 10$ & $\mathbf{   690}$ & \phantom{0}$\mathbf{\pm}$\phantom{0} & $\mathbf{ 12}$ \\
$\mathbf{100}$ & $   470$ &  \phantom{0}$\pm$\phantom{0} & $ 14$ & $   629$ &  \phantom{0}$\pm$\phantom{0} & $ 12$ & $   439$ &  \phantom{0}$\pm$\phantom{0} & $  8$ & $   726$ &  \phantom{0}$\pm$\phantom{0} & $ 15$ & $\mathbf{   522}$ & \phantom{0}$\mathbf{\pm}$\phantom{0} & $\mathbf{ 25}$ \\
$\mathbf{1000}$ & $   298$ &  \phantom{0}$\pm$\phantom{0} & $  8$ & $   367$ &  \phantom{0}$\pm$\phantom{0} & $  6$ & $   261$ &  \phantom{0}$\pm$\phantom{0} & $  4$ & $   426$ &  \phantom{0}$\pm$\phantom{0} & $  7$ & $\mathbf{   283}$ & \phantom{0}$\mathbf{\pm}$\phantom{0} & $\mathbf{  5}$ \\
$\mathbf{10^4}$ & $   128$ &  \phantom{0}$\pm$\phantom{0} & $  2$ & $   160$ &  \phantom{0}$\pm$\phantom{0} & $  2$ & $   116$ &  \phantom{0}$\pm$\phantom{0} & $  1$ & $   181$ &  \phantom{0}$\pm$\phantom{0} & $  2$ & $\mathbf{   115}$ & \phantom{0}$\mathbf{\pm}$\phantom{0} & $\mathbf{  2}$ \\
\\[-5.0pt] \hline
{\bf Total (0.1--}$\mathbf{10^4 M_{\oplus})}$
& \multicolumn{3}{c}{  1636}& \multicolumn{3}{c}{  2131}& \multicolumn{3}{c}{  1499}& \multicolumn{3}{c}{  2687}& \multicolumn{3}{c}{{\bf   1806}}\\
\hline
\end{tabularx}

\\{\bf Notes:} As \autoref{cassan_table}, but using the event rate (${\sim}2.4$) and detection efficiency compromise ($1.475$) scalings that were used in the \wfirst-AFTA final report~\citep{Spergel2015}. These yields should be considered obsolete, and are only presented to aid comparison with the previous \wfirst\ reports.
\label{cassan_table_report}
\end{table*}

\autoref{cassan_table} presents our fiducial estimate for the planet yield of \wfirst. To compute these yields we multiply the raw yields from \autoref{raw_table} by the following, fiducial, form of the planet mass function, 
\begin{equation}
\frac{\dd^2 N}{\dd \log \mpl \dd \log a} = \left\{\begin{array}{ll}
0.24\ \text{dex}^{-2} \left(\frac{\mpl}{95\mearth}\right)^{-0.73} & \text{if} \mpl\ge 5.2\mearth,\\
2\ \text{dex}^{-2} & \text{if} \mpl< 5.2\mearth,
\end{array}\right.
\label{fiducialmf}
\end{equation}
which we will refer to hereafter as the fiducial mass function. This mass function is based on the power-law bound planet mass function measured by \citet{Cassan2012} from microlensing observations, but saturated at a value of 2 planets per star below a mass of $5.2\mearth$, which is roughly where \citet{Cassan2012} lost sufficient sensitivity to measure the mass function. For context, this saturation value is comparable to the planet density of the inner solar system between ${\sim}0.2$ and $2$ AU. We discuss the impact on our results of more recent determinations of the mass ratio function in \autoref{discuss}. To estimate the total number of planets \wfirst\ will detect we integrated the resulting numbers using the trapezoidal rule; for the Cycle 7 design this results in an expected total yield of ${\sim}1400$ planets, noting that it does not include free-floating planets which would add a few hundred to this figure~\citep{Spergel2015}. This yield is smaller than \kepler's total yield, but of the same order of magnitude, suggesting that a similar precision in demographics will be achievable.

\begin{figure}
\includegraphics[width=\columnwidth]{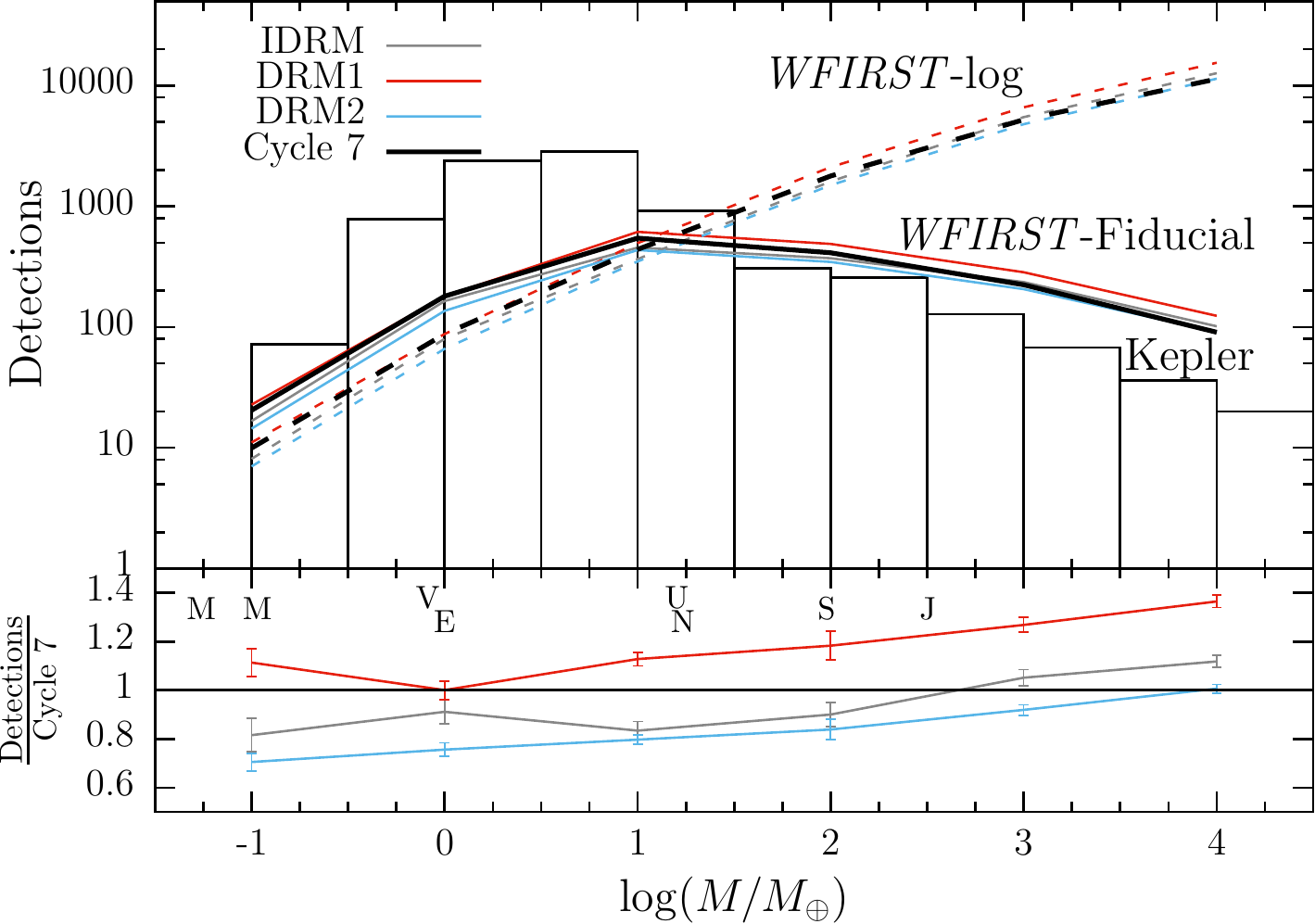}
\caption{Number of detections as a function of planet mass for each \wfirst\ design and for two different mass functions. Also shown is the distribution of Kepler candidates assuming a mass-radius relation of $(M/\mearth)=(R/R_{\earth})^{2.06}$~\citep{Lissauer2011}. Letters indicate the masses of Solar System planets. The lower panel shows the yields of the DRM designs relative to that of Cycle 7.}
\label{yieldplot}
\end{figure}

\autoref{yieldplot} shows the simulation yields of Tables~\ref{raw_table} and \ref{cassan_table} graphically, and compares them to \kepler. To estimate the \kepler\ mass function we applied the simplistic mass-radius $\mpl/\mearth = (R_{\rm p}/\rearth)^{2.06}$ relation of \citet{Lissauer2011} to the \kepler\ candidates with \texttt{koi\_score} $>0.5$.\footnote{https://exoplanetarchive.ipac.caltech.edu/} The first point to note is the large numbers of planets that \wfirst\ is expected to detect, using any of the designs and across a wide range of masses. As we discuss later, the cold planet mass function below ${\sim}10\mearth$ is almost completely unconstrained at present, meaning that \wfirst\ will add at least two orders of magnitude in mass sensitivity beyond current knowledge.

The lower panel of \autoref{yieldplot} compares the relative yields of the previous \wfirst\ designs to the current Cycle 7 design as a function of planet mass. Except for the highest mass planets, the Cycle 7 design outperforms the IDRM and DRM2 designs in terms of planet detection yield, though note that DRM2 had a significantly shorter total survey duration. The DRM1 design has a ${\sim}10$~percent higher yield than the current Cycle 7 design at low masses, but a larger difference at large planet masses. The comparison of only the planet detection yields between Cycle 7 and the ${\sim} 1$-m class designs is, however, unfair. Cycle 7's 2.4-m mirror enables factor of ${\sim}2$ improvements in the measurement of host and planet masses through lens flux, image elongation and color-dependent centroid shifts~\citep{Bennett2007}, which the planet detection yield alone does not account for. We will discuss these measurements in \autoref{properties}.

We have not shown the AFTA yields in \autoref{yieldplot} because we consider them to be unrealistic. The significant drop in yield from the AFTA design to Cycle 7 deserves some discussion, though. While there were many changes between the designs, one change dominates the reduction in yield. This was the adoption of more accurate estimates of the slew and settle time of the spacecraft, and not a result of descopes of mission hardware. For the AFTA designs, we had assumed values for slew performance that were the same as for the smaller IDRM and DRM designs, which was unrealistic given the larger mass and moment of inertia associated with a larger mirror, secondary mirror support structure, and spacecraft bus. Our AFTA results are further unrealistic, because we applied the ${\sim}0.4\degr$ slew time to all slews between fields, when some slews will be longer. These optimistic assumptions for the previous designs, if corrected, would likely result in a less optimal field layout and a reduction in yields, though perhaps relatively smaller than the drop from AFTA to Cycle 7. If it were possible to use attitude control systems that provide significantly faster slew performance (e.g., control moment gyros instead of reaction wheels), then significant gains in the yield of the \wfirst\ microlensing survey could be realized.

\subsection{Sensitivity to moon-mass objects}\label{lowmass}

To trace out the approximate limits of the sensitivity of \wfirst\ to low-mass planets, as well as wide and close separation planets, we ran simulations on a grid in planet mass and semimajor axis over the ranges $-2\le\log(\mpl/\mearth)<4$ and $-2\le\log(a/\text{AU})<2$. We required that the events have impact parameters $|\uzero|<3$ and times of closest approach in the range $0\le\tzero<2011$~days. The full details of the computations, are described in \autoref{curvecomp}, including details of how false positive detections due to numerical errors are accounted for. We note that the simulations were conducted using the parameters of the AFTA design, but we used the analytic estimate method described in \autoref{analytics} to predict the yield of the Cycle 7 design from the AFTA design simulations. Other than the change in field layout, there is only a small change in exposure time for the analytic approximation to account for, so we expect the uncertainty due to this conversion to be small.

\begin{figure*}
\includegraphics[width=\textwidth]{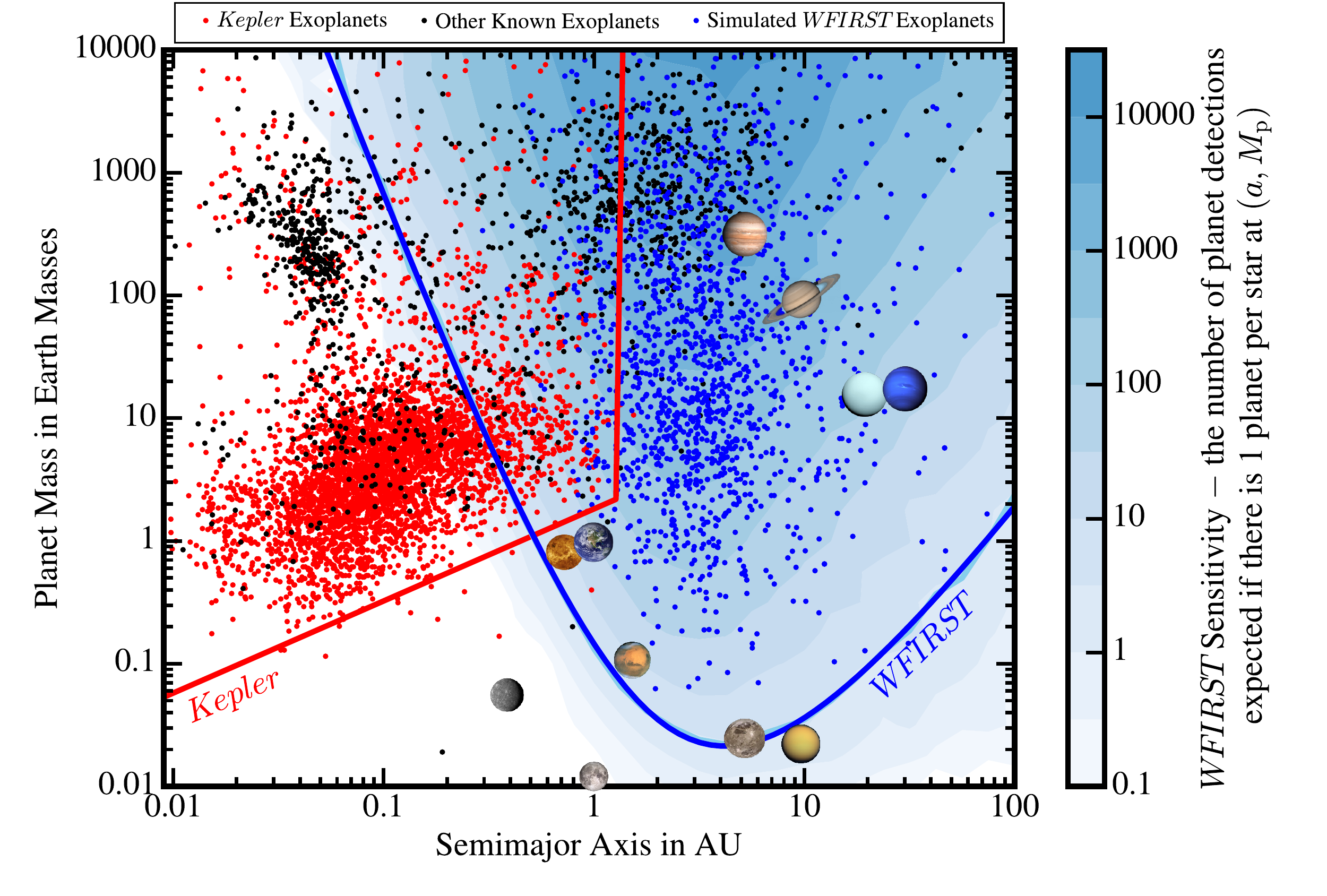}
\caption{Comparison of the \wfirst\ Cycle 7 design sensitivity to that of \kepler\ in the planet mass-semimajor axis plane. The red line shows an approximation of the \kepler\ planet detection limit based on \citet{Burke2015}. Blue shading shows the number of \wfirst\ planet detections during the mission if there is one planet per star at a given mass and semimajor axis point; this is directly proportional to the average detection efficiency. The thick, dark blue line is an functional fit to the 3-detection per mission contour, while the lighter blue line barely visible beneath it is the actual contour. Red dots show Kepler candidate and confirmed planets, black dots show all other known planets extracted from the NASA exoplanet archive \citep[accessed 28 Feb. 2018][]{Akeson2013}. Blue dots show a simulated realization of the planets detected by the \wfirst\ microlensing survey, assuming our fiducial planet mass function (\autoref{fiducialmf}), though note that in constructing this sample of simulated detections we did not simulate planets smaller than $0.03\mearth$ or with semimajor axis less than $0.3$~AU. Solar system bodies are shown by their images, including the satellites Ganymede, Titan, and the Moon at the semimajor axis of their hosts. Images of the solar system planets credit to NASA. The data and scripts used to make this plot are available at \url{https://github.com/mtpenny/wfirst-ml-figures}.}
\label{sensitivity}
\end{figure*} 

\begin{figure}
\includegraphics[width=\columnwidth]{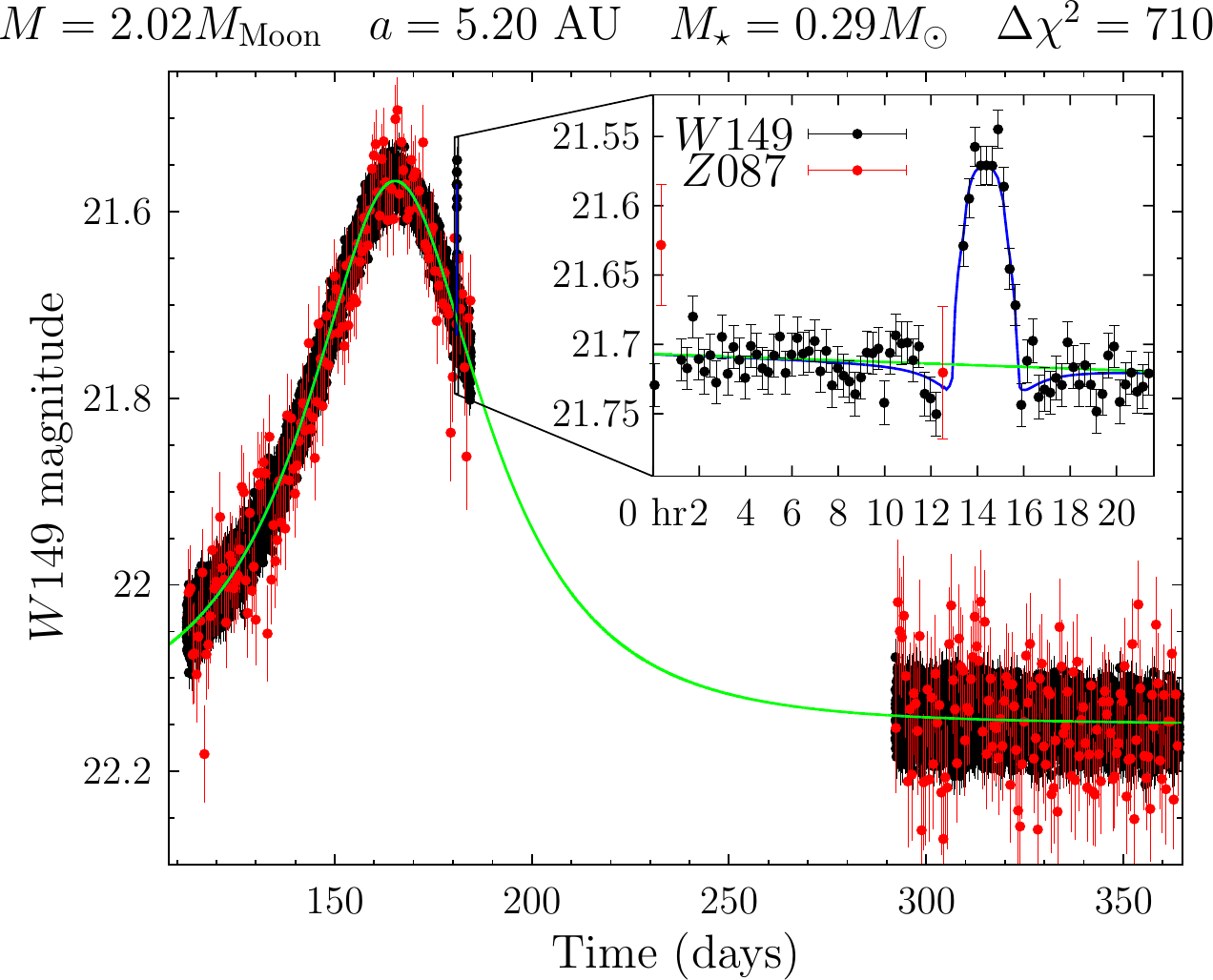}
\caption{Example lightcurve of a $0.025$-$\mearth$ (Ganymede-mass) bound planet detection from simulations of the AFTA design. Black and red data points are in the $W149$ and $Z087$ filters, respectively. The blue curve shows the underlying ``true'' lightcurve and the green line shows the best fit single lens lightcurve.}
\label{lightcurves}
\end{figure}

\autoref{sensitivity} shows the results of the grid computation. The shading shows the planet detection rate (in units of planets per full survey) at each mass-semimajor axis point. {\it The blue sensitivity curve plots the specific point in the mass-semimajor axis plane where 3 planet detections can be expected in the course of the mission if there is one such planet per star.} The sensitivity curve is also a line of constant detection efficiency~\citep[e.g.,][]{Peale1997}.
We found that the sensitivity curve is very well approximated by an analytic function
\begin{equation}
\log (\mpl/\mearth) = \alpha + \beta\log (a/\text{AU}) + \gamma\sqrt{\delta^2+[\log (a/\text{AU})-\epsilon]^2},
\label{senscurve}
\end{equation}
where $a$ is the semimajor axis and the parameters take the following values: $\alpha = -3.90$, $\beta = -1.15$, $\gamma=3.56$, $\delta=0.783$, $\epsilon=0.356$. The analytic function (bright blue) is plotted above the sensitivity curve data (pale blue) in the figure, but the sensitivity data curve is almost invisible underneath the analytic fit. 
It is interesting to note that seven of the eight Solar System planets fall within the \wfirst\ sensitivity curve, with only Mercury outside. However, in place of Mercury, when we place the Solar System's moons on the diagram at the orbital separations of their host planets, the Galilean moon Ganymede lies just within the sensitivity curve. \autoref{lightcurves} shows an example lightcurve of such a Ganymede-mass exoplanet that \wfirst\ could detect. If the Solar System moons were placed at the minimum point of the analytic curve ($a=4.2$~AU, $M=0.021\mearth$), Ganymede ($0.025\mearth$) and Titan ($0.023\mearth$) would be above the curve and Callisto ($0.018\mearth$), Io ($0.015\mearth$) and the Moon ($0.012\mearth$) would be below it; all other known Solar System bodies have masses less than $0.01\mearth$. At $1\mearth$, the sensitivity curve stretches from $0.5$~AU to $70$~AU. Removing the constraint that the impact parameter of the host star's microlensing event be $|\uzero|<3$ results in there being no upper limit on the semimajor axis at which AFTA has sensitivity to Earth-mass planets. However, most of these more distant planets would be seen as free-floating planet candidates due to undetectable magnification from the source star~\citep[see][for a discussion of constraining the presence of host stars in free-floating planet candidate events]{Henderson2016-char}.

The shading in \autoref{sensitivity} does not accurately represent the distribution of planets that we can expect to detect with \wfirst, only \wfirst's sensitivity to planets. \wfirst\ has a high-detection efficiency for large-mass planets, but these have consistently been shown to be rare~\citep[e.g.,][]{Cumming2008,Howard2010,Cassan2012,Fressin2013,Shvartzvald2016,Suzuki2016}. To give a better idea of the distribution of planet detections \wfirst\ will detect, we simulated a survey with planets populated from our fiducial mass function. A realization from this simulation is shown as the blue dots in \autoref{sensitivity}. Note that the lower mass limit for this simulation was $\mpl=0.03\mearth$ and the smallest semimajor axis was $0.3$~AU, so a small number of the most extreme planets that could be detected will be missing from this realization. In adding these simulated planets to the plot, we do not account for the uncertainty in measurements of either $\mpl$ or $a$. We expect the majority of \wfirst's planets to have mass measurements from either lens-detection measurements or parallax, so given the large range of masses covered, uncertainties in $\mpl$ would not result in a significant change of appearance, and features in the planet mass function should be easily distinguishable. With a mass measurement also comes a measurement of the physical Einstein radius, and so the physical projected separation of the planet and star. This must be deprojected to the actual semimajor axis, which will result in a substantial uncertainty in the estimated value of $a$ relative to the range of $a$, though for some planets orbital motion measurements can better constrain $a$~\citep[e.g.,][]{Bennett2010,Penny2011-om}. Our simulations do not include an eccentricity distribution, so we can not estimate the uncertainty associated with this deprojection.

\begin{figure*}
\includegraphics[width=0.49\textwidth]{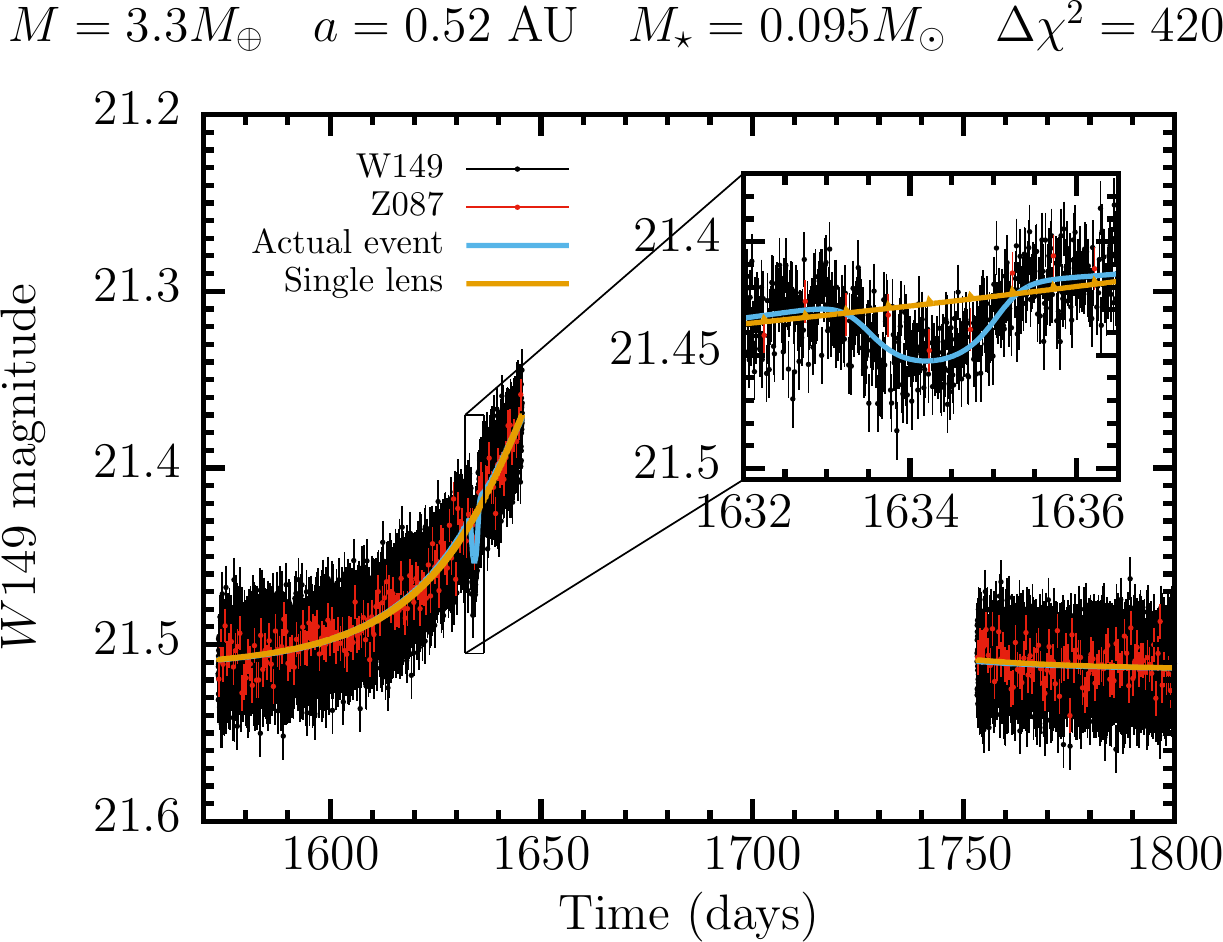}
\includegraphics[width=0.49\textwidth]{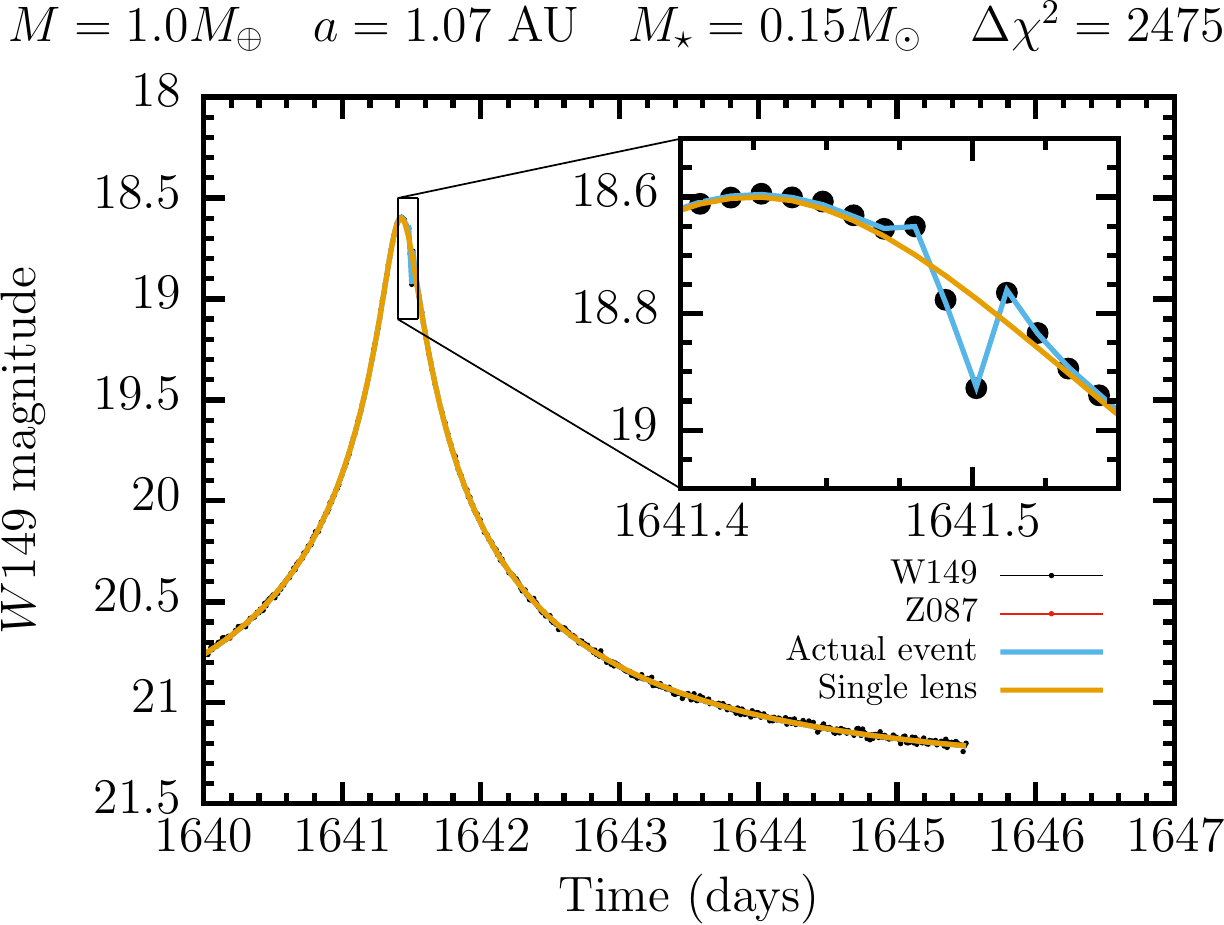}\\
\includegraphics[width=0.49\textwidth]{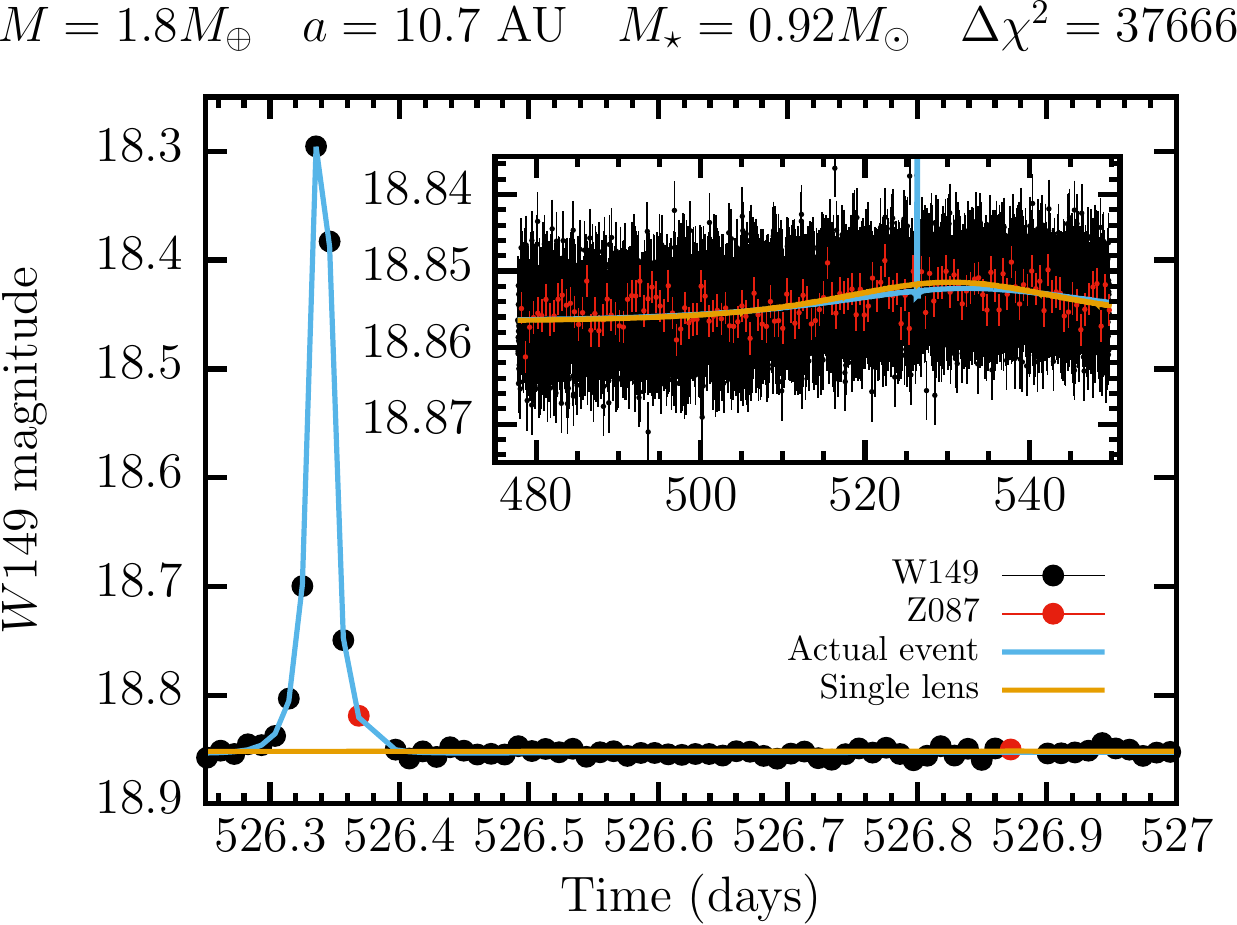}
\includegraphics[width=0.49\textwidth]{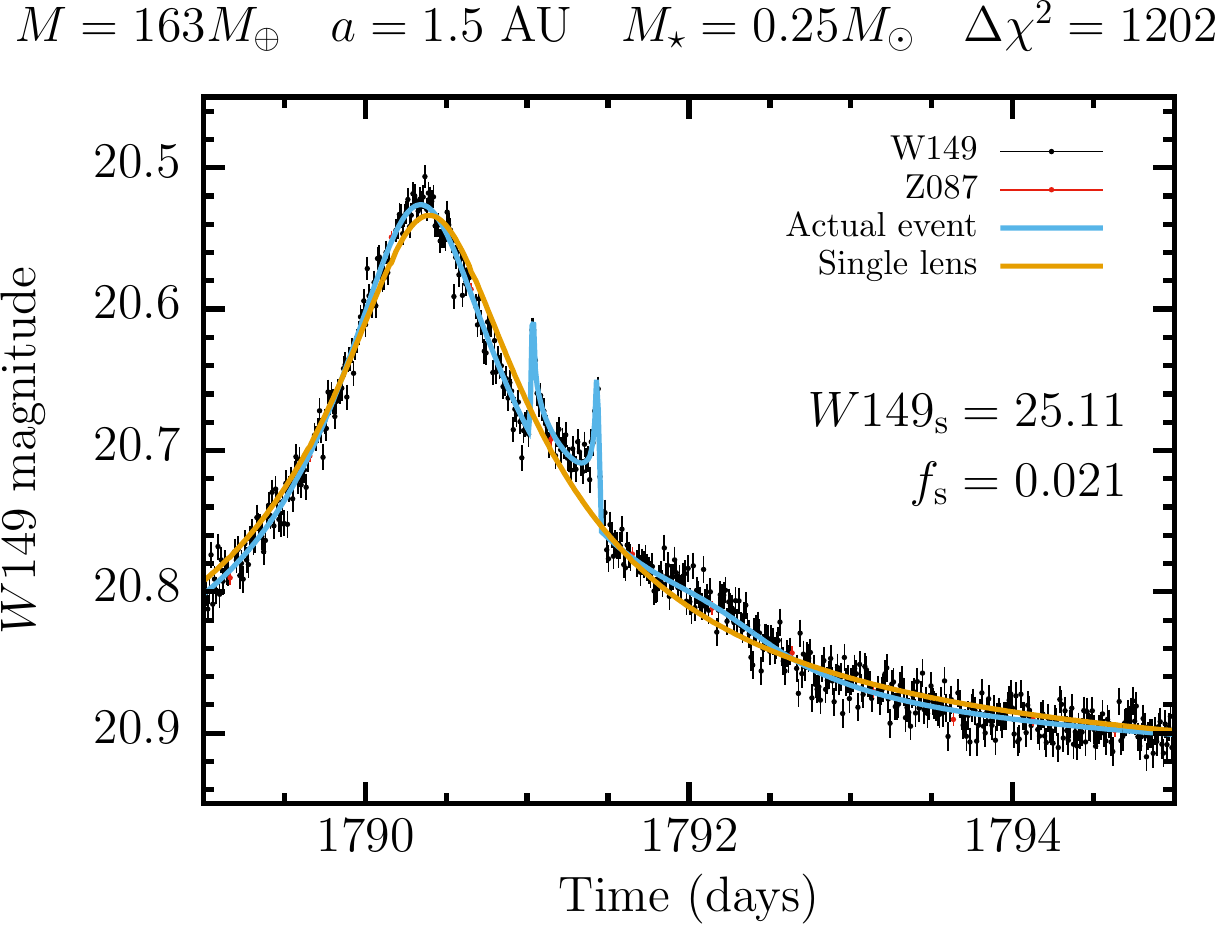}\\
\includegraphics[width=0.49\textwidth]{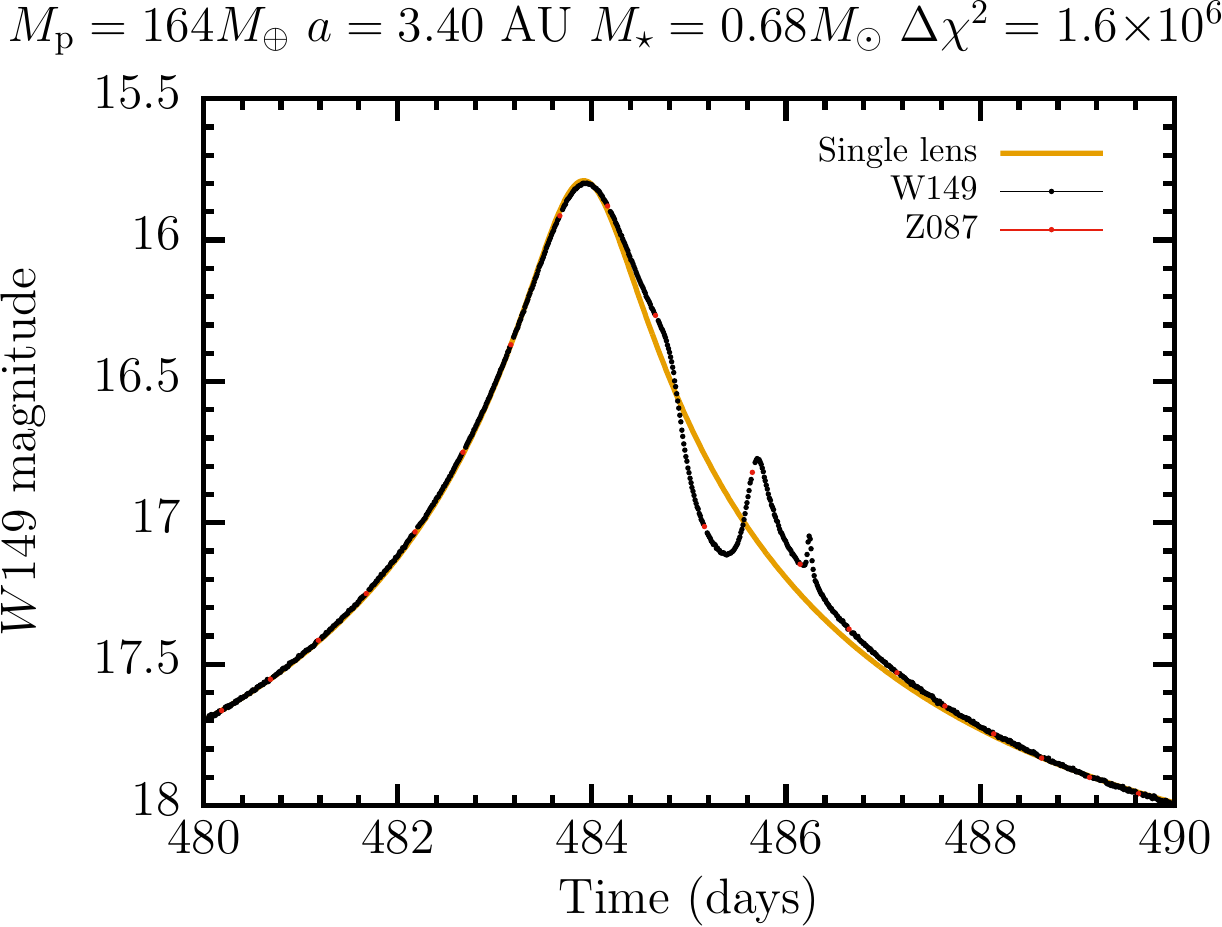}
\includegraphics[width=0.49\textwidth]{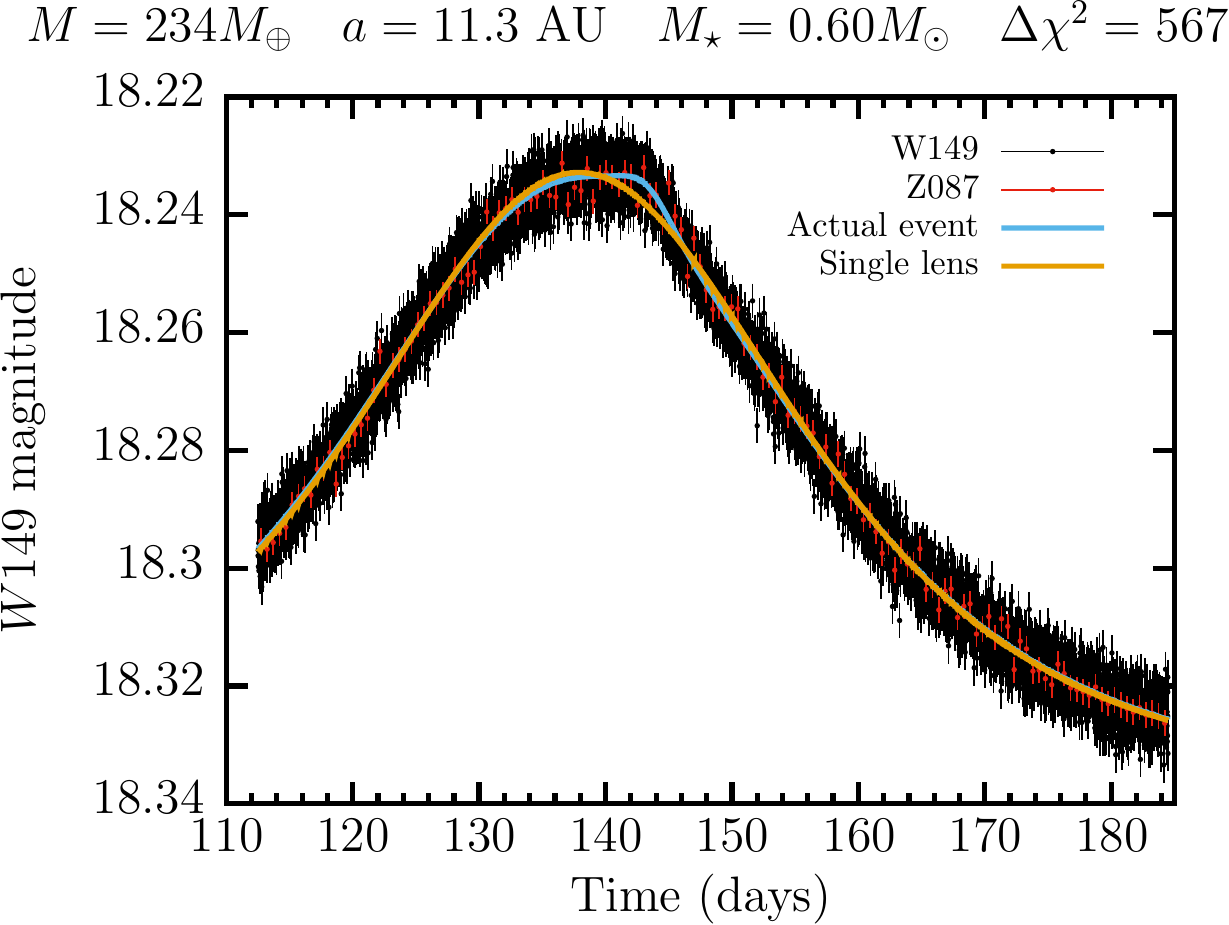}
\caption{Examples of simulated \wfirst\ lightcurves using the Cycle 7 design parameters, chosen to display the some of the variety of lightcurve features that \wfirst\ will detect, or challenges that will impact the analysis of events. The examples demonstrate a lightcurve with missing peak data due to \wfirst's limited observing seasons (\emph{top left}), an Earth-mass planet orbiting a $0.15\msun$ star at $1$~AU (\emph{top right}), a wide orbit planet with a very low amplitude host microlensing event (\emph{middle left}), an event with a very faint source and high blending (\emph{middle right}), a high signal to noise detection of a massive planet (\emph{bottom left}), a low signal to noise detection of a massive planet on a wide orbit (\emph{bottom right}).}
\label{exlc}
\end{figure*}

\subsection{Properties of the {\it WFIRST} microlensing events}\label{properties}

\begin{figure}
\includegraphics[width=\columnwidth]{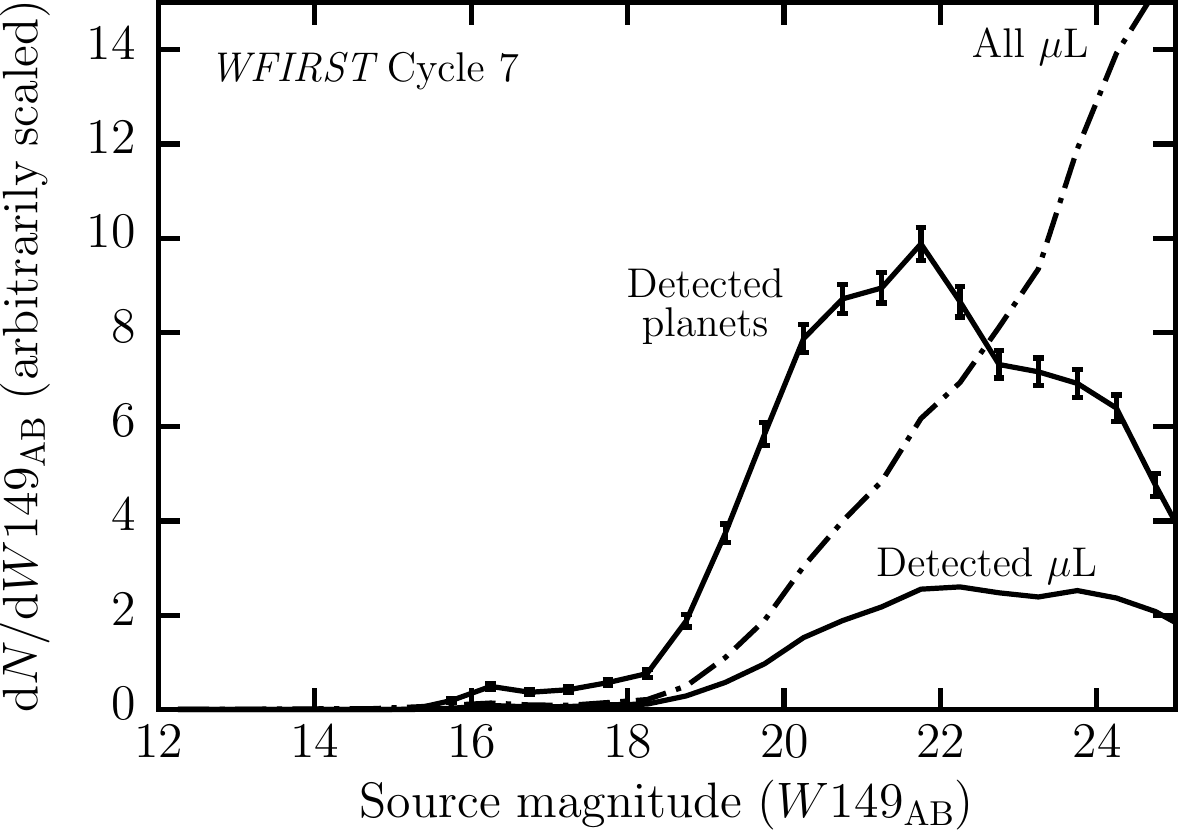}
\includegraphics[width=\columnwidth]{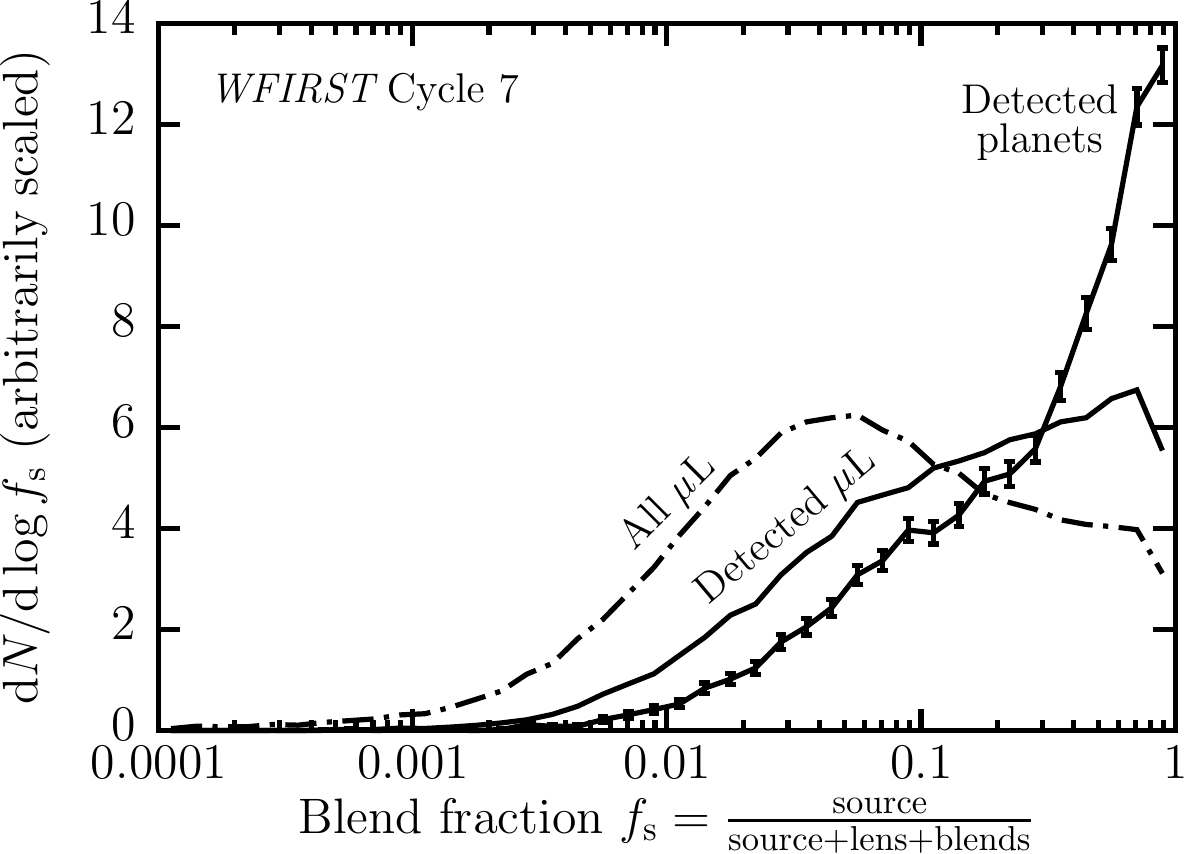}
\caption{\emph{Top}: The distribution of source magnitudes for different subsets of the simulated microlensing events. ``All $\mu$L'' is the distribution of all simulated microlensing events,  ``Detected $\mu$L'' is the distribution of events with flux variation detected at $\Delta\chi^2>500$ above a flat baseline, and ``Detected planets'' is the distribution for events with $1\mearth$ planet detections. Each distribution is on its own arbitrary scale. \emph{Bottom:} The distribution of the fraction of baseline flux contributed by the source, $f_{\rm s}$. Again, each distribution has been arbitrarily scaled.}
\label{sourcemag}
\end{figure}

The \wfirst\ microlensing survey will search for microlensing events from fainter sources than are observed from the ground, and its resolution will be sufficient to at least partially resolve lenses and sources over the course of the 5 year mission, so it is important to consider the properties of the sources, lenses and blending. \autoref{sourcemag} plots the distribution of source magnitudes and blending of the microlensing events that the Cycle 7 design will be able to observe. In both plots we show the distribution for every microlensing event that occurs in the \wfirst\ field on sources brighter than $H_{\mathrm{Vega}}=25$, regardless of whether it will be detected or not (labeled all $\mu$L), the events that cause a microlensing event that is detectable as a $\Delta\chi^2>500$ deviation from a flat baseline (labeled detected $\mu$L), and the events with detectable Earth-mass planets (labeled detected planets).

While the number of all microlensing events per magnitude keeps rising beyond a magnitude of $W149=25$, the number of detectable microlensing events exhibits a broad peak between $W149=22$ and 24, before beginning to fall. For planet detections, the source magnitude distribution peaks between $W149\approx20$--$22$, but only begins to fall rapidly fainter than $W149\approx 24$.

The lower panel of \autoref{sourcemag} shows the distribution of blending. The blending parameter $\blendfs$ is the ratio of source flux to total flux in the photometric aperture when the source is unmagnified. We have measured $\blendfs$ in the same $3\times 3$ pixel aperture we have used for photometry. Being as the input source magnitude distribution continues to rise towards faint magnitudes, the majority of microlensing events we simulated were significantly blended (i.e, $\blendfs<0.5$), despite \wfirst's small PSF. This is also the case for microlensing events that \wfirst\ will detect, though the distribution $\dd N/\dd\log\blendfs$ does peak above $\blendfs=0.5$. For events with planets detected, the distribution of $\blendfs$ is much more skewed towards small amounts of blending $\blendfs\sim 1$, but there remains a significant tail with large amounts of blending.

\begin{figure}
\includegraphics[width=\columnwidth]{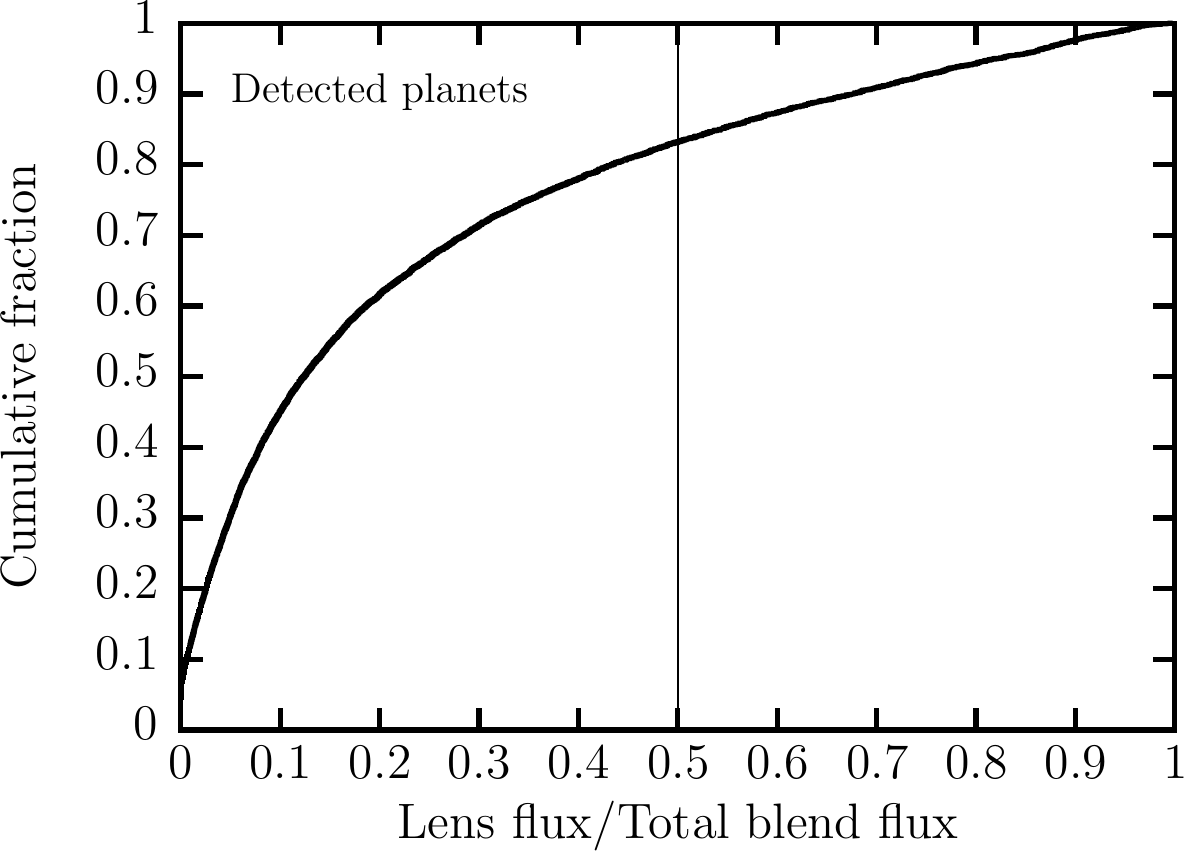}
\caption{Cumulative distribution of the contribution of the lens flux to the total blend flux in a $3\times3$~pixel aperture around events with detected planets. The vertical line shows the point at which the lens provides more than $50$\% of the flux.}
\label{lensblend}
\end{figure}

Figure~\ref{lensblend} shows the relative contribution of the lens star to the total blend flux for detected planets. In fewer than $20$\% of events with planet detections does the lens flux dominate the blended light. However, the majority events will have a lens flux within a factor of 10 of the total blend flux. Without knowing the nature of the blended light (i.e., is it due to the PSF wings of bright stars, or nearby fainter stars), it is difficult to say more about how this added confusion will affect host mass measurements via lens detection~\citep[see, e.g.,][]{Bennett2007,Henderson2014-ao,Bennett2015}.

\begin{figure}
\includegraphics[width=\columnwidth]{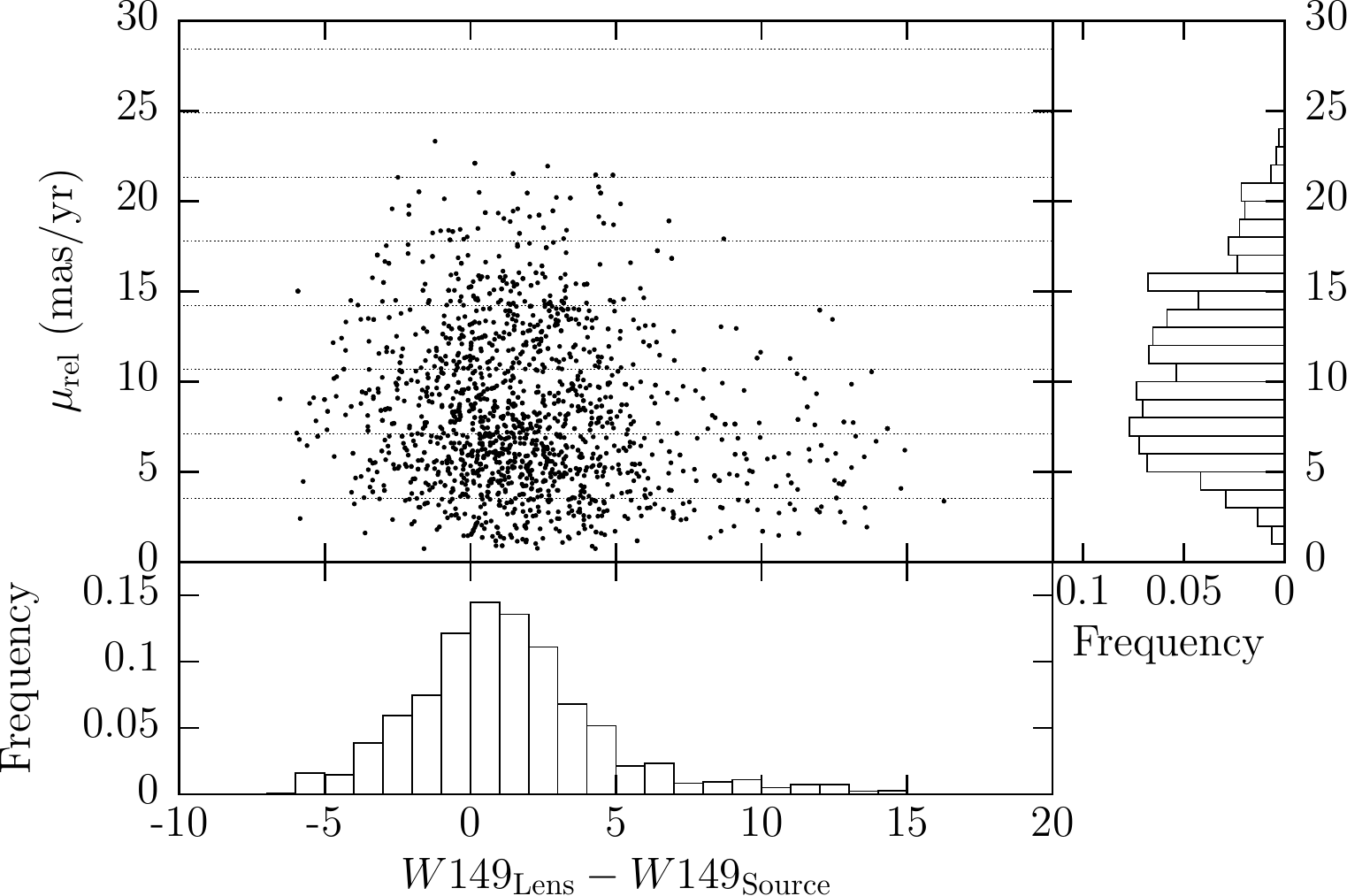}
\caption{Plot of the lens-source relative proper motion plotted against the lens-source brightness contrast for the simulated planet sample from \autoref{sensitivity}. Horizontal lines are spaced vertically by the proper motion required for the separation to change by ${\sim}0.1\times$ the PSF FWHM in 4.5 years, the spacing between the first and last microlensing season. Histograms on each axis show the marginalized distributions. Note that $\murel$ has not been corrected for the BGM1106's high proper motion dispersions as discussed in \autoref{kinematics}.}
\label{contrast}
\end{figure}

\autoref{contrast} shows the joint distribution of the lens-source brightness contrast and the lens-source relative proper motion for the simulated sample shown in \autoref{sensitivity}. These are the two intrinsic properties of the event that have the largest impact on whether \wfirst\ will be able to characterize the lens through detection of lens flux and motion relative to the source. The proper motions are not corrected in any way for the overly large proper motion dispersions in the BGM1106's bulge stars, which cause $\murel$ to be ${\sim}15$--$25$\% too large (see \autoref{kinematics} for a detailed discussion). Regardless of this, it can be seen that the majority of source-lens pairs will separate by ${\sim}10$~percent of the FWHM of the PSF in the course of the mission, and few cases will separate by more than half the PSF FWHM. By carefully modeling the PSF and the motion of the source and lens, it will be possible to measure lens fluxes and lens-source separations when the separation is significantly smaller than the PSF~\citep[see, e.g.,][]{Bennett2007,Henderson2014-ao,Bennett2015}. We leave it to future work to simulate these measurements.

\begin{figure}
\includegraphics[width=\columnwidth]{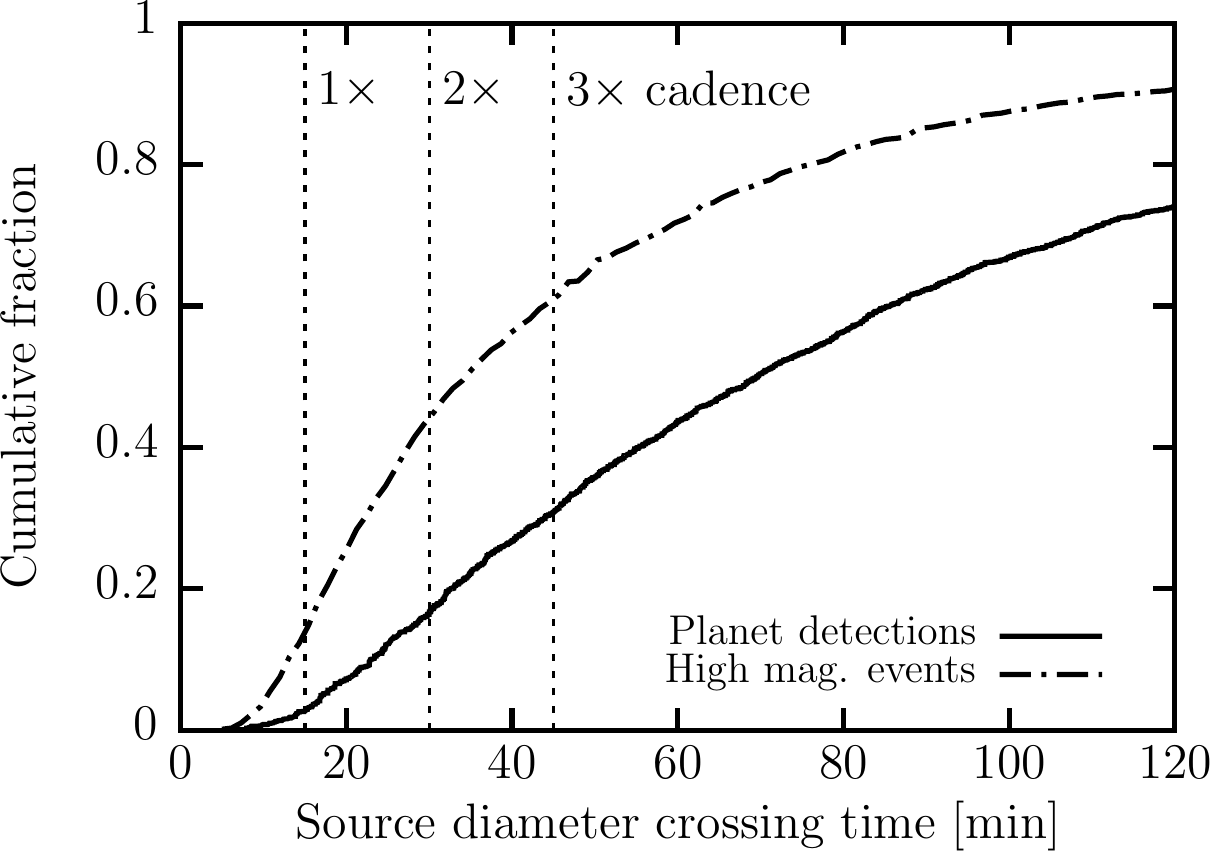}
\caption{Cumulative distribution of source diameter crossing times $2t_*$ for events with $1\mearth$ planet detections and high-magnification events with $|\uzero|<0.05$. Vertical lines indicate multiples of the 15-min observing cadence. Most events will have at least one measurement per source diameter crossing time.}
\label{sourcecross}
\end{figure}

The final property of \wfirst\ events we shall examine is the source diameter crossing time $2t_*$, where $t_*$ is the time taken for the source (relative to the lens) to traverse its radius. This is important to consider, because, in addition to the requirement of sampling the planetary perturbation, the cadence must be chosen to also sample any finite source effects, which can be used to measure the angular Einstein radius~\citep[e.g.,][]{Nemiroff1994,Yoo2004}. With \wfirst\ being sensitive to fainter, and thus smaller, source stars than any previous microlensing survey, there is a possibility that assumptions about the required cadence overlooked this fact. It is especially important to consider as \citet{Chung2017} have identified a degeneracy in measuring $2t_*$ for some single lens events when only a single measurement is affected by finite source effects. \autoref{sourcecross} shows the cumulative distribution of $2t_*$ for events with $1\mearth$ planet detections, and for high magnification events with $|\uzero|<0.05$, compared to the cadence of the \wfirst\ microlensing survey observations. Over 80\% of events with planet detections will have at least two points per $2t_*$, and over 70\% will have three points per $2t_*$. This is likely an underestimate because we expect $2t_*$ to be longer than simulated due to the overestimated velocity dispersion of the BGM1106's bulge. High-magnification events detected by \wfirst\ will tend to have shorter $2t_*$ because they can be detected on fainter source stars on average. Even so, ${\sim}60$\% of events will be guaranteed two observations per $2t_*$ and more than this can expect two measurements more often than not in any given $2t_*$-long time interval. Free-floating planet events will have brighter sources than high-magnification stellar microlensing events on average, so we conclude that for most events $15$~min cadence is sufficient to avoid the possible source radius measurement degeneracy \citep[identified by][]{Chung2017} when there are only a small number of photometric measurements over the part of the lightcurve affected by finite source effects.

\section{Evaluating the effect of changes to mission design on planet yields}\label{tradeoffs}
The design of any space mission must balance capabilities with cost. \wfirst\ straddles the boundary between a targeted, single-goal mission for which a focused set of hardware can be optimized within a relatively constrained parameter space, and a general purpose observatory where the breadth of capabilities should be optimized. Combined, \wfirst's primary missions present a relatively broad scope for optimization, though the synergies between the observational requirements of each survey and current economic considerations constrain this scope considerably. In this section we consider the effect of changes in the design of the spacecraft and the survey it carries out on the overall planet yields. We begin by outlining how changes can be quickly estimated analytically in \autoref{analytics}, present the results of two trade study simulations that we use to test the analytic estimates in Sections~\ref{bandtrade} and \ref{bgtrade}, and then apply the analytic estimates to optimize the field choice for the \wfirst\ Cycle 7 design in \autoref{fieldopt}. 

\subsection{Analytic estimates of the change in yield}\label{analytics}

\begin{figure}
\includegraphics[width=\columnwidth]{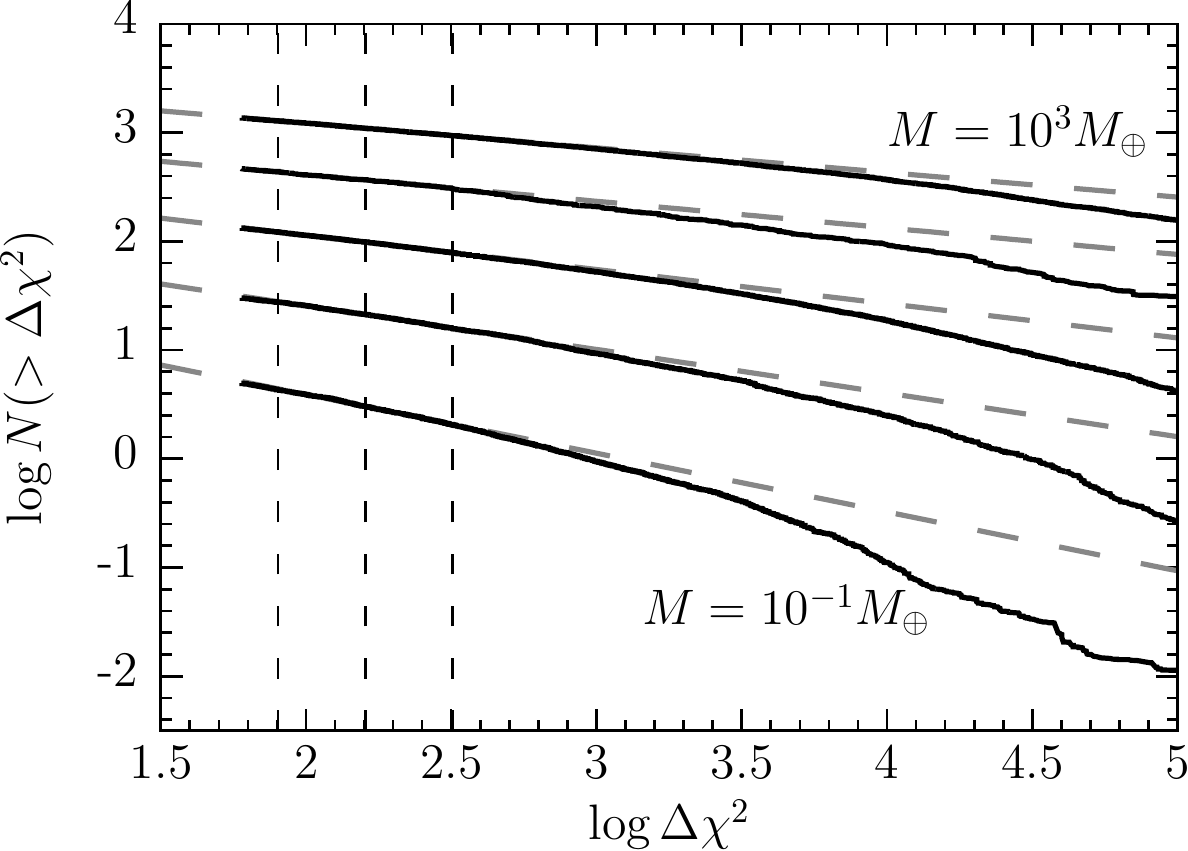}
\caption{The cumulative distribution of $\Delta\chi^2$ for different planet masses (solid black lines) ranging from $0.1$ to $1000\mearth$ in factor-of-ten steps. Gray dashed lines show the power law fits to the cumulative distributions over a factor of two above and below the adopted $\Delta\chi^2=160$ threshold (as indicated by the vertical dashed lines). The slopes of the fits are listed in \autoref{chi2alpha}}.
\label{nvdc2}
\end{figure}

\begin{table}
\caption{Power law slopes fitted to the cumulative $\Delta\chi^2$ distributions}
\begin{tabular}{lccccc}
\hline
Mass ($\mearth$)& \multicolumn{5}{c}{$\alpha$} \\
 & IDRM & DRM1 & DRM2 & AFTA & Cycle 7\\
\hline
$0.1$ & $0.674$ & $0.473$ & $0.520$ & $0.513$ & $0.534$ \\
$1$ & $0.420$ & $0.364$ & $0.355$ & $0.366$ & $0.399$ \\
$10$ & $0.324$ & $0.290$ & $0.296$ & $0.310$ & $0.313$ \\
$100$ & $0.315$ & $0.246$ & $0.241$ & $0.265$ & $0.245$ \\
$1000$ & $0.268$ & $0.212$ & $0.212$ & $0.223$ & $0.227$ \\
$10000$ & $0.201$ & $0.168$ & $0.151$ & $0.193$ & $0.204$ \\
\hline
\end{tabular}

\label{chi2alpha}
\end{table}

It is possible to estimate the effect of a change in the design of hardware or survey strategy on the total planet yield without performing a full simulation. The only detection criteria we use for bound planets is a cut on $\Delta\chi^2$. Therefore, the distribution of $\Delta\chi^2$ combined with a model of how $\Delta\chi^2$ changes with design can be used to estimate a change in yield. The cumulative distribution of the number of planet detections with $\Delta\chi^2$ greater than a threshold $X$, $N(\Delta\chi^2>X)$, can be approximated locally by a power law \citep[e.g.,][]{Bennett2002} 
\begin{equation}
N(\Delta\chi^2>X) \propto X^{-\alpha},
\label{dchi2plaw}
\end{equation}
as can be seen in \autoref{nvdc2}; the fitted slopes are listed in \autoref{chi2alpha}. The slope of the power law $\alpha$ is a function of both the planet mass and semimajor axis of the planetary companion. The range of validity of the approximation can extend by more than an order of magnitudes in $\Delta\chi^2$ in some cases, though for planets close to the edges of \wfirst's parameter space it becomes increasingly inaccurate. Finally, if one knows the ratio of the new $\Delta\chi^2$ to the old
\begin{equation}
\delta = \Delta\chi_{\mathrm{new}}^2/\Delta\chi_{\mathrm{old}}^2,
\end{equation}
then from \autoref{dchi2plaw} the estimate of the yield for the new design $N_{\rm new}$ is simply
\begin{equation}
N_{\mathrm{new}} \approx N_{\mathrm{old}}\delta^{\alpha}.
\label{newyield}
\end{equation}
To increase the range of validity of the approximation a higher order polynomial could be fit to the local cumulative $\Delta\chi^2$ distribution in $\log$-$\log$ space, or the full $\Delta\chi^2$ distribution could be used to directly evaluate the change in yield that corresponds to a given change in $\Delta\chi^2$; we assess these options in \autoref{fieldopt}.

We can calculate $\delta$ for two scenarios by computing the signal-to-noise ratios ($SNR$) for each scenario, because if the signal is constant $\Delta\chi^2\propto(SNR)^2$. The signal-to-noise ratio $SNR$ of each photometric measurement is
\begin{equation}
SNR = \frac{N_{\mathrm{s}}}{\sqrt{N_{\mathrm{s}}+N_{\mathrm{BG}}+\sigma_{\mathrm{det}}^2+\sigma_{\mathrm{sys}}^2}},
\label{snr}
\end{equation}
where $N_{\mathrm{s}}$ is the number of photons from the source, $N_{\mathrm{BG}}$ is the number of detected photons from blended stars and smooth backgrounds, $\sigma_{\mathrm{det}}$ is the uncertainty on the total number of counts due to readout noise, dark current, thermal photons from the spacecraft and other detector effects, and $\sigma_{\mathrm{sys}}$ is a systematic error component, which in this paper we assume is a small constant ($0.001$) multiplied by $N_{\mathrm{s}}$. Each element of the $SNR$ will scale differently as parameters of the telescope or survey change, and in general will be a function of magnification that will be different for each microlensing event. It will therefore be difficult to estimate the average ratio of $\Delta\chi^2$s because $SNR$ will not be a simple function of the changing parameters. However, each of the $N$ terms in the equation scale linearly with the photon collection efficiacy of each scenario, so if both of the $\sigma^2$ components are small and there is nothing to change the ratio of source to background between the different scenarios, the ratio of signal to noise will scale as the square root of photon collection efficacy and so $\delta$ will scale linearly. This will be the case if the only the exposure time were to change, but not if the mirror diameter were changed, because in the latter case the mirror diameter affects the resolution and hence the ratio of source to background photons will change from event to event.

The concept can be illustrated by an example. We may be interested in how the planet yield would be affected if the observation cadence was halved, keeping all else fixed for this example. On average, halving the number of data points will halve the $\Delta\chi^2$. However, halving the cadence allows the exposure time to be increased by a factor of $(2\texp+\tohead)/\texp=2.73$ for \afta, which assuming that systematic errors and detector noise are negligible, increases the per-exposure signal to noise by $\sqrt{2.73}=1.65$. Overall then the $\Delta\chi^2$ will increase by a factor of $\delta = 1.65^2/2=1.37$. Without worrying about the normalization of the power law in \autoref{dchi2plaw}, we can estimate that relative to the fiducial case, halving the cadence will result in a yield that is a factor $N_{\mathrm{cadence}/2}/N_{\mathrm{cadence}} \approx 1.37^{\alpha}$ times larger than the fiducial yield. So, for the examples shown in \autoref{nvdc2}, halving the cadence would result in an increase in yields of $100$\percent$\times (1.37^{\alpha}-1) \approx 9$\percent\ for $100\mearth$-mass planets where $\alpha=0.270$. For Earth-mass planets where $\alpha=0.379$, the increase would be $13$\percent.

Note, however, that the above example also presents a cautionary tale. In our pursuit of larger $\Delta\chi^2$, we have neglected the role of sampling. In order for each planet detection to be useful for demographic studies, we not only need to detect a deviation from a single-lens microlensing lightcurve at a specified significance, we also need to be able to fit the lightcurve with a unique planetary model and be sure that the deviation is not caused by systematics in the data. By halving the number of data points over a potentially short-lived planetary deviation, we have significantly degraded our power to reject systematic errors and to constrain our lightcurve model. These effects could ultimately reduce the number of useful detections by a factor larger than the increase in detections due to the improvement in the average $\Delta\chi^2$. We must therefore take care to not over interpret any of the results in this section. To properly assess the impact of changes in design therefore requires even more detailed simulations than we have conducted here, that assess not only the detection significance but also the level to which events can be characterized. However, with sensible restrictions on the survey design in place to ensure that detected events will be well characterized (such as restrictions on the minimum cadence), these simulations and analytic estimates can provide useful insight into the effect of design trade-offs.

\subsection{Testing Analytic Estimates: Bandpass}\label{bandtrade}

The filter bandpass primarily effects the amount of light that reaches the detector, but also influences the width of the PSF. Therefore, the actual effect of the bandpass on the photometric precision will change from event to event due to the differing colors of the source, lens, background and any blended stars, as well as differing amounts of reddening. 

\wfirst's HgCdTe detectors can be designed to have a specified red cutoff wavelength, with options in the range of $2.0$--$2.4$~microns considered for \wfirst. Longer cutoff wavelengths allow for larger total throughputs in the wide microlensing band, as well as a longer wavelength baseline for its more standard filters~\citep{Green2012}. However, this added capability comes at the cost of additional spacecraft and instrument cooling that is needed to reduce the thermal background emitted by the mirrors and other components in the optical path. 

To test the impact of bandpass on the planet yield of the \wfirst\ microlensing survey, we ran simulations of DRM1 with $2.0$ and $2.4$~$\mu$m cutoff detectors, using $W149$ and $W169$ filters, respectively. This scenario is also ideal for quantitatively testing the validity of our analytic relative yield estimates. It should be possible to approximate the effect of the bandpass to zeroth order by just considering the total photon throughput for a source with a spectrum that is representative of sources that will be observed. To test this assumption, we ran an additional simulation that used the $W169$ filter, but the $W149$ zeropoint magnitude, which we will refer to as $W169^{\prime}$.

\begin{table}
\caption{Relative yield of $1\mearth$ planets for DRM1 comparing $W149$ and $W169$ bandpasses.}\label{bandpasssim}
\begin{tabularx}{\columnwidth}{lXXX}
\hline
Simulation name & $W149$ & $W169^{\prime}$ & $W169$ \\
\hline
Zeropoint (mag) & 26.377 & 26.377 & 26.636 \\
Filter & $W149$ & $W169$ & $W169$ \\
\hline
Simulation & 0.910 & 0.915 & 1 \\
Prediction & 0.901 & 0.917 & --- \\
\hline
\end{tabularx}
{\bf Notes}: Yields are given relative to the standard $W169$ simulation. The $W169^{\prime}$ simulation uses the $W169$ filter, but a total throughput (zeropoint) equal to the $W149$ simulation. The filter determines the PSF, brightness of stars relative to the zeropoint magnitude and the surface brightness of the zodiacal light.
\end{table}

The relevant parameters and relative yields of these simulations are listed in \autoref{bandpasssim}. All other parameters between the three simulations are identical. We note that we should have reduced the amount of thermal noise in the $W149$ simulation, but neglected to do so. Fortunately, this mistake has essentially no effect on the result because blended light and sky background dominate over the detector noise for detector noise levels this low. The $W149$ bandpass results in a drop in yield of $9.0$~percent relative to the $W169$ bandpass. Because each simulation used the same set of simulated events, the statistical uncertainty in the relative yields is significantly smaller than the $1$~percent Poisson uncertainty on the total yield.

We predicted the relative yield using the analytic formalism from \autoref{analytics}. If we ignore detector and systematic noise sources, and assume that both blend stars and source stars have the same color, then the ratio of $\Delta\chi^2$ between $W149$ and $W169$ will depend only on the difference in zeropoints and the assumed $W149-W169$ color, which we take to be $W149-W169=0.052$, which is the median of source colors for the planets detected in the $W169$ simulation. Therefore, $\log\delta=-0.4(0.259+0.052)=-0.12$, and the predicted reduction in yield for $W149$ relative to $W169$ is $(\delta^{\alpha}-1)=9.9$~percent for the $1$-$\mearth$ DRM1 value of $\alpha=0.364$, which is in reasonable agreement with the simulation's value of $9.0$~percent. This estimate does not account for the reduced blending in $W149$ thanks to the narrower PSF, which might explain why the analytic estimate predicts a slightly larger drop in yield than the simulation produces. The same calculation for the $W169^{\prime}$ simulation that only changes the zeropoint of the $W169$ simulation, predicts a drop in yield of $8.3$~percent compared to the actual simulation result of $8.5$~percent, and is thus a much closer match. The $W169^{\prime}$ simulation shows that the majority of the change in yields is due to the change in zeropoint associated with a change in bandpass. However, because of the relatively shallow slope of the $\Delta\chi^2$ distribution, even a relatively significant change in bandpass (${\sim}40$~percent) results in a significantly smaller change in planet yield (${\sim}10$~percent), at least for Earth-mass planets. While we have only simulated the impact for Earth-mass planets, having validated the analytic approximation of the yield change, we can apply it to other planet masses. For example, we predict that switching to a $2.0$~$\mu$m cutoff for DRM1 would reduce $0.1$-$\mearth$ yields by $13$~percent.

\subsection{Testing Analytic Estimates: Background light}\label{bgtrade}

\begin{figure}
\includegraphics[width=\columnwidth]{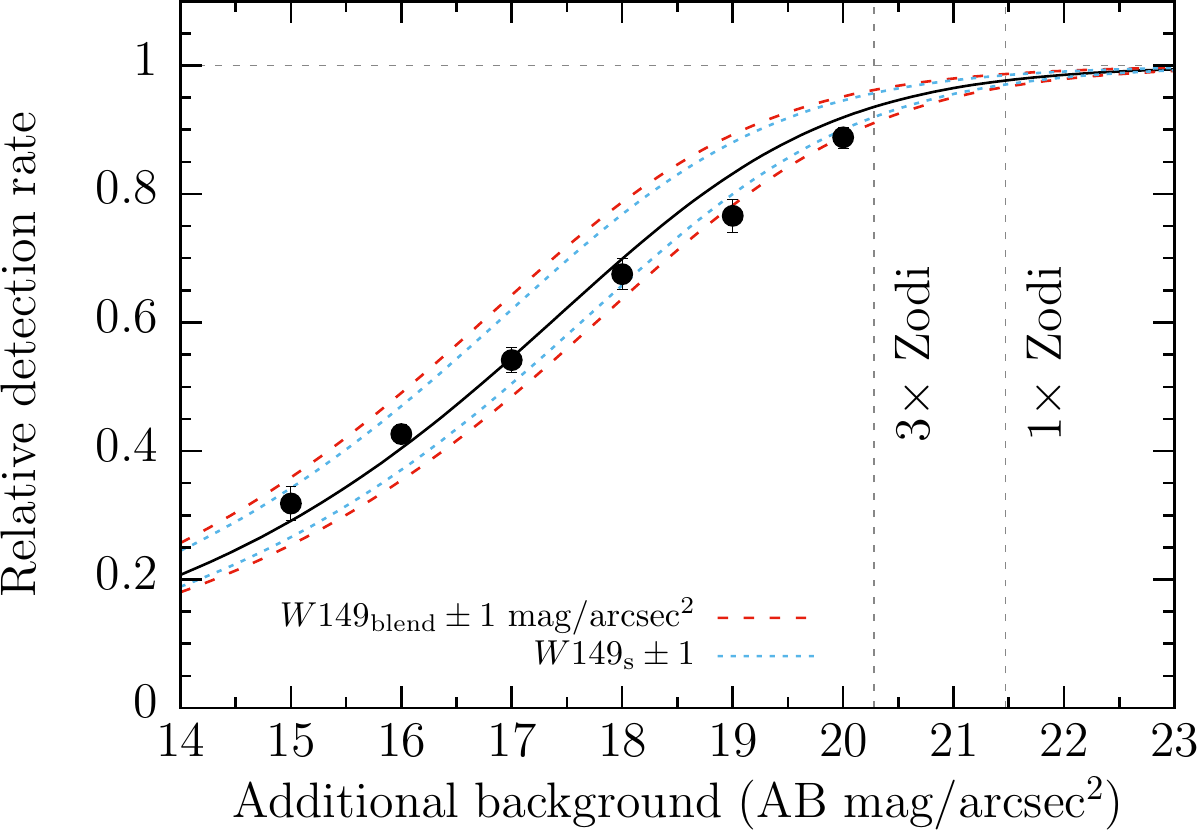}
\caption{Relative yield of $1$-$\mearth$ planets in the presence of elevated smooth backgrounds. Data points are the results of our simulation and the solid black line is our analytic prediction adopting values of the average surface brightness of blend stars $W149_{\rm blend}=19.2$~mag~arcsec$^{-2}$ and the average source star brightness $W149_{\rm s}=22$. Red and blue dashed and dotted lines show the impact of changing the values of these parameters by $\pm 1$~mag~arcsec$^{-2}$ or $\pm 1$~mag, respectively.}
\label{highbg}
\end{figure}

While there is nothing that one can do to control the amount of astrophysical background light that enters the telescope from diffuse backgrounds, it is nevertheless important to consider the effect that varying levels of background have on the yield of the survey. To investigate the impact of the zodiacal light -- the dominant smooth background -- and its variations over the course of the \wfirst\ survey we implemented a time- and position-dependent model of the zodiacal light that is described in \autoref{zodi}. We found the impact of variations in the zodiacal light over the course of \wfirst's 72-day seasons to be negligible on the overall yields. We also investigated the impact of adding additional smooth backgrounds to the images. This can be used to estimate the impact of observing when the moon lies near to the microlensing fields (if \wfirst\ is in a geosynchronous Earth orbit, as was baselined for early versions of the 2.4-m design). \autoref{highbg} shows the yield of Earth-mass planets in $0.3$--$30$~AU orbits as a function of the surface brightness of the added background relative to the case with just zodiacal light (which is $W149 \approx 21.5$~mag~arcsec$^2$). The yield drops steadily as the background surface brightness increases, but the drop is not severe unless the additional background is very bright. In order to cut the yields in half, the additional background must exceed the zodiacal light by a factor of ${\sim 80}$.
This test is also relevant to increased thermal noise backgrounds for differing telescope and instrument operating temperatures, and the choice of whether to include a cold pupil mask on the $W149$ filter.

To test the analytic estimate, we assumed that a typical source will have a magnitude of $W149_{\rm s}=22$ (the peak of the detected microlensing event source magnitude distribution, see \autoref{sourcemag}) and that the influence of blended stars would be a smooth background of $W149_{\rm blend}=19.2$~mag~arcsec$^{-2}$, based on summing up the flux of stars in BGM1106 with magnitudes $H_{\mathrm{Vega}}>15$ \citepalias[which was arbitrarily chosen to be the boundary of one of our BGM1106 catalogs, see Table~1 of][]{Penny2013}. Each simulated scenario used the same set of microlensing events, meaning that the uncertainty on the relative yield was significantly smaller than the Poisson uncertainty on any of the individual simulations. To compute the uncertainty on the relative yields, we split the simulated sample into 10 parts and computed the variance in relative yield measured for each of the subsamples.

There is good qualitative agreement between the analytic estimate and the simulation results, but, moving from right to left in the plot, the relative detection rate for analytic estimate falls off less quickly than the simulations until providing a better match at backgrounds brighter than $W149=18$~mag~arcsec$^{-2}$. Additionally, our analytic estimate depends on two model parameters (the typical source magnitude and the effective surface brightness of unresolved and blended stars). We appear to have chosen their values well, but the choices we made were somewhat arbitrary, and other choices could have been justified. This demonstrates that care needs to be taken when using analytic estimates without simulations to anchor them.

\subsection{Applying analytic estimates: Optimizing Field Selection for Cycle 7}\label{fieldopt}

More accurate models of the mass, inertia and reaction wheel complement of the spacecraft led to significantly longer estimates of the slew and settle time for the spacecraft in Cycle 7. This prompted us to perform a more detailed accounting of the survey's overheads and a re-optimization of the number and placement of the survey's fields. This process relied on analytic yield change estimates to quickly assess the yield for a large number of potential exposure times as described below. To simplify the optimization process, we constrained the cadence to be fixed at 15 minutes, and optimized only for the yield of Earth-mass planets in a broad range of orbits (i.e., with $a$ logarithmically distributed between $0.3$ and $30$~AU).

\begin{figure}
\includegraphics[width=\columnwidth]{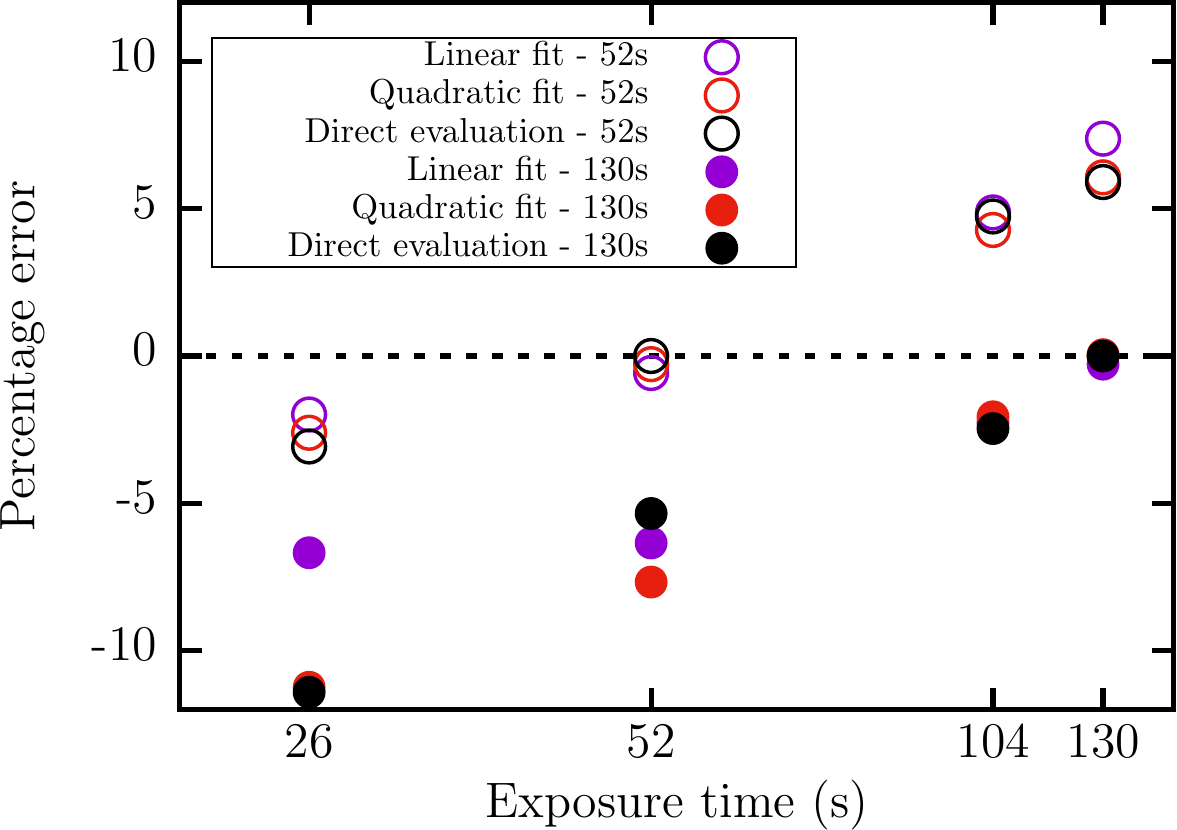}
\caption{The fractional error of analytic yield estimates compared to actual simulated yields for simulations with different exposure times, and all else held fixed (including cadence). Analytic estimates of the yield were computed based on the cumulative $\Delta\chi^2$ distributions of simulations run using 52~s (open circles) and 130~s exposure times (filled circles) for the main $W149$ survey observations. The $\Delta\chi^2$ distribution was modeled by linear fits to the cumulative $\Delta\chi^2$ distribution, quadratic fits to the distribution, and direct evaluation of the distribution (indicated by point color).}
\label{chi2test}
\end{figure}

Before beginning the optimization process, we took heed of our above warning to treat the analytic yield change estimates carefully. To test the accuracy of the analytic yield estimates, we simulated identical surveys using the Cycle 7 mission parameters at four different exposure times spanning the expected range of values to be seriously considered in the optimization exercise: $26, 52, 104$ and $130$~s. We used the $\Delta\chi^2$ distributions of the $130$~s and $52$~s simulations to predict the yield of each other simulation. We used both a linear (i.e., power law), and a quadratic fit to $\log N(\Delta\chi^2>X)$ vs $\log \Delta\chi^2$, as well as a direct evaluation of the cumulative $\Delta\chi^2$ distribution. The results are shown in \autoref{chi2test}. No particular method of estimating the change in yield showed consistently better accuracy. The error in the approximation grew by ${\sim}5$\% for every factor of two change in exposure time, whether the $52$~s or $130$~s simulations were used as the basis of the approximation. This level of inaccuracy is a reasonable price to pay for the computational cost savings the method provides.

Unlike with the previous designs, we evaluated the cumulative slew time for a given set of fields using estimates of the cumulative slew, settle and detector reset time as a function of slew angle provided by the \wfirst\ project office. This calculation was done for a large number of candidate field layouts with varying geometries and numbers of fields, with the best path being selected by brute force solution of the traveling salesman problem. The exposure time was divided evenly between the number of fields from the remainder of 15 minute cadence minus the total slew time. We constructed a map of the $1$-$\mearth$ planet detection rate per unit area for each candidate layout's exposure time $t_{\rm exp}$ by direct evaluation of the $52$~s simulation's cumulative $\Delta\chi^2$ distribution at $X=52 \text{s}/t_{\rm exp}$ individually for each sight line in the map. The total number of detections for a each layout was then evaluated using polygon clipping~\citep{gpc} to estimate the fraction of the area represented by each sight line that falls within a given chip of each field.

\begin{figure}
\includegraphics[width=\columnwidth]{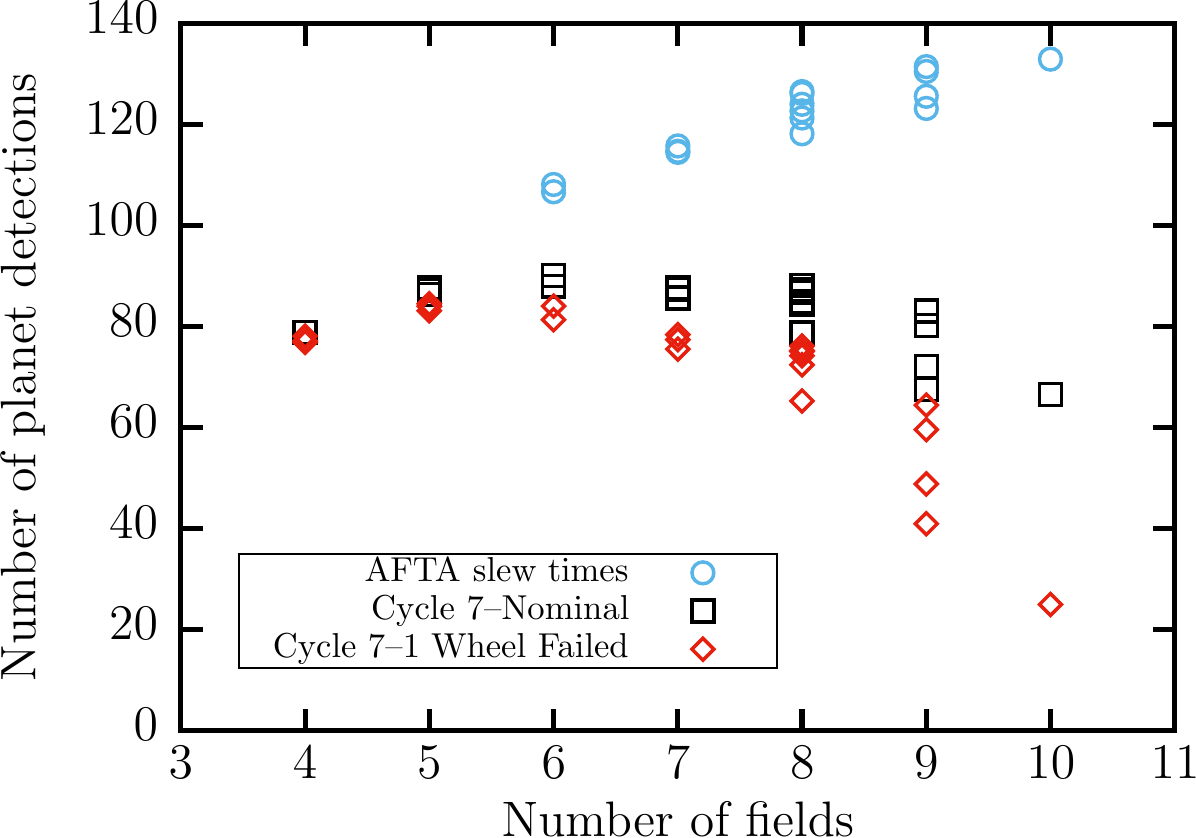}\\
\includegraphics[width=\columnwidth]{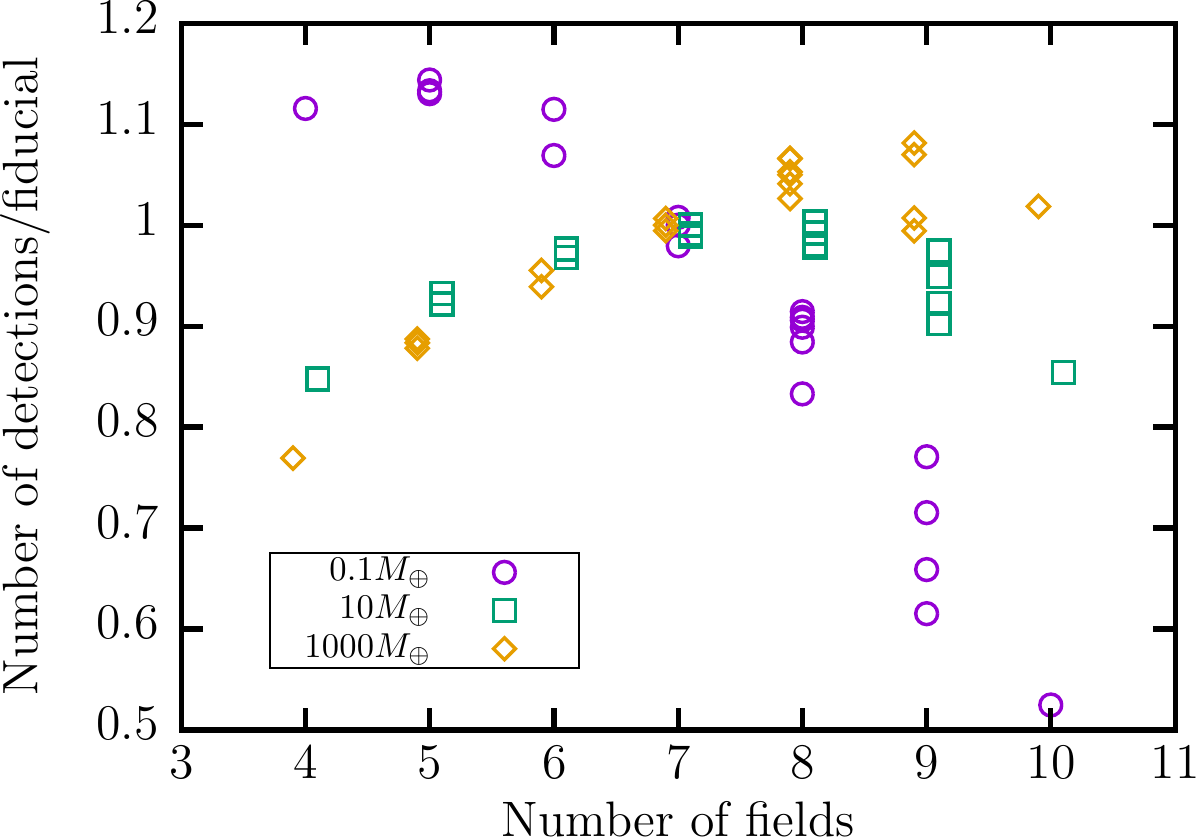}\\
\caption{\emph{Top}: $1$-$\mearth$ planet yield as a function of the number of fields for simulations of the Cycle 7 design, the Cycle 7 design with a failed reaction wheel, and the overly-optimistic AFTA slew times. Several possible layouts are considered for each number of fields. \emph{Bottom}: The planet yield relative to the adopted field layout (see \autoref{fieldplacement}) as a function of the number of fields for the nominal Cycle 7 slew times and for $0.1$-, $10$-, and $1000$-$\mearth$ planets. Different numbers of fields would maximize the yield for different planet masses.}
\label{nfields}
\end{figure}

The results of the optimization exercise are shown in the top panel of \autoref{nfields} for three different slew time versus slew angle profiles: the first is a constant slew, settle and reset time of $38$~s independent of slew distance (this was assumed in the AFTA simulations), and the other two profiles are for \wfirst\ Cycle 7 with all of its reaction wheels operational and with one reaction wheel inoperable. For each number of fields, we consider several possible layouts. For the AFTA slew times, the optimum number of fields was 10 or larger (we did not consider layouts with more than 10 fields). For more realistic Cycle 7 slew times there is a broad optimum of between 5 and 8 fields. We adopt a slightly sub-optimal field layout of 7 fields (shown in \autoref{fieldplacement}) to allow for some margin in yields. Note that the relatively coarse resolution of our event rate map, uncertainties in the yields for each sight line, and use of the analytic yield estimates prevent accurate determination of the true optimum field layout within the broad optimum we find. With one reaction wheel in-operational the optimum number of fields would be fewer, at 5 or 6. This optimum is somewhat sharper than the optimum for all wheels operational. The factor of ${\sim}2.2$ reduction in estimated slew performance between AFTA and Cycle 7 results in a large reduction in planet yield of ${\sim} 50$\%, which accounts for essentially all of the difference between the AFTA and Cycle 7 yields that we find. The optimum number of fields will depend on planet mass, as shown in the lower panel of \autoref{nfields} shows the yield as a function of field number for $0.1$-, $10$-, and $1000$-$\mearth$ planets.

\section{Discussion}\label{discuss}

The statistical power of an exoplanet survey to infer demographics is directly related to the expected yield of the survey assuming a given exoplanet population. The ability to accurately estimate the survey yields is therefore an important input into mission design. Nevertheless, yield predictions have numerous potential sources of uncertainty, and it is just as important to understand these. We therefore devote this section to summarizing the sources of uncertainty in our yield predictions and suggesting ways in which this uncertainty can be reduced.

The sources of uncertainty in our results can be broken down into three broad categories. The first is due to our ability to simulate how the spacecraft collects data and how it will be processed. The second is due to our ability to measure and model the astrophysical components that produce microlensing events, i.e., the Galaxy and its stellar populations. Finally, our assumptions of the planetary population impacts the mission yields that we predict.

\subsection{Simulation Uncertainties}

A principle concern when building a simulation is the balance between realism and computational cost. One is invariably forced to make compromises on the former in order to obtain a manageable run time. By building simulated images of the \wfirst\ microlensing fields, combined with a model of smooth backgrounds, our simulations should reasonably capture all significant sources of photon noise. Our simulation of the detectors is somewhat simplistic; we assume that \wfirst's HAWAII-4RG detectors behave like CCDs, which is probably reasonable given that photon noise always dominates over read noise for the microlensing fields and their exposure times. The most significant form of uncertainty in the data collection and processing category is the processing element. Our simulations perform a simple aperture photometry noise calculation and limit its precision with a constant 1~millimagnitude term added in quadrature. Ultimately, the use of a small, fixed, unweighted aperture should result in an underestimate of the achievable photon-noise limited photometric precision, which helps to offset our inability to simulate all of the imperfections between photons entering the telescope and a photometric measurement. However, there are a number of additional steps between measuring photometry and declaring a planet detection that could be significantly affected by sources of systematic noise (instrumental and/or astrophysical) that we do not simulate. Estimating the impact of these sources of noise is challenging, and would likely require a full end-to-end simulation. Realistic noise simulations based on lab tests of H4RG detectors~\citep[e.g.,][]{Rauscher2015} will help in this regard. We note that the impact of any change in the simulation of photometry, insofar as it changes the photometric precision by a uniform scaling factor, can be estimated by changing the $\Delta\chi^2$ threshold for declaring a detection.

Another important uncertainty in the simulations is the ability to convert a detection, i.e., a signal above the $\Delta\chi^2$ threshold, into a bone fide planet with measured parameters. This process can be affected by various discrete and continuous degeneracies~\citep[see][for a review]{Gaudi2012} that can lead to ambiguity between planetary interpretations and stellar binary lenses and sources as the cause of the lightcurve anomaly. These ambiguous events can be dealt with by Bayesian probability accounting, but naturally add uncertainty to any inferences, especially if they constitute a significant fraction of the potential planet sample. Reassuringly, \citet{Suzuki2016} found only 1 out of 23 events in their systematically selected sample of planets had an ambiguous binary lens interpretation, and 6 out of 23 events suffered a close-wide degeneracy that impacted the measurement of the projected separation $s$, but did not significantly affect the mass ratio $q$. The improved photometry possible from space will likely resolve some fraction of the degeneracies and ambiguities that are seen in ground-based data, but the more subtle features detectable in space-based data may introduce a higher fraction of ambiguous events; e.g., in a simulation of a high-cadence microlensing survey with uniform photometry \citet{Zhu2014} found 55\% of planets were detected without caustic crossings. Additionally, in this work we have not simulated the measurement of planet and host masses that observations from space enable. For the MPF mission design, with pixels more than twice the size of \wfirst's Cycle 7 design, \citet{Bennett2010} estimate that more than half of the planetary events will have better than 10\% mass measurements via some form of direct detection of lens light. These estimates do not account for the potential contamination of the measurement by either bound stellar companions to the source or lens~\citep{Henderson2015-ao,Koshimoto2017-950}

\subsection{Galactic model uncertainties}

The microlensing event rate and the properties of the microlensing events will depend on the distribution of stars in the Galaxy, their kinematics and their masses. No model of the Galaxy will be able to fully capture its complexities, and so our estimates will have some degree of uncertainty due to any shortcomings of the model. In \autoref{ratecorr} we estimated the corrections necessary to match the microlensing event rate of the model. In this section we examine further the possible causes for the BGM1106's under-prediction of event rates and the impact of any model uncertainties or errors on the properties of the microlensing events. We note that this discussion only applies specifically to version 1106 of the \besancon\ model that we have used in this paper.

\subsubsection{Bar angle}

In BGM1106 the Galactic bulge is modeled as a triaxial bar with an angle of $12.5\degr$ to the Sun-Galactic center line and a major axis scale length of $1.63$~kpc. This is in contrast to modeling of the distribution of red clump giants found in the OGLE and VVV surveys, which find a bar angle ${\sim}30\degr$ and a major axis scale length of ${\sim}0.7$~kpc~\citep{Stanek1994,Rattenbury2007,Wegg2013,Cao2013}. With no distance information, both models can reproduce the 2-d distribution of bulge stars on the sky (note that $1.63$~kpc~$\sin 12.5\degr \approx 0.7$~kpc~$\sin 30\degr \approx 0.35$~kpc). Red clump stars are standard candles~\citep{Stanek1994}, and so can trace out the third dimension of the bulge density distribution. Along the line of sight $(\ell,b)=(1.25,-2.65)$ the BGM1106 predicts a distance modulus dispersion of $0.36$~mag for bulge stars, which is much larger than the value of $0.20$~mag that \citet{Nataf2013} measure from OGLE clump giants after subtracting in quadrature an intrinsic magnitude dispersion of $0.09$~mag and an extinction dispersion of ${\sim}0.11$~mag (the total observed magnitude dispersion of the red clump is therefore $0.24$~mag). \citet{Simion2017}, working with the Galaxia code that implements the 2003 \besancon\ model~\citep{Sharma2011}, found that slightly smaller bar angles of $20$--$25\degr$ provided the best fit to VVV red clump counts, but that there was some degeneracy between the bar angle and the red clump dispersion due to sources other than distance dispersion.

We found in \autoref{starcounts} that the BGM1106 only slightly under-predicts bulge star counts, so the principle impact of the bar angle is only to spread bulge stars along the line of sight. This was confirmed by comparing the red clump star counts of \citet{Nataf2013} as a function of $(\ell,b)$ to the total stellar mass of the BGM1106 bulge.

The line of sight distribution of lenses and sources affects the distribution of microlensing event properties. The Einstein radius depends on the relative distances of the source and lens as 
\begin{equation}
\re = \sqrt{\frac{4G}{c^2}M\dl(1-\dl/\ds)},
\end{equation}
where $G$ and $c$ are the gravitational constant and speed of light, respectively, and $\dl$ and $\ds$ are the lens and source distances, respectively. The larger line of sight distance dispersion for the BGM1106 relative to that measured will result in ${\sim}15$\% larger Einstein radii for events with bulge lenses and bulge sources (bulge-bulge lensing), and a smaller impact on events with disk lenses and bulge sources (bulge-disk lensing). Event timescales will be larger by the same degree, and the ratio of bulge to disk lenses will also be increased. However, with only a maximum effect of ${\sim}15$\%, the impact will be relatively minor.

\subsubsection{Bulge kinematics}\label{kinematics}

In addition to the bar angle, the kinematics of the bulge stars are not in agreement with measurements. We compared the predicted BGM1106 proper motions to {\it Hubble} Space Telescope (HST) proper motion measurements in the ${\sim}11$~arcmin$^2$ SWEEPS field~\citep{Clarkson2008}. \citet{Clarkson2008} separated bulge and disk populations by selecting stars above the bulge main sequence turn off where the disk's main sequence stars were well separated from the giant branch, which is dominated by bulge stars. For this comparison, we combined the BGM1106 catalogs of the two sight lines closest to the SWEEPS field at $(\ell,b)=(1\fdg1,-2\fdg7)$ and $(1\fdg35,-2\fdg7)$ to improve the statistics for these bright stars. We roughly mimicked the \citet{Clarkson2008} selection in our BGM1106 catalogs by selecting stars with $15.95<I_{\mathrm{AB}}<17.95$ and assigned those with $(I-J)_{\mathrm{AB}}<0.5$ to the blue (disk proxy) population and those with $(I-J)_{\mathrm{AB}}>0.58$ to the red (bulge proxy) population. The two catalogs combined represent stars drawn from a solid angle of $1.44$~arcmin$^2$, and there were a total of $37$ stars in the blue disk proxy sample and $105$ stars in the red bulge proxy sample. 

\citet{Clarkson2008} measured their proper motions in an arbitrary reference frame, so we can only compare the blue-red proper motion offsets and the proper motion dispersions. They find an offset between the blue and red population proper motions of $(\Delta\mu_{\ell},\Delta\mu_b) = (3.24\pm 0.15,-0.81\pm 0.12)$~mas~yr$^{-1}$, where the $\Delta$ represents blue minus red. For the BGM1106 proper motions we find $(\Delta\mu_{\ell},\Delta\mu_b) = ( 3.53\pm 0.65,-0.12\pm 0.32)$, which are largely consistent with each other.

\citet{Clarkson2008} do not report the proper motion dispersions they measure, but we are able to extract them from their Figure~21, finding $(\sigma_{\ell},\sigma_b)=(2.2,1.3)$~mas~yr$^{-1}$ for the blue (disk) population and $(\sigma_{\ell},\sigma_b)=(3.0,2.8)$~mas~yr$^{-1}$ for the red (bulge) population. Individual proper motion uncertainties in the HST data are likely below $0.3$~mas~yr$^{-1}$ for each star, so have a negligible impact on the measured dispersions, and the sample size is larger than our comparison sample so the statistical uncertainty in the estimates of the HST proper motion dispersions will be insignificant in our comparison.
For the blue BGM1106 population we find proper motion dispersions of $(\sigma_{\ell},\sigma_b)=(2.47\pm 0.29,1.11\pm 0.13)$, which is largely consistent with the HST measurements in both axes. For the red BGM1106 population we find $(\sigma_{\ell},\sigma_b)=(5.19 \pm  0.36,2.64 \pm  0.18)$; the latitudinal dispersion is consistent with the HST measurements, but the longitudinal dispersion is too large by a factor of $1.73 \pm 0.12$.

\begin{figure}
\includegraphics[width=\columnwidth]{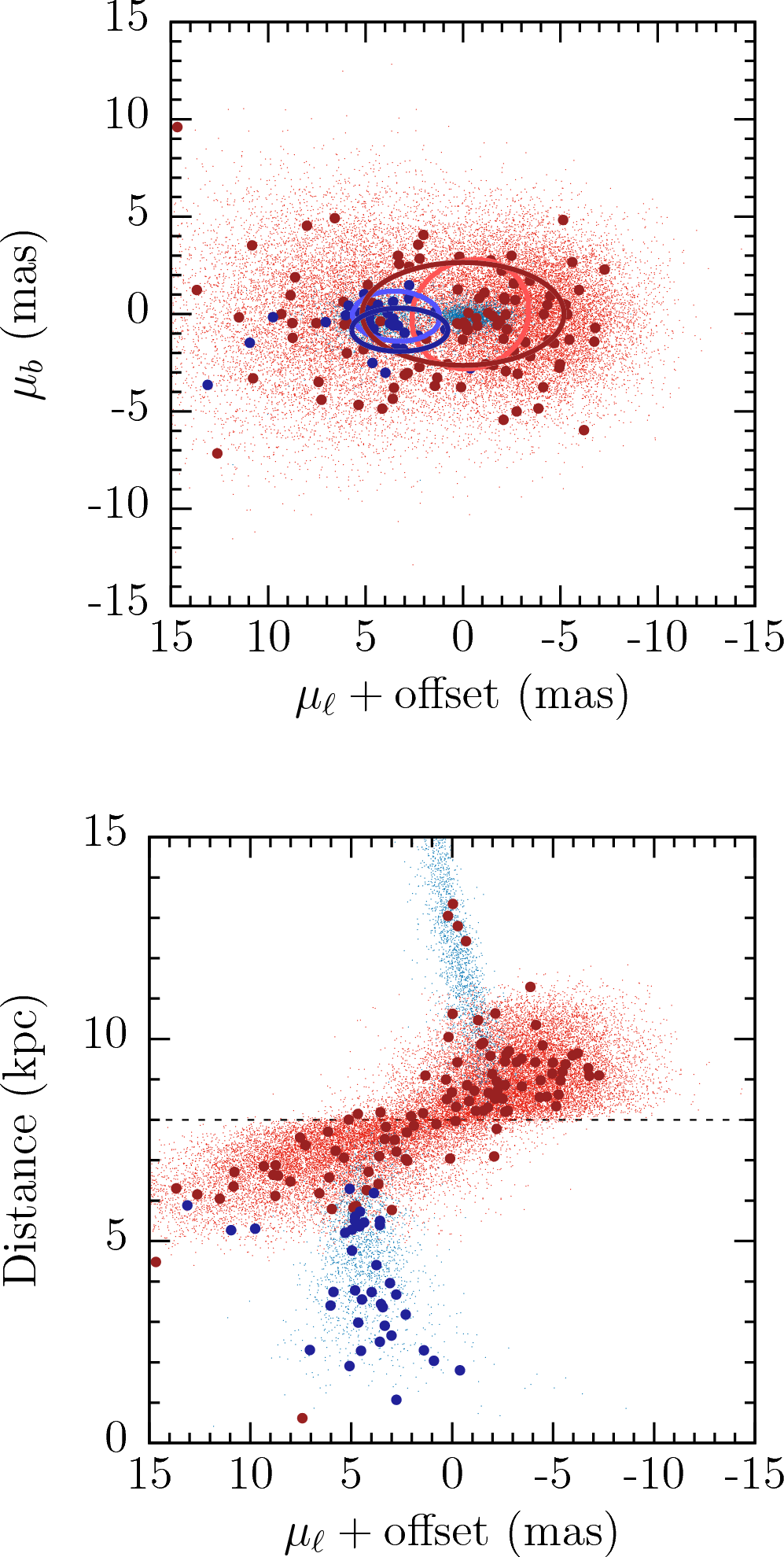} 
\caption{Comparison of BGM1106 model proper motions with those measured using \hst\ by \citet{Clarkson2008}. \emph{Top}: Proper motion vector point diagram. Small points are all potential source stars ($H_{\rm Vega}<25$) from the BGM1106 (light blue are disk stars, while light red are bulge stars). Larger blue and red points are BGM1106 stars belonging to disk and bulge proxy populations, respectively, with selections designed to replicate those of \citet{Clarkson2008}. Light blue and red lines are the $1$-$\sigma$ proper motion dispersion contours measured by \citet{Clarkson2008} for disk and bulge proxy stars, respectively, while the darker lines are the same $1$-$\sigma$ dispersion contours, but for the BGM1106 disk and bulge proxy stars.  \emph{Bottom}: Distance versus $\ell$ proper motion for the same BGM1106 stars as in the top panel. The CMD cuts do a good job of isolating disk and bulge populations, though with some cross contamination. The BGM1106 disk proper motions match the data well, but the bulge proper motions have a dispersion that is too large the $\ell$ direction. The dotted line shows the distance of the Galactic center in the model.}
\label{clarksoncomp}
\end{figure}

We can get an idea for the cause of the discrepancy between the BGM1106 and the data by looking at the proper-motion vector point diagram and the longitudinal proper motion plotted against distance, both shown in \autoref{clarksoncomp}. The first thing to notice is that the color-magnitude selection does a good job separating near-side disk stars from bulge stars, admittedly with a small degree of cross-contamination between the disk and bulge populations. Also, the selected stars do not appear to be a significantly biased subset of the underlying population. This means that we cannot ascribe the discrepancy to a difference in the selection of the stars for each population between the observation and the model. The random velocity dispersions in the BGM1106 bulge population are $(\sigma_U,\sigma_V,\sigma_W)=(113,115,100)$~km~s$^{-1}$, which correspond to proper motions dispersions of the order of $3$~mas~yr$^{-1}$ at a distance of $8$~kpc, i.e., enough to account for all of the measured HST value of $\sigma_{\ell}$. In addition to the dispersion component, the BGM1106 bulge stars also have an additional solid body rotation component, rotating at $40$~km~s$^{-1}$~kpc$^{-1}$. The combination of the longer bar and its small angle lead to a range of solid body rotation velocities from ${\sim}-60$~\kms\ on the near end of the bar to ${\sim}+50$~\kms\ as the sight line leaves the far side. This results in a range of $-1.9$ to $+1.1$~mas~yr$^{-1}$ in proper motion, that would corresponds to an additional dispersion of ${\sim}0.87$~mas~yr$^{-1}$ to be added in quadrature. Finally, inspection of the BGM1106's $V$-component velocities (in a $UVW$ system) as a function of distance suggests that in addition the solid body rotation and the random dispersions there is an additional component similar to the rotation curve of a stellar disk in a dark matter halo. This results in a ${\sim}300$~km~s$^{-1}$ offset between the mean velocities of bulge stars that are ${\sim}1$~kpc from the Galactic center on opposite sides. This causes a rapid change of ${\sim}8$~mas~yr$^{-1}$ in the mean proper motion of stars as function of distance at the distance of the Galactic center.  This can be seen in the lower panel of \autoref{clarksoncomp} as an offset in proper motions at a distance of ${\sim}8$~kpc. This offset results in additional ${\sim}4$~mas~yr$^{-1}$ to be added in quadrature to the proper motion dispersion. Each of the three sources of dispersion combined in quadrature result in a dispersion of $5.1$~mas~yr$^{-1}$, which is consistent with the value of $\sigma_{\ell}=5.19\pm0.36$ we measured from the catalogs.

The addition of a potential quasi circular velocity component to the bulge stars' velocities appears to be in error, because we expect the bulge stellar population to be pressure supported with a sub-dominant cylindrical rotational component to provide the pattern speed of the bar. In the model, however, the quasi circular velocity component together with the range of bulge star Galactocentric distances, leads to this velocity component dominating the longitudinal proper motion dispersion. The dichotomy in near-side bulge versus far-side bulge velocities imposed by a large circular velocity also increases the event rate of events with far bulge sources and near bulge lenses, which have the largest Einstein radii, so will increase the mean Einstein radius of bulge-bulge lenses somewhat. The increased longitudinal proper motion dispersion of bulge stars will also increase the relative event rate of bulge-disk lensing. In all cases, the BGM1106's event timescales will be shorter than would be produced by more realistic kinematics.

\subsubsection{Bulge initial mass function}\label{bulgeimf}

As discussed already in \autoref{starcounts}, the BGM1106's mass function in the bulge differs from typically assumed mass functions~\citep[e.g.,][]{Kroupa2001} in both its shallow slope $\dd N/\dd M \propto M^{-1.0}$ and its high lower mass cut off of $0.15\msun$. We found that replacing the mass function with something more reasonable, such as from \citet{Sumi2011}, a \citep{Kroupa2001} slope of $M^{-1.3}$ between $0.08$ and $0.7\msun$ and a slope of $M^{-0.5}$ between $0.01$ and $0.08\msun$, would improve the match of the shape of the BGM1106's luminosity function to measurements from \citep{Calamida2015}. \citet{Wegg2017} find that a similar mass function, when combined with an $N$-body bulge model fit to infrared star counts and radial velocity distributions, can simultaneously fit both microlensing optical depths and timescale distributions, though the slope of the brown-dwarf mass function in the bulge remains uncertain ($-0.65\pm0.89$). We note that we have not considered the effects of age or metallicity that can affect star counts, especially for evolved stars, or in the case of metallicity, M-dwarfs as well.

Adding stars and brown dwarfs below $0.15\msun$ to the BGM mass function would increase the optical depth and event rate per star for bulge-bulge lensing by factors of ${\sim}1.9$ and ${\sim}3.4$, respectively. The disk's mass function in the BGM1106 has a more typical slope, and extends down to $0.08\msun$ but adding brown dwarfs would increase the optical depth and event rate for bulge-disk lensing, also, but to a lesser degree than for bulge-bulge lensing. It is therefore likely that the form of the mass function can explain a significant amount of the factor of ${\sim}2.1$ under-prediction of the event rate by the BGM1106, the factor of ${\sim}1.8$ under-prediction of the optical depth, and its shallower slope of the luminosity function. Adding low-mass stars to the mass function would also act to decrease the mean event timescale~\citep{Awiphan2016}. 

\subsubsection{Extinction}

The BGM1106 uses the 3-d extinction model of \citet{Marshall2006} to provide extinction as a function of distance, and the \citet{Cardelli1989} reddening law with $R_V=3.1$ to convert extinctions in the $K_s$-band to other wavelengths. \citet{Schultheis2014} assessed the performance of various 2- and 3-d extinction maps and found that generally 3-d extinction maps were accurate, but failed along certain site lines. Numerous studies have shown that the reddening law towards the bulge differs from the $R_V=3.1$, with $R_V{\sim}2.5$ more typical~\citep[e.g.][]{Nataf2013}. It is even possible that the reddening law deviates from a power law in the $1$--$2$~$\mu$m range~\citep{Hosek2018}.

Despite this uncertainty, it is likely that the impact of errors in the extinction will only have a small impact on the predicted yields. This is simply because the total extinction across our fields is only $A_H{\sim}0.6$, though it does reach $A_H{\sim}1$ in small parts of the fields closest to the plane. Therefore, even a large fractional error corresponds to a relatively small absolute error. We can use the analytic framework in \autoref{analytics} to estimate the impact. Errors in the extinction will affect all stellar noise (source and blends) equally. A $33$\% under-estimate of the extinction, or $0.2$~mag, reduces flux and $\Delta\chi^2$ by 17\%. Using the slope of the $\Delta\chi^2$ distribution from \autoref{chi2alpha}, $\alpha=-0.399$ for $1$-$\mearth$ planets and the Cycle 7 design, the under-prediction of extinction  would results in an over-prediction of the Earth-mass planet yield of ${\sim}7$\%. Averaged over the whole proposed fields, an error this large seems unlikely. However, we note that the relatively coarse extinction map we have used (resolution $0\fdg25\times0\fdg25$) may have impacted our field optimization. 

\subsubsection{Impact on {\it WFIRST}'s planet yield}

In \autoref{ratecorr} we have adopted an event rate scaling to match measured microlensing event rates per red clump source and the number of faint sources. This correction factor will be valid to first order because the majority of \wfirst's sources will be in the bulge, like the red clump stars. Therefore, the issues raised with the model in this section should not affect our predictions for the number of microlensing events \wfirst\ will detect. However, they will affect the properties of the events, which may impact the detection efficiency.

The lack of low-mass stars and brown dwarfs means the mean lens mass is too large, and at fixed planet mass the mass ratio will be too small. This has the effect of reducing our detection efficiency per event slightly at fixed planet mass. The BGM1106's bulge kinematics result in timescales that are too short, which results in reduced planet detection efficiency due to shorter planetary anomalies. The increase in Einstein ring radii due to the elongated bulge will increase timescales slightly, but not enough to counteract the effect from kinematics. 

In addition to the average detection efficiency, the mass function and bar angle issues cause the BGM1106 to overestimate average Einstein radius. This means that its peak sensitivity to planets will be at slightly smaller semimajor axis than indicated by \autoref{sensitivity}.

We have concluded that the BGM1106 probably under-predicts the number of source stars because of its shallow mass function slope. However, we corrected for the under-prediction by simply multiplying the event rate for all source stars, which likely has the effect of over-correcting for bright stars, brighter than $F814W_{\rm Vega}\sim 22$. Roughly two thirds of $1\mearth$ planet detections come from events with source stars above this boundary, which would imply that the an over-correction of bright source stars leads to an over estimate of ${\sim}14$\% in Earth-mass planet yield, and probably a larger overestimate for the most difficult to detect planets. However, if we had estimated our event rate scaling using the MOA all star event rates instead of  the extended red clump event rates (see \autoref{gammacomp}) and only used data in the range of Galactic latitudes where \wfirst\ will probably observe, we would have derived a larger correction.

In summary, the issues with the Galactic model after corrections act to both increase and reduce the detection efficiency or number of events by mostly small factors. To a certain degree then we can expect the effects of different signs to cancel, and the associated uncertainties to grow in quadrature. It is likely therefore that a single large uncertainty will dominate over smaller uncertainties. The quantity with the largest uncertainty is probably the microlensing event rate, due to the relatively low signal to noise ratio when subdivided down to square-degree scales, and the need to extrapolate closer to the Galactic plane. We reiterate, however, that while there will remain significant uncertainties in the absolute yield predictions, the relative yields between designs simulated with the same methodology and common parameters and assumptions will be much less uncertain.

\subsection{Planet Population Uncertainties}

The uncertainty in our assumptions about the population of planets is large, and some regions of the parameter space are completely unconstrained. A major goal of the survey, after all, is to detect and measure the occurrence rate of the cold planet population that can not be conducted by any other method, or by microlensing observations from Earth. Nevertheless there are measurements of planet abundances in the regions of \wfirst's sensitivity from microlensing surveys~\citep{Sumi2010,Gould2010,Cassan2012,Shvartzvald2016,Clanton2016,Suzuki2016}, and near it from transits~\citep[e.g.,][]{Burke2015}, radial velocity~\citep[e.g.,][]{Cumming2008, Johnson2010, Bonfils2013, Montet2014} and direct imaging~\citep{Nielsen2010,Chauvin2015,Bowler2015}. These observational constraints allow us to anchor extrapolations into the regions that are unexplored. 

Our fiducial joint mass-semimajor axis occurrence distribution (in many places we have simply referred to this as our fiducial mass function) assumes a broken power law in mass, and a log-uniform distribution in semimajor axis. The high-mass end of the mass function was chosen to match the estimate of the mass function from \citet{Cassan2012} based on microlensing searches. This measured mass function only extends down to ${\sim}5\mearth$, so we chose to saturate the power law at a value of $2$~planets per dex mass per dex semimajor axis per star at $5.2\mearth$ to prevent overly optimistic predictions of large numbers of low mass planets. Since adopting this as our fiducial mass function, several studies have advanced upon the \citet{Cassan2012} result on which it was based. While we have chosen to retain the fiducial mass function for consistency and easy comparison with the past \wfirst\ reports, it is worth examining what impact that adopting another joint mass-semimajor axis function would have. 

\begin{figure}
\includegraphics[width=\columnwidth]{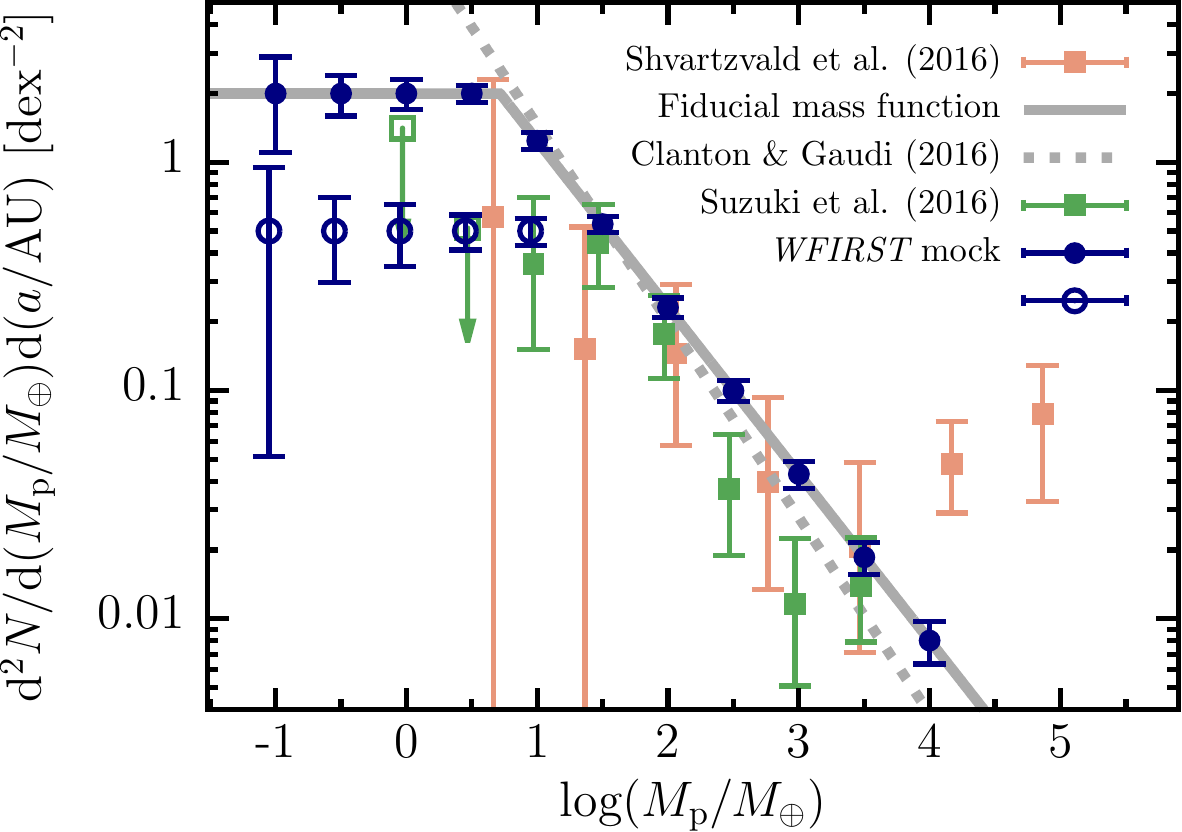}
\caption{Comparison of our fiducial mass function (solid gray line) to the latest measurements based on microlensing data. Salmon and green points with error bars show the planet occurrence as a function of mass ratio converted into planet masses, assuming a $0.5\msun$ host star, from \citealt{Suzuki2016} and \citealt{Shvartzvald2016}, respectively, and the dotted gray line shows a fit to microlensing, radial velocity trends, and direct imaging data by \citet{Clanton2016}. Dark blue data points show the mass function measurement precision of a mock survey by \wfirst\ (Cycle 7 design) in 0.5~dex mass bins assuming only half of the planet detections can be used; filled points follow the fiducial mass function, while open points follow a mass function that saturates at an occurrence rate a factor of 4 lower than the fiducial mass function.}
\label{massfnplt}
\end{figure}

\citet{Suzuki2016} analyzed a larger and more homogeneously selected \citep[compared to][]{Cassan2012} set of microlensing planet detections from the MOA survey. \citet{Shvartzvald2016} also studied the mass ratio function with a smaller sample of events. Both mass ratio function measurements are shown in \autoref{massfnplt} in comparison to our fiducial mass function. To convert mass ratio to planet mass we assumed a host mass of $0.5\msun$. \citet{Suzuki2016} found evidence for a turnover in the planet mass {\it ratio} function at $q\approx1.7\times10^{-4}$, which corresponds to a planet mass of ${\sim}30\mearth$ assuming a mean host mass of $0.5\msun$. This turnover appears to be confirmed by an independent analysis using very different methods~\citep{Udalski2018}, but the exact shape of the turnover, and the steepness of the subsequent decline below this mass, are very poorly constrained. In a bin centered at $\mpl\approx 3\mearth$ \citet{Suzuki2016} place a $95$\% upper limit on the mass function of $0.5$~planets per dex$^2$, about a factor of four below the value of our saturated fiducial mass function. At ${\sim}30$--$100\mearth$ the fiducial mass function is a good match to the \citet{Suzuki2016} data, but at larger masses it again overestimates the measurements, but to a lesser degree than at small masses. All together, this suggests that our total yields will be overestimated somewhat, and the yields at low masses could be overestimated significantly.

It is important to recognize that the planet yields assuming a fiducial mass function that we estimate in this study are only a useful proxy for the true value of the survey which is planet sensitivity, or its power to measure the planet occurrence rate. This sensitivity is independent of the assumed mass function, and, as can be seen from \autoref{sensitivity}, the sensitivity will extend down to ${\sim}$few$\times10^{-2}\mearth$. In this sense the sensitivity range is the point at which no detections of planets of mass $\mpl$ during the mission ceases to be an interesting constraint on planet occurrence. In \autoref{massfnplt} we show that even with a mass function saturation value a factor of $4$ lower than our fiducial mass function, \wfirst\ will detect a sufficient number of planets to measure the mass function in $0.5$~dex bins to below $1\mearth$, and it would set interesting upper limits on the planet occurrence at masses below this. In these estimates we assumed that only half of \wfirst's planets would be utilized in this mass function measurement. Over its entire range it will provide measurements of the mass function with far greater precision than is currently possible from ground-based surveys. While not a direct comparison, \citet{Henderson2014-kmt} predict that KMTNet, during its nominal 5 year survey, would find a factor of ${\sim}16$ fewer 1-$\mearth$ planets than \wfirst\ will find in its nominal (Cycle 7) survey.

In addition to uncertainties in the planet mass function, there is even greater uncertainty in the form of the planet occurrence as a function of semimajor axis near to \wfirst's peak sensitivity. \citet{Clanton2014} have shown that that there is at present very little overlap in the sensitivity regions of current microlensing and radial velocity (RV) surveys, but radial velocity surveys tend to show an increase in planet occurrence with log semimajor axis or log period~\citep[e.g.,][]{Cumming2008,Bonfils2013} that appears to be consistent with microlensing occurrence rates when extrapolated~\citep[see, e.g.,][]{Gould2010,Suzuki2016}. Results from \kepler\ show a similar rising trend for large planets, but a shallow decline in occurrence beyond $P{\sim}10$~d for planets smaller than Neptune~\citep[e.g.][]{Petigura2018}; whether these trends continue throughout \wfirst's region of sensitivity is unconstrained at present. At larger orbital separations, \citet{Clanton2016} find that a cut-off in planet occurrence at ${\sim}20$~AU is required to remain consistent with microlensing, radial velocity trends, and direct imaging results. 

\subsection{Future Improvements}

To make further progress on estimating the yields of the \wfirst\ survey requires work on several fronts. The most critical need is observational, due to the long lead time necessary to observe, analyze and interpret new data. Advances in simulations are also needed to better understand the relative importance of mass measurements in optimizing \wfirst's fields and observing strategy.

Observationally, the most important measurement to make is of the microlensing event rate in the potential \wfirst\ fields, in the infrared and to a depth as close as possible to that achievable by \wfirst. Such a measurement will also test the ability of event rate models based only on star counts~\citep{Poleski2016} to predict event rates closer to the Galactic plane. This requires an infrared microlensing survey in order to penetrate the dust near the Galactic plane and reach sources throughout the bulge and into the far side of the disk. Such a survey is underway using the UK Infrared Telescope~\citep[UKIRT,][]{Shvartzvald2017,Shvartzvald2018}, but VISTA, an infrared telescope with a better location for bulge observations and a field of view that is ${\sim}3$-times larger than UKIRT's, is not currently conducting observations optimized for microlensing~\citep[though note that the Vista Variables in the Via Lactea (VVV) survey has discovered a number of microlensing events][]{Navarro2017}. In the time it takes the UKIRT survey to build up enough events, progress can be made by analyzing the full data sets of MOA and OGLE surveys. Currently the study measuring optical depths and event rates with the largest number of events analyzes 474 MOA events from the 2007 and 2008 seasons~(\citealt{Sumi2013}, \citealt{Sumi2016}; \citealt{Wyrzykowski2015} only provides the timescale distribution of a larger sample of 3718 events from the OGLE-III survey, and not optical depth and event rate measurements). The current phase of the MOA survey has now been operating for over a decade, and additionally, the OGLE-IV survey has been discovering ${\sim}2000$ events a year since 2011. Analysis of the full data sets of both these surveys would enable measurements of the event rate at much higher spatial resolution than is now possible.

In addition to the event rate, a better understanding is needed of the source magnitude distribution in the infrared. Deep luminosity functions in the $I$ band are available at latitudes of $b\approx -6, -4$ and $-2.7$~\citep{Zoccali2000,Holtzman1998,Calamida2015}, and in the $J$ band at $b\approx-6$~\citep{Zoccali2000}. \wfirst\ will probably observe much closer to the plane, and in addition to bulge stars, there will be a significant contribution from stars in the near and far disk, which will have very different event rates. Understanding the break down of components will require new, deep, infrared magnitude distributions and proper motion measurements from {\it Hubble} and/or the {\it James Webb Space Telescope} (JWST), ideally for several sight lines in the potential \wfirst\ fields. Images produced from special high-density mode scans with {\it Gaia}\footnote{\url{https://www.esa.int/spaceinimages/Images/2017/08/Gaia_sky_mapper_image_near_the_Galactic_centre}} can cover a much larger area than {\it HST} or {\it JWST}, but so far have only been carried out in Baade's window, too far from the expected \wfirst\ fields. Every high-resolution image taken in the \wfirst\ fields before launch can will also provide a ``precovery'' data set for lens mass measurements, extending the baseline over which PSF elongation and color-dependent centroid shifts~\citep[e.g.,][]{Bennett2007} can be measured. A small sample of these could be used as ground truth for the \wfirst\ measurements over the $4.5$~yr mission baseline. An ambitious survey of the entire ${\sim}2$~deg$^2$ \wfirst\ microlensing survey fields, in a similar manner to the Panchromatic Hubble Andromeda Treasury (PHAT) survey~\citep{Dalcanton2012}, would require ${\sim}1500$ {\it HST} pointings, or ${\sim} 750$ {\it JWST} NIRCAM pointings, and would provide immense legacy value for what will become one of the most intensely observed patches of sky~\citep{Yee2014b}. For example, optical photometry from {\it HST} could provide photometric metallicity estimates for every source and a large fraction of lens stars. {\it JWST} 3.6 and $4.5$~$\mu$m imaging could be used to measure star-by star extinctions using the Rayleigh-Jeans color excess~\citep[e.g.,][]{Majewski2011}. Both {\it JWST} and {\it HST} will have roughly twice the angular resolution of \wfirst\ and so their imaging can assist in cases where local, random stellar over densities hamper the interpretation of \wfirst\ images. The improved resolution and increased time baseline would vastly improved measurements of proper motions for all stars in the \wfirst\ field, which would help in the measurement of parallaxes from \wfirst's microlensing data~\citep[]{Gould2015}

There remains significant room for improvement in Galactic models. Fully simulating a \wfirst-like survey requires all the features of a population synthesis model, in order to understand not just the event rates, but the properties of the lenses and sources. The \besancon\ model is presently the only publicly accessible population synthesis model that incorporates kinematics. The latest version of the model\footnote{\url{http://modele2016.obs-besancon.fr/}} has some changes relative to the version we used here, but many of the problems we have identified with it remain (e.g., the small bar angle, too-fast kinematics). The publicly accessible \besancon\ model also lacks any flexibility to adjust model parameters, which is important for maintaining a model in agreement with burgeoning data sets and for understanding the propagation of model uncertainties to yields. The TRILEGAL model~\citep{Girardi2005,Vanhollebeke2009} provides some flexibility to adjust structural parameters, but the publicly available version does not include kinematics.\footnote{\url{http://stev.oapd.inaf.it/cgi-bin/trilegal}} The publicly available Galaxia code~\citep{Sharma2011} implements a version of the \besancon\ model and can also accommodate $N$-body models, potentially providing the necessary flexibility. New, more flexible versions of the \besancon\ model are under development, that improve the evolutionary tracks, add flexibility of the IMF, star formation history, and bar angle~\citep[e.g.,][]{Czekaj2014,Lagarde2017}. {\sc GalMod} is a another new population synthesis model accessed via web forms with significant flexibility in its model parameters~\citep{Pasetto2018}. 

There is also significant work needed on microlensing simulations to better understand the information that \wfirst\ will be able to measure for each planet it finds. This is especially the case for host mass measurements, which will be possible though one or more of the techniques: detecting the host as it separates from the source and measuring image elongation, color-dependent centroid shifts or directly resolving the lens~\citep[e.g.,][]{Bennett2007,Henderson2015-ao,Bhattacharya2017}, measuring the microlensing parallax with or without finite source measurements~\citep[e.g.,][]{Yee2013,Yee2015pieflux,Bachelet2018}, or even measuring astrometric microlensing~\citep[]{Gould2014apie}. The error budget of these measurements is likely to be dominated by systematic errors, and so more detailed end-to-end simulations of the stacking, photometry and astrometry pipelines are likely necessary in order to fully understand \wfirst's capabilities. These simulations will then allow the optimization of \wfirst's survey for characterized planets and not just the total number of detected planets. It will also be important to understand how \wfirst\ performs for more exotic planetary systems (e.g., multiplanet systems, \citealt{Zhu2014}, circumbinary planets \citealt{Luhn2016}, exomoons~\citealt{Liebig2010}, etc.) and for rejecting possible false positive detections caused by, for example, binary source stars~\citep[e.g.,][]{Gaudi1998bs}.

\section{Conclusion}\label{conclusions}

We have performed detailed simulations of several potential designs of the \wfirst\ mission in order to estimate the planet detection yield of its microlensing survey. We derived a correction factor to apply to microlensing event rates computed using the \besancon\ model in order to normalize event rates to those measured by microlensing surveys. Having done so, we estimate that the most recent \wfirst\ design (Cycle 7) will be able to detect ${\sim} 180$ Earth-mass planets and ${\sim}1400$ cold exoplanets in total. For Earth-mass planets, its sensitivity will extend from ${\sim} 1$~AU outwards, and will have a wider range of sensitivity at higher masses. The lower limit of \wfirst's sensitivity for planets in suitable orbits extends down to the mass regime of the solar system's moons, e.g., Ganymede. The mission will fulfill the goals assigned to its microlensing component by the 2010 decadal survey committee to ``determine how common Earth-like planets are over a wide range of orbital parameters.'' However, significant observational, Galactic modeling, and simulation work still needs to be done in order to optimize and fully understand the yields of the survey.

\facility{Exoplanet Archive}

\software{Matplotlib~\citep{matplotlib}, NumPy~\citep{numpy}, SciPy~\citep{scipy}, Astropy~\citep{astropy}, gnuplot, WebbPSF~\citep{WebbPSF}, General Polygon Clipper library~\citep{gpc}, MATLAB package for astronomy and astrophysics~\citep{Ofek2014}.}


This paper is dedicated to the memory of Neil Gehrels, who, as WFIRST Project Scientist, gracefully sheparded this mission from its formal initiation in 2016, until his untimely death. We miss him and his leadership dearly. We would also like to thank members of the \wfirst\ Microlensing Science Investigation Team, as well as Jeff Kruk, Dave Content, Kevin Grady, and many others in the \wfirst\ program office for their support.  We would like to thank Chris Hirata and Jay Anderson for their practical help and discussions.  Finally, we would like to single out David Bennett, who has been a constant source of input and guidance as we have developed our models and worked on this paper.  We recognize that, without his tireless efforts, a microlensing survey on WFIRST would not exist. 

This work was performed in part under contract with the California Institute of Technology (Caltech)/Jet Propulsion Laboratory (JPL) funded by NASA through the Sagan Fellowship Program executed by the NASA Exoplanet Science Institute. MTP and BSG were supported by NASA grants NNX14AF63G and NNG16PJ32C, as well as the Thomas Jefferson Chair for Discovery and Space Exploration. SM was supported by NSFC grants No. 11333003, 11390372, and 11761131004.

\section*{\phantom{0}}
\phantom{0}

\bibliographystyle{mn2e}
\bibliography{libraryshort,apj-jour}

\clearpage
\newpage
\appendix

\section{Improvements to \mabuls{}}\label{mabulsimprovements}
\subsection{Custom bandpasses}\label{customfilters}

\wfirst\ will perform its microlensing survey using a wide infrared filter, covering the full range of detector sensitivity in order to maximize photon count rates. The \besancon\ model provides stellar magnitudes in several standard photometric systems, computed from stellar atmosphere models~\citep{Robin2003}, but it is not easy to compute magnitudes in additional bands. We solved the problem of calculating magnitudes in custom bandpasses by producing what amount to smoothed spectral energy distributions (SEDs) by interpolating between the different available pass bands, and then integrating the product of the smoothed SED with the system throughput curve for each desired custom filter.

The BGM only outputs a limited number of stellar magnitudes out of the whole range available -- in the catalogs we were using $R, I, J$ and $H$ -- so we found it necessary to supplement these with synthesized $K$ and $L$ magnitudes in order to completely cover and bracket the \wfirst\ detector sensitivity range. The magnitudes in these bands were synthesized by assuming that the star was emitting as a black body, and that extinction followed the extinction law listed in the BGM1106 header data, namely $A_K/A_V=0.118$ and $A_L/A_V=0.0$. The smoothed SEDs were interpolated using radial basis functions (RBFs).

\begin{figure}
\includegraphics[width=0.5\columnwidth]{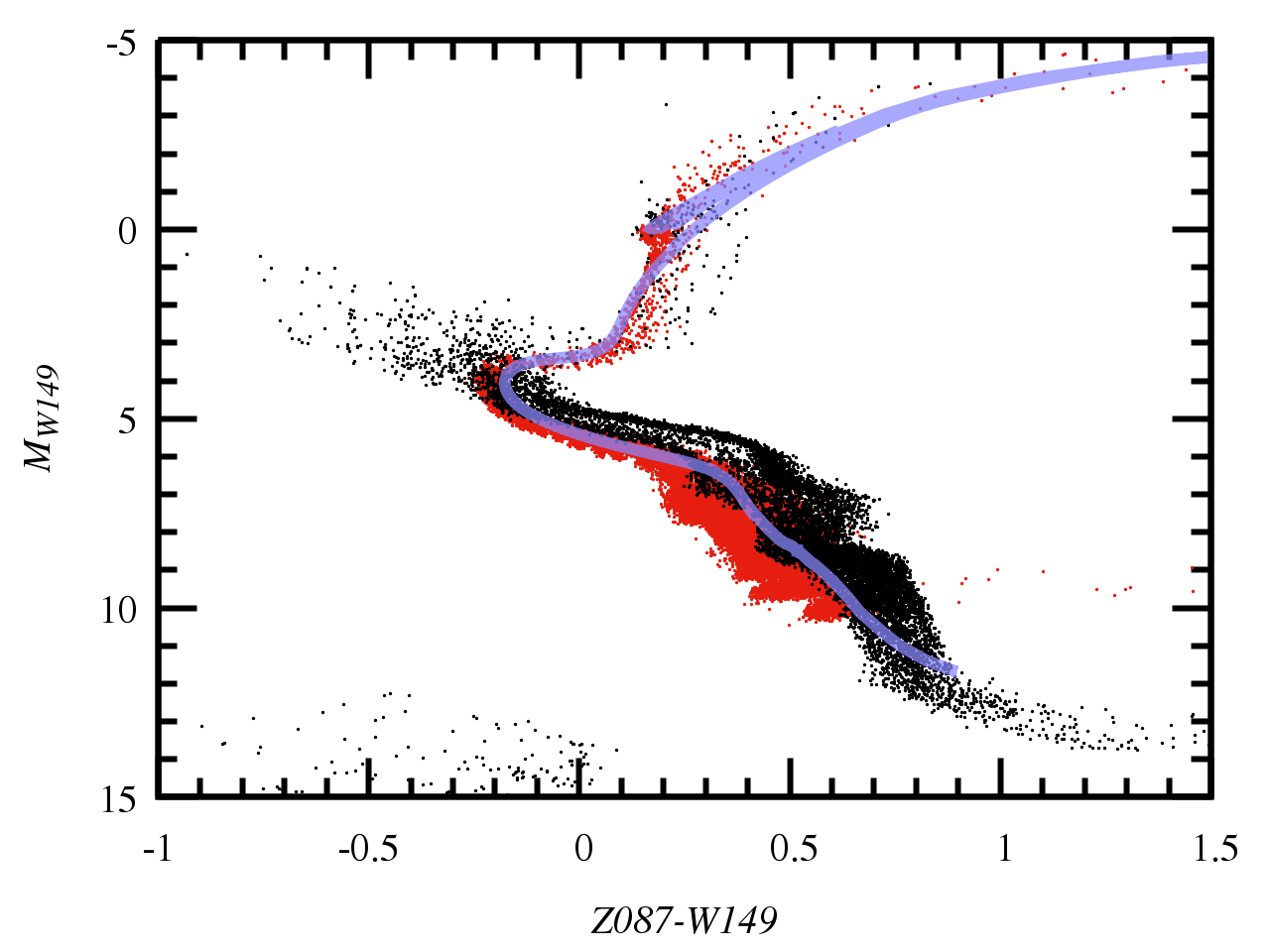}
\caption{The color-absolute magnitude diagram in the principle \wfirst\ microlensing filters, $W149$ and $Z087$. Shown are stars from three BGM1106 sightlines at $\ell=-0.4\degr$ and $b=-3.2, -1.95,$ and $-0.7\degr$, combined, with disk stars plotted with black dots and bulge stars with red dots. The evolutionary tracks and synthetic colors differ between the disk and bulge stars as described by \citet{Robin2003} and references therein. The blue line shows a MIST version 0.3 \citep{Choi2013mist} isochrone for a 10~Gyr, [Fe/H]$=0.0$ population computed using \wfirst\ filter profiles, shifted by $\Delta(Z087-W149)=-0.05$, demonstrating that our scheme for computing \wfirst\ colors and magnitudes works reasonably well. To aid conversion to apparent magnitudes, the distance modulus to the bulge population is approximately $14.5$, and the extinction and reddening will typically be $A_{W149}{\sim}0.5$ and $E(Z087-W149){\sim}0.5$ in the expected \wfirst\ microlensing fields.}
\label{cmd}
\end{figure}

To test of synthetic \wfirst\ magnitudes we combined catalogs from 3 BGM sightlines ($\ell=-0.4\degr$ and $b=-3.2, -1.95,$ and $-0.7\degr$) and mean extinctions to bulge stars of $A_V=2.35, 3.62,$ and $14.37$, respectively. Colors and magnitudes were dereddened using values of $A_{W149}/A_V=0.225$ and $E(Z087-W149)/A_V=0.208$ chosen to align the red clump of each field. These values compare to $0.210$ and $0.295$, respectively, computed at the central wavelength of the filters using the \citet{Cardelli1989} extinction law with $R_V=3.1$; we can expect some difference due to the very wide bandpass of the W149 filter. \autoref{cmd} shows the resulting color-absolute magnitude diagram. The dereddening does not work perfectly, with the turnoff location of bulge stars differing by ${\sim}0.03$ mag in $Z087-W149$ for the different values of extinction. We note that disk stars tend to be redder than the bulge stars, likely due to the different stellar evolution and synthetic photometry models used for each population as described by \citet{Robin2003} and references therein. We also compared the dereddened BGM stars to a MIST version 0.3 isochrone computed using \wfirst\ bandpasses \citep[][]{Choi2013mist} for a 10 Gyr, solar metallicity population. Subtracting 0.05~mag from $Z087-W149$ MIST isochrone brings it into good alignment with the main sequence BGM1106 stars, though the color of the turn-off and giant branch disagree by ${\sim}0.05$~mag in opposite directions; it is possible that differences in stellar evolution codes or the filter transmission curves used for the MIST isochrones could cause these problems. Overall, it appears that our scheme for computing magnitudes in the \wfirst\ bandpasses is reasonable, and should not inject errors significantly larger than the theoretical uncertainties associated with the choice of isochrones.

\subsection{Zodiacal light model}\label{zodi}

Outside the Earth's atmosphere, in the near infrared, the brightest diffuse background is caused by the Zodiacal light -- light from the Sun scattered off interplanetary dust grains. In \citetalias{Penny2013} we assumed this was constant, taking the mean value at the times that \euclid\ could possibly conduct a putative bulge microlensing survey~\citepalias[see][for details of such a survey]{Penny2013}. However, the level of the zodiacal background varies as a function of the elongation of the target fields relative to the Sun, an effect that may become important for long observing seasons. For this reason we incorporate a full-sky model of the zodiacal light, removing the need for the \mabuls\ user to calculate the average level of the zodiacal light in their required bandpasses.

The zodiacal light brightness in a given bandpass is calculated by integrating the RBF interpolated zodiacal light spectrum \citep[as provided by][including solar elongation dependent color terms]{Leinert1998} over the throughput curve of the bandpass. The spatial dependence of the zodiacal light is calculated by RBF interpolation of the map provided by \citet{Leinert1998}.

\subsection{Faster photometry routines}\label{photrout}

In \citetalias{Penny2013} we performed photometry on a pixel-by-pixel noise realization of each image at each epoch. This was computationally expensive, and in certain circumstances was the primary bottleneck of the computation. To speed up the photometry we implemented a routine that takes as input a noiseless realization of the baseline image, accounts for blending, and returns a simple function to compute the photometric signal and uncertainty as a function of magnification. With this, only a single realization of the noise on the photometric data point is needed. The routine also solves for the magnification at which saturation is reached in one of the pixels of the aperture, allowing saturation to be identified accurately without building a realization of an image. 

\subsection{Improved observer-centric velocities/timescales}\label{obstE}

In order to include parallax effects in lightcurves we now compute geocentric, or more accurately observer-centric, microlensing event timescales.  In \citepalias{Penny2013}, we used heliocentric velocities to compute event timescales. Due to the observer's motion about the Sun (typically the Earth's, which ranges from -30 to 30~km~s$^{-1}$ projected onto the sky), the relative source-lens velocity will change over the course of the year. This compares to the ${\sim}200$--$1000$~km~s$^{-1}$ projected velocities of the source and lens~\citep[see, e.g.,][]{CalchiNovati2015}, implying a typical modification of the timescales by ${\sim}3$--$15$~percent. However, for both \euclid\ and \wfirst\, microlensing observations will be made at or near quadrature, meaning that the projected velocity of the Earth will be close to zero. The effect on planet yields of using the improved timescales is therefore likely to be very small. Some of the simulations presented here use the improved timescales, while others were completed before the improved timescale calculation was implemented. Full details are given by \citet{Penny2017}.

\section{Computing the mass-semimajor axis sensitivity curve or ``The making of Figure~9''}\label{curvecomp}
In order to compute the sensitivity curve shown in \autoref{sensitivity} required computing the planet detection rate on a grid of planet mass and semimajor axis, spaced by 0.25 and 0.125~dex, respectively. Obtaining reasonably accurate results is computationally intensive, with the required computation increasing as one over the square of the detection efficiency in order to achieve equal Poisson statistical uncertainties. With $\tzero$ drawn from any point in the 5 year mission, and a ${\sim}24$\% observing duty cycle, the detection efficiency is ${\sim}10^{-5}$ at the 3-detection line. This implies that a 10~percent statistical error would require ${\sim}100/10^{-5}=10^7$ lightcurves to be generated at each grid point near the sensitivity curve, with most of these showing no detection. In order to make the computation tractable we developed the CROIN parametrization~\citep{Penny2014}, which is used to generate only lightcurves where there is a reasonable probability of a planet detection. This coordinate system is centered on the planetary caustic(s), and the region around the caustic that contains a detectable planetary signature is a circle of radius $r_{\rm c}(s,q)$, whose analytic functional dependence on the projected separation relative to the Einstein radius $s$ and mass ratio $q$ is derived empirically by \citet{Penny2014}. Only source trajectories with impact parameters relative to the planetary caustic $u_{\rm c}<r_{\rm c}$ are simulated. We used the CROIN parametrization for planet masses $M\le 10\mearth$ and for larger masses with $\log (a/\text{AU}) \ge 1.125$, reducing the number of required lightcurve computations by more than two orders of magnitude. When using the CROIN parametrization we still require that the main-event impact parameter and peak time obey $-3\leq\uzero<3$ and $0\leq\tzero<2011$.  

For the low-mass planets that \wfirst\ is sensitive to, computing the lightcurve is not a trivial operation and is prone to numerical errors. For its speed, we primarily relied on a contour integration code~\citep{Gould1997, Dominik1998} written by S. Mao. This solves the complex 5$^{\text{th}}$-order binary lens polynomial at many points around the source circumference and then links the resulting solutions into a number of potentially merging images. The coefficients of the polynomial have additive terms of the order of 1, and various combinations of powers of $q$ and $s$. For the lowest mass planets we consider, $q\sim10^{-8}$, so we suspect that the numerical errors are a result of catastrophic cancellations in parts of the calculation where this is difficult if not impossible to avoid. The vast majority of errors are caught by error handling routines, and when this occurs the lightcurve is passed to a much slower but more robust inverse ray shooting routine~\citep[e.g.][]{Kayser1986}, but occasionally errors slip past the error handing routines.

\begin{figure*}
\includegraphics[width=0.49\textwidth]{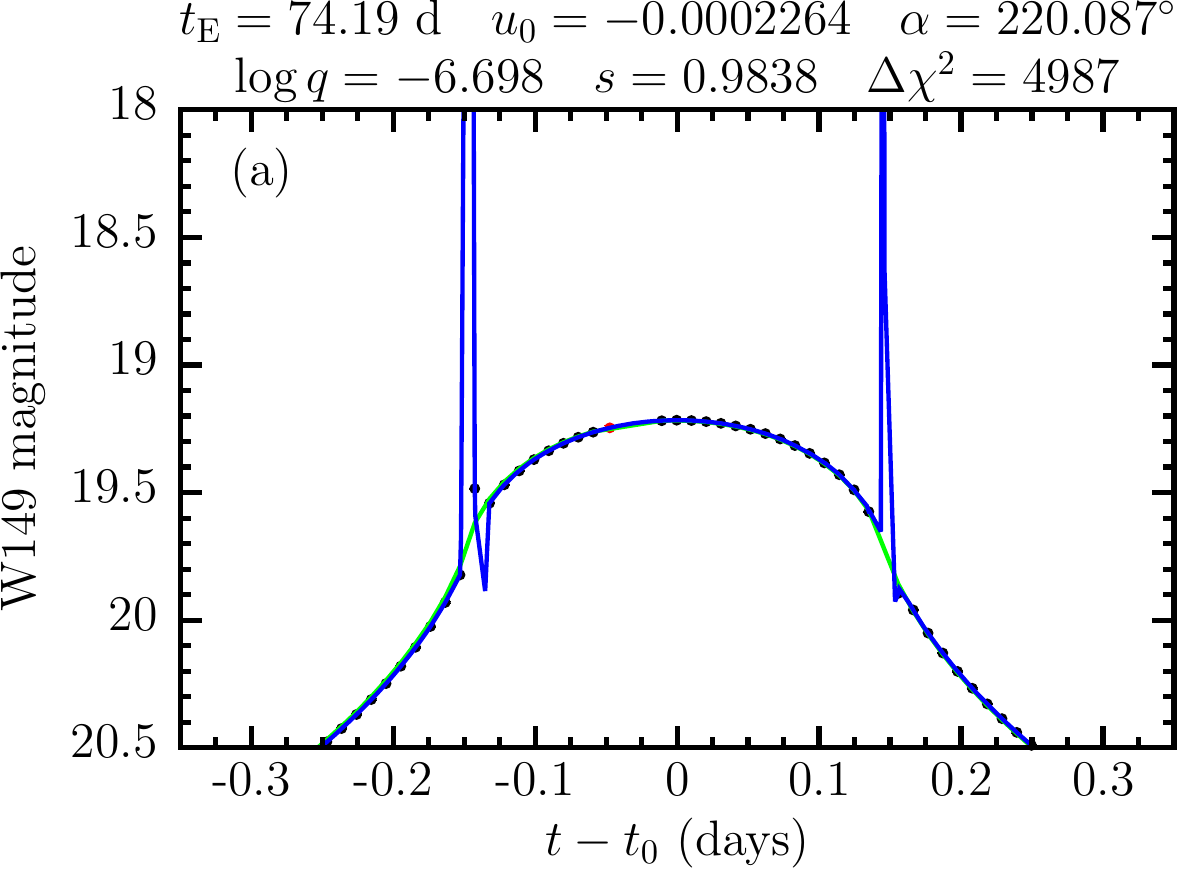}\hspace{0.02\textwidth}
\includegraphics[width=0.49\textwidth]{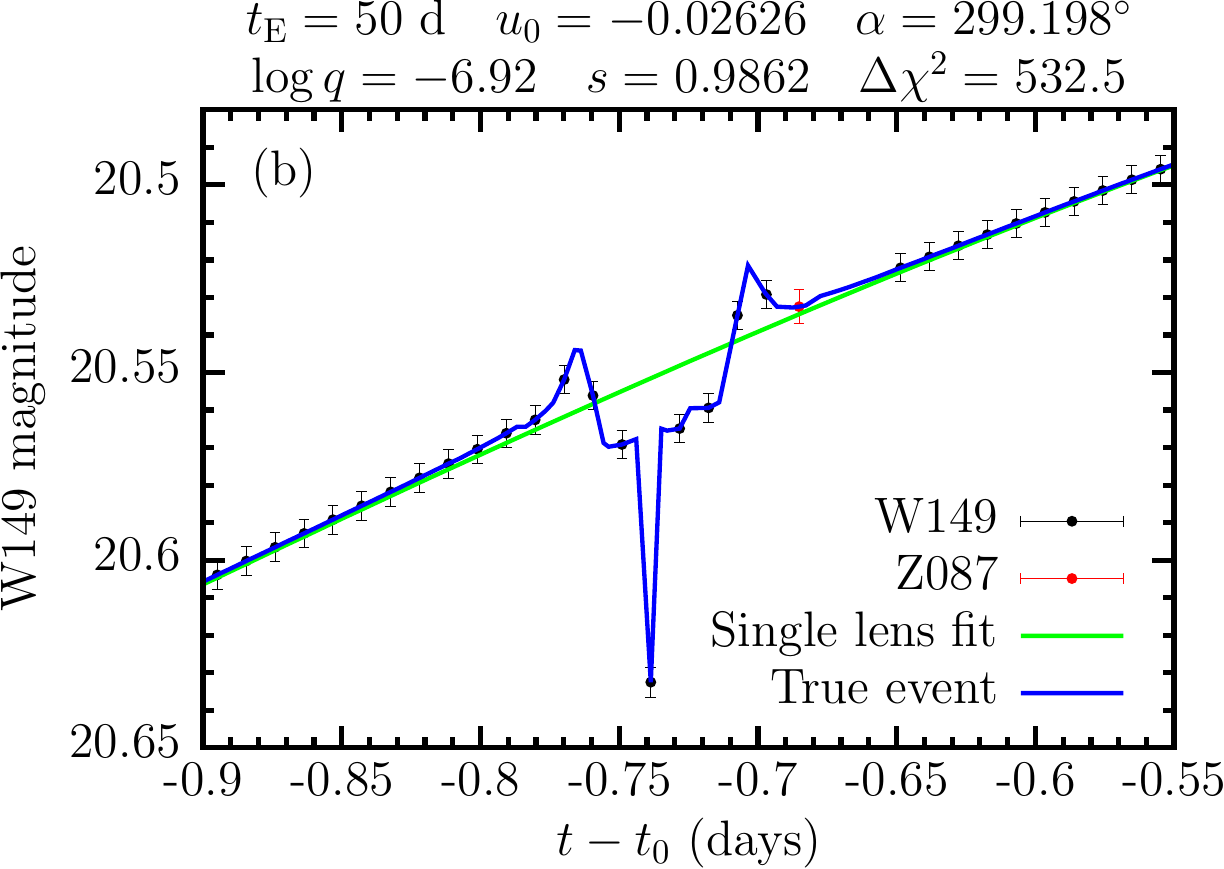}\vspace{12pt}\\
\includegraphics[width=0.49\textwidth]{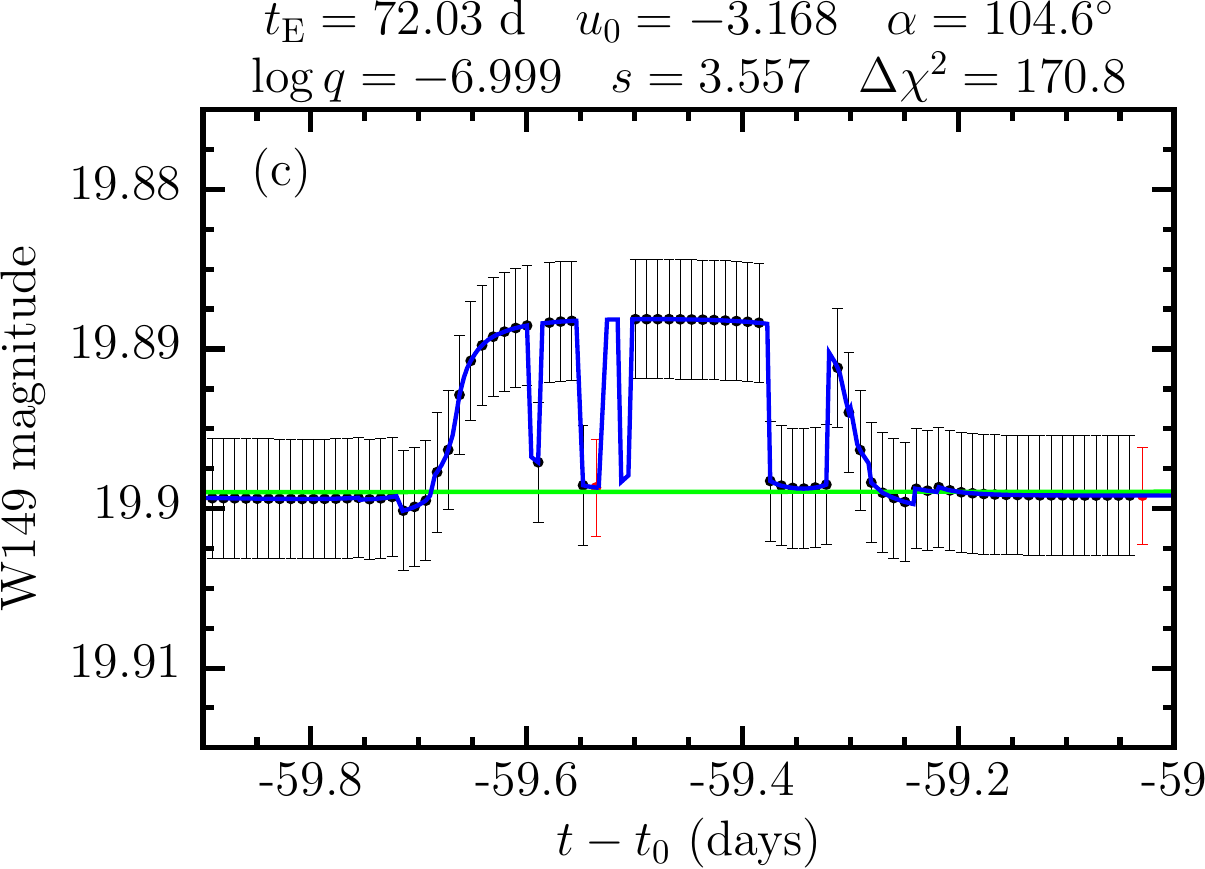}\hspace{0.02\textwidth}
\includegraphics[width=0.49\textwidth]{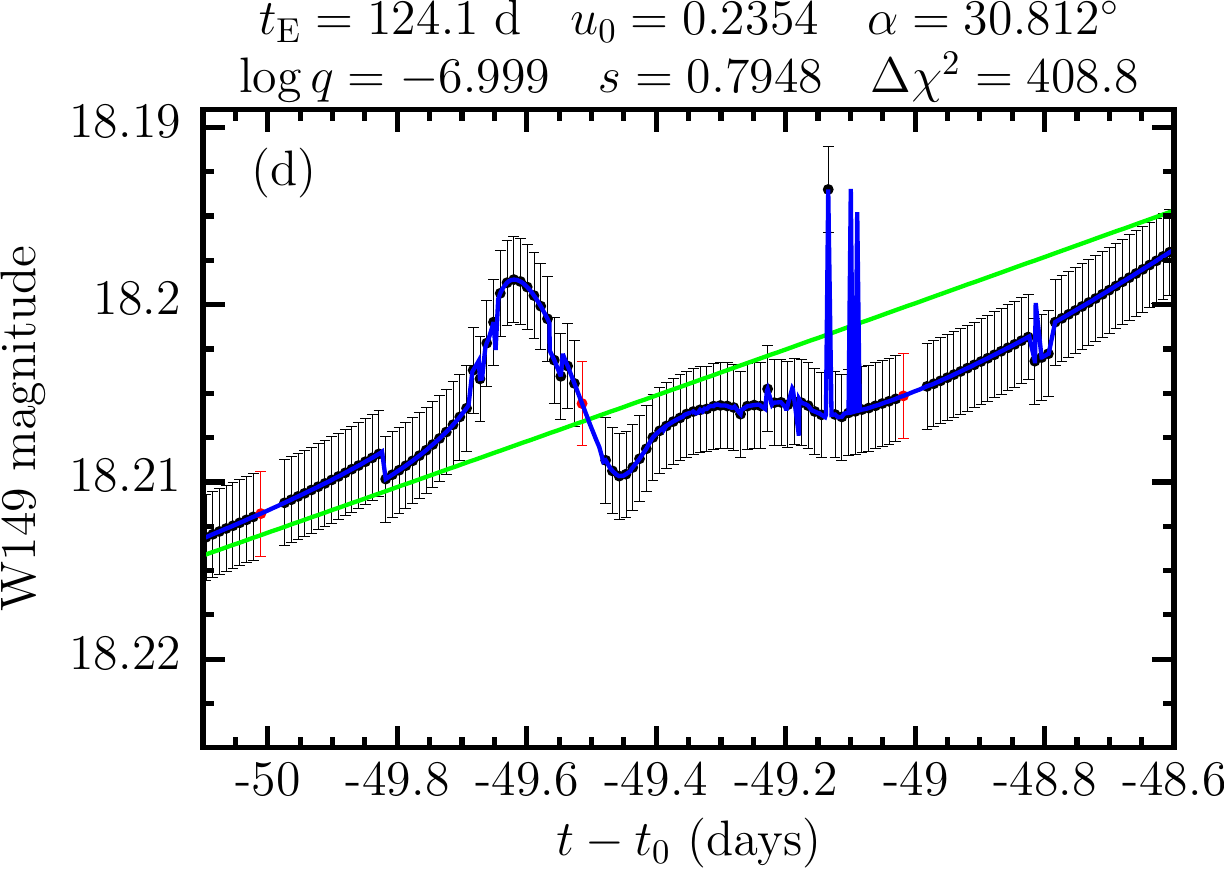}
\caption{Examples of lightcurves exhibiting numerical errors that were not caught by error reporting routines. Each type of error is discussed in the text. Data points with error bars show the predicted magnitude and uncertainty of measurements without any noise, the blue line shows the simulated event, including numerical errors, and the green line shows the best fit single lens model. Parameters for each event shown are listed above each plot; $\alpha$ is the angle subtended by the source trajectory relative to the binary axis.}
\label{badlc}
\end{figure*}

In order to make sure that these numerical errors were not significantly affecting results we visually inspected a large sample of the lightcurves of the lowest mass exoplanet detections. We found examples of errors that \emph{did} cause false positive detections and \emph{could} cause false negative detections. As our simulations only output the lightcurves of a sample of planet \emph{detections}, we are not able to assess the degree to which our predicted yields are reduced by false negatives. As we correct for false positives (see below) and not for false negatives our planet yield predictions at the lowest masses are likely to be conservative. \autoref{badlc} shows four examples that represent the overwhelming majority of the errors we found -- two are false positives, and two were errors that did not affect the designation of the event as a detection, but had the errors been more severe one of these would have resulted in a false negative. 

The first example (a) shows a type of false positive that only occurs in high-magnification events where the source is resolved by the magnification structure surrounding the host star. Most of these examples had only a single discrepant data point. However, due to the high photometric precision that high-magnification events enable, the discrepant point causes a large $\Delta\chi^2$.

The second example of false positive (b) is one of a more general class where a significantly discrepant data point (or several) occurs during a planetary anomaly. They can be either positive or negative deviations, and are typically sharp changes relative to the source crossing time. The events with these numerical errors are only classified as false positives if removal of the discrepant points would move the event below the $\Delta\chi^2$ threshold.

A potential false negative non-detection (c), which we call a ``finite source drop-out,'' occurs for events with wide-separation planets that show a small amplitude top-hat planetary signature due to extreme finite source effects. They are caused when two images near the planet are both incorrectly flagged as false solutions to the the lens equation, leaving three valid solutions instead of five.\footnote{The binary lens equation will have either 3 or 5 solutions depending on whether the source is inside or outside a caustic, but the complex 5$^{\text{th}}$-order polynomial that can be formed by rearranging the lens equation always has 5 solutions. The validity of each solution is checked by inverse shooting a ray from each candidate solution and checking the proximity of the ray to the source position from which it should have originated.} In principle it is possible for all the data points during a planetary anomaly to experience this problem, in which case the event would not count as a planet detection and the simulation would not output the lightcurve. It is therefore impossible to estimate the number of these occurrences by our current method of inspection. However, we can guess that the number is likely small because most instances of drop-outs show only a small fraction of the top hat dropping out. 

The final example (d) of numerical errors is likely very similar to the finite source drop-out but occurs for all separations. Again, the magnification is artificially reduced, but usually by an insignificant amount that does not affect the event's status as a detection or non-detection.

\begin{figure}
\includegraphics[width=\columnwidth]{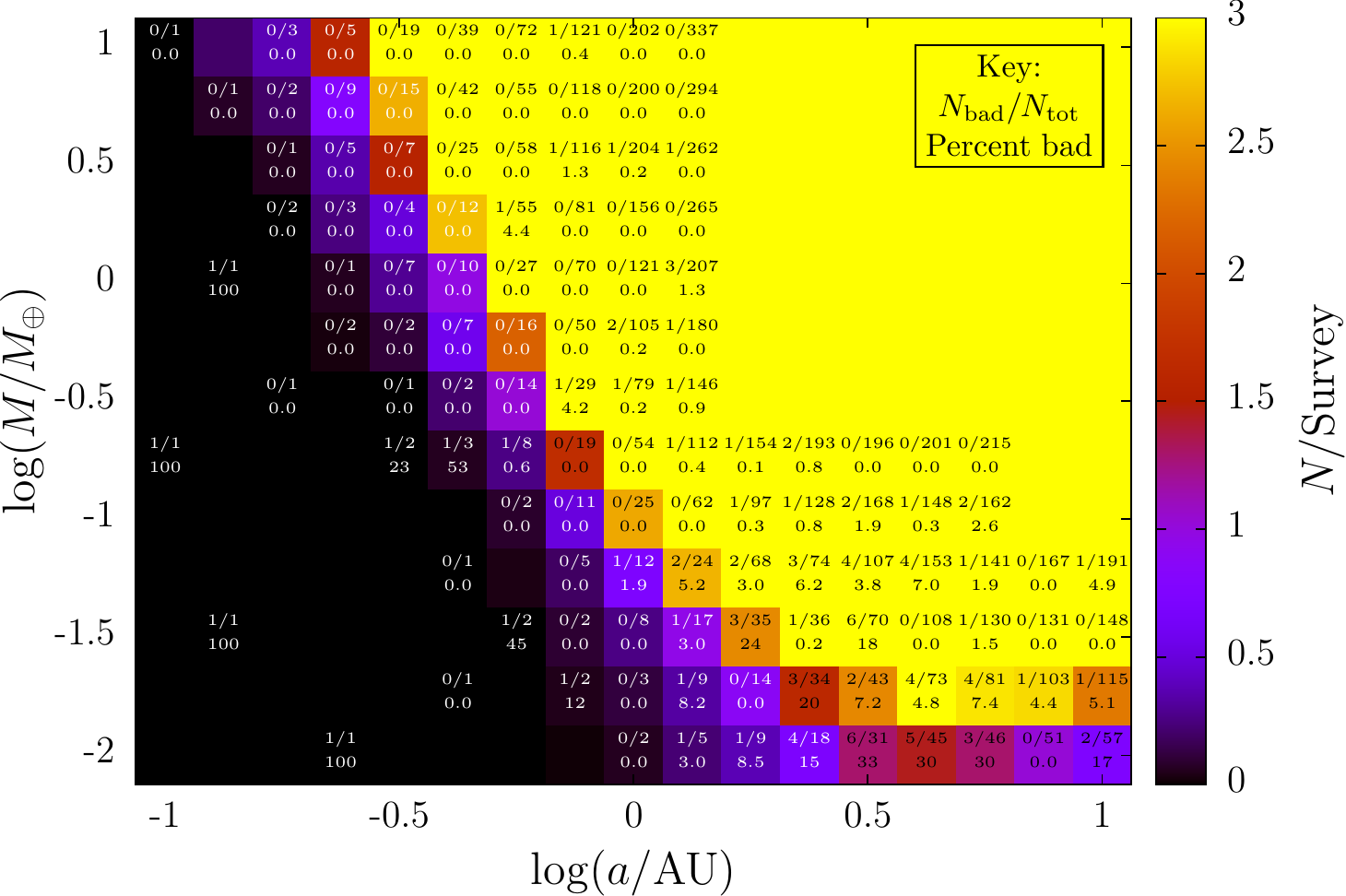}
\caption{Fraction of false positive planet detections for points on the planet mass-semimajor axis grid. In each cell the fraction is the number of false positive lightcurves over the number of inspected lightcurves. The number below that is the percentage of false positives, weighted by the event rate as described in \autoref{sims}. The shading is a linear scale that saturates to yellow at 3 detections during the survey, before correction for the false positive rate.}
\label{fpgrid}
\end{figure}

\begin{figure}
\includegraphics[width=0.5\columnwidth]{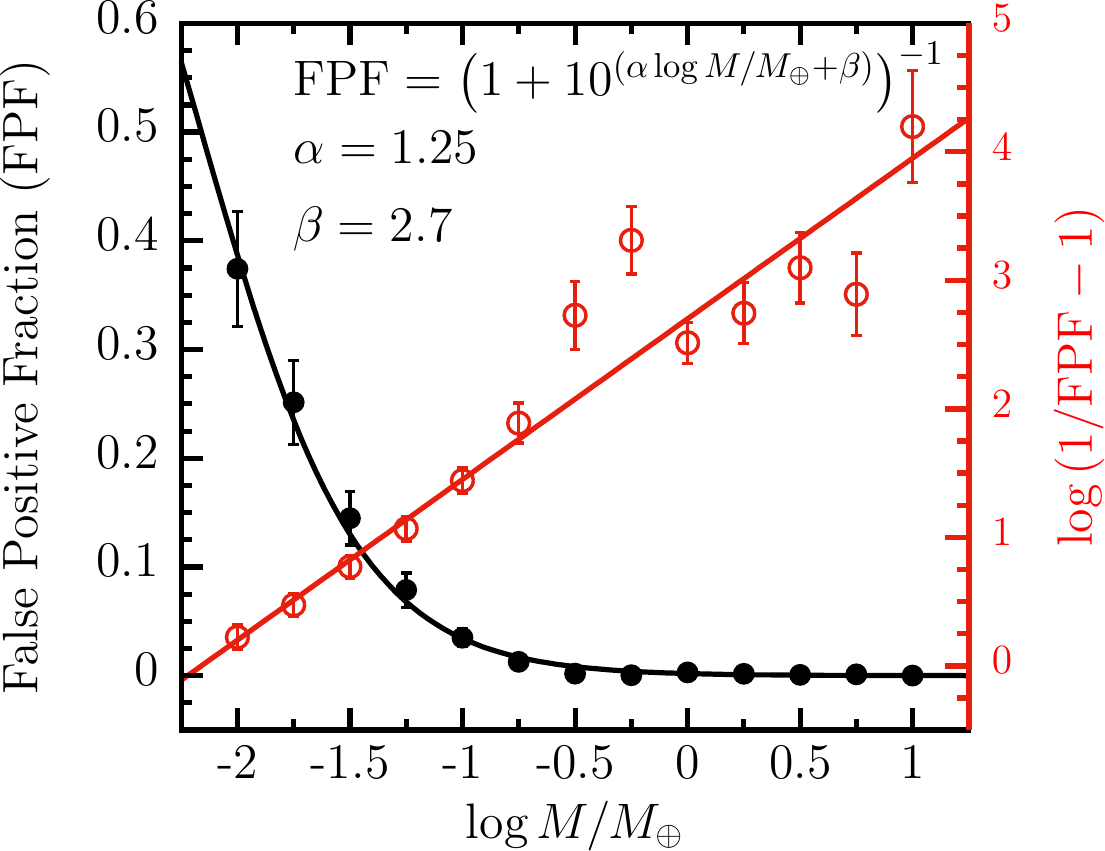}
\caption{Average rate-weighted false-positive fraction as a function of planet mass. Filled black points show the rate-weighted false-positive fraction as measured by inspecting lightcurves. The open red points show the same data but showing the quantity $\log(1/\text{FPF}-1)$ plotted against the axis on the right. The lines show the same best fit analytic model.}
\label{fpfit}
\end{figure}

The false positive fraction becomes significant as the planet mass decreases, so it is important that we correct for it. \autoref{fpgrid} shows the event-rate weighted (see \autoref{sims}) false positive fraction for points on the planet mass-semimajor axis grid, along with the number of events inspected and the number of events that were false positives. The initial intention was to correct each point on the grid individually by its own false positive rate, but there were not enough lightcurves output by the simulation for this to be accurate. Instead, being as the false positive rate seemed to be relatively independent of semimajor axis (within the large error bars), we took the rate weighted average of all semimajor axes at each value of the mass. The false positive fractions can be transformed to a form that is roughly a single power law in mass
\begin{equation}
\log\left(\frac{1}{\text{FPF}}-1\right) = \alpha\log(M/\mearth) + \beta,
\label{fpeq}
\end{equation}
where $\text{FPF}$ is the false positive fraction and $\alpha=1.25$ and $\beta=2.7$ are the best fit linear regression parameters. This is shown in \autoref{fpfit}.

\begin{figure}
\includegraphics[width=0.5\columnwidth]{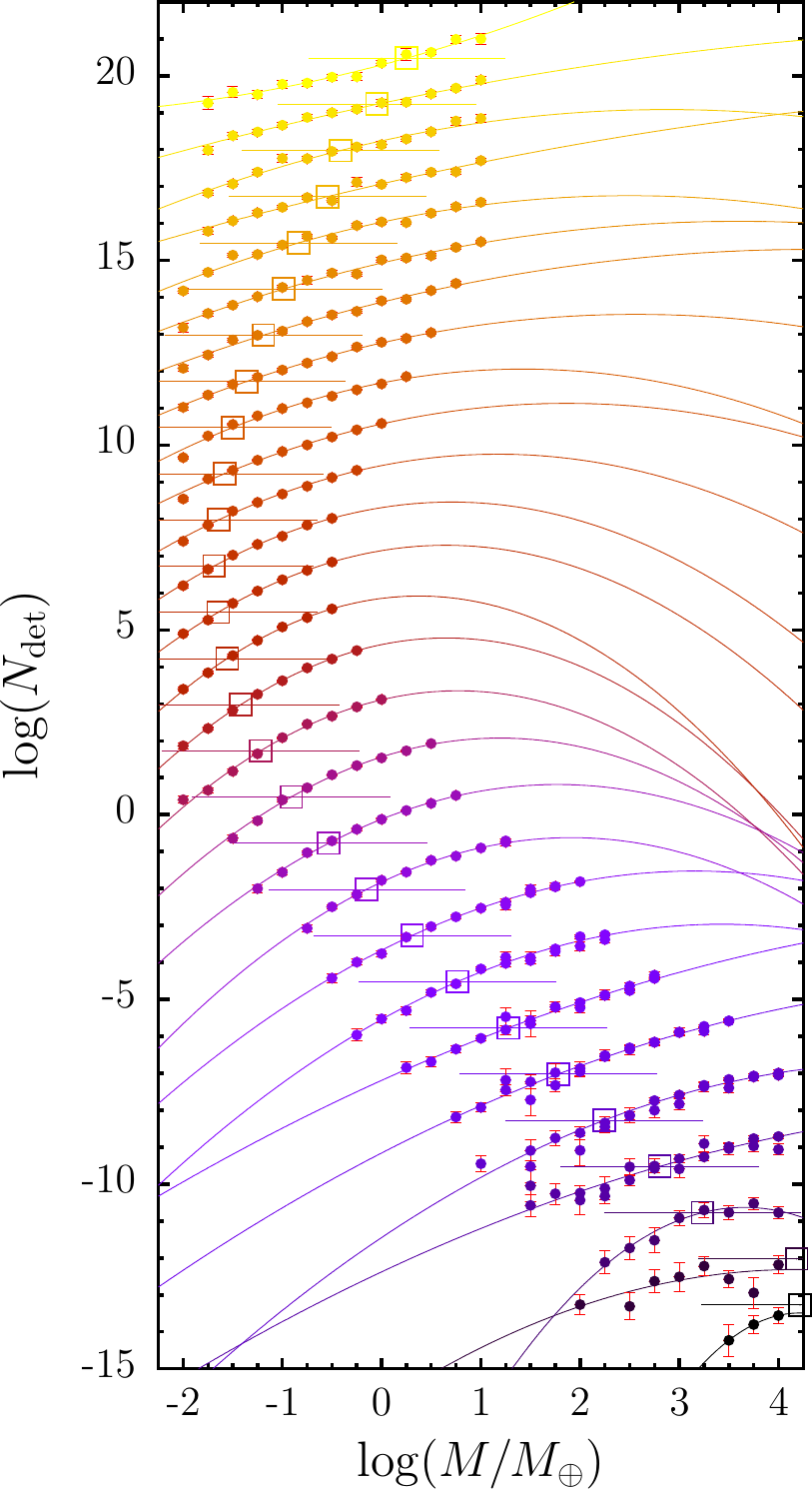}
\caption{False-positive-corrected planet detection rates plotted as a function of mass for each of the semimajor axis grid points. The detection rates have been shifted vertically by an amount $10\log(a/\text{AU})$, and are also color-coded by semimajor axis with black corresponding to $a=0.01$~AU and yellow to $a=100$~AU. Curved lines are quadratic fits to the points with detection rates within a factor of $30$ of the $3$~detections per survey threshold that is plotted in \autoref{sensitivity}. For each semimajor axis we plot the $3$-detection threshold as a horizontal line and an open square point marking the position at which the quadratic curve crosses the threshold.}
\label{fitmap}
\end{figure}

Returning to the problem of estimating \wfirst's sensitivity in the mass-semimajor axis planet, we correct the gridded planet detection rates using \autoref{fpeq}. For a given semimajor axis, to find the mass at which the planet detection rate crosses the 3 detection threshold we fit a quadratic polynomial in $\log(M/\mearth)$ to the log of the planet detection rates that are within a factor of 30 of the $3$~per survey threshold, and then solve the resulting quadratic equation (taking care to select the appropriate root). We repeat this process for each semimajor axis grid point as is shown in \autoref{fitmap}. We excluded the points with $\log(a/\text{AU})=-2$ and $-1.875$ from the analysis because our grid did not extend to high enough masses to properly bracket the point at which 3 detections are expected.

\end{document}